\documentclass[a4paper,11pt]{article}
\usepackage[compatibility=false]{caption}

\usepackage{jheppub}
\usepackage{slashed}
\usepackage{subfigure}
\usepackage{xcolor}
\usepackage{booktabs}
\usepackage{adjustbox}
\usepackage{dsfont}
\usepackage{ulem}
\usepackage{extarrows,cancel}
\usepackage{mathtools}
\usepackage{bm}       
\usepackage{bbm} 
\usepackage{multirow}
\usepackage[T1]{fontenc}
\usepackage{array}
\usepackage{tabularx}
\usepackage{verbatim}
\usepackage{url}
\usepackage{orcidlink}
\usepackage{lineno}

\usepackage{float}

\usepackage{tikz}
\usepackage{tikz-feynman}
\usetikzlibrary{decorations.pathmorphing}
\usepackage{amsmath}
\usepackage{subcaption} 
\usetikzlibrary{arrows.meta, decorations.markings}
\usetikzlibrary{positioning}
\tikzfeynmanset{
	fermion/.style={
		/tikz/postaction={
			/tikz/decoration={
				markings,
				mark=at position 0.6 with {\arrow{Triangle[length=6pt, width=4.5pt]}},
			},
			/tikz/decorate=true,
		},
	},
	fermion-open/.style={
		/tikz/postaction={
			/tikz/decoration={
				markings,
				mark=at position 0.6 with {\arrow{Triangle[open,length=6pt, width=4.5pt]}},
			},
			/tikz/decorate=true,
		},
	},
}
\makeatletter
\gdef\@fpheader{}
\makeatother

\newcommand{\MPP}{M_\phi^2}
\newcommand{\MSbar}{\overline{\text{MS}}}

\newcommand{\epsabc}{\epsilon^{\alpha\beta\gamma}}
\newcommand{\BtoDL}{\bar{B}_{(s)}^0 \to D_{(s)}^{(*)+} L^-}
\newcommand\Bdecay[3][\bar{B}^0]{#1 \to #2 #3}
\newcommand{\asdpi}{\frac{\alpha_s}{4\pi}}

\newcommand{\lamL}{\lambda^L_{ij}}
\newcommand{\lamR}{\lambda^R_{ij}}
\newcommand{\lamLx}{\lambda^{L*}_{kl}}
\newcommand{\lamRx}{\lambda^{R*}_{kl}}

\newcommand{\QaSLLq}{\tilde{\mathcal{Q}}_1^{SLL}}
\newcommand{\QbSLLq}{\tilde{\mathcal{Q}}_2^{SLL}}
\newcommand{\QcSLLq}{\tilde{\mathcal{Q}}_3^{SLL}}
\newcommand{\QdSLLq}{\tilde{\mathcal{Q}}_4^{SLL}}
\newcommand{\QaSLRq}{\tilde{\mathcal{Q}}_1^{SLR}}
\newcommand{\QbSLRq}{\tilde{\mathcal{Q}}_2^{SLR}}
\newcommand{\QaSLLqga}{\tilde{\mathcal{Q}}_{\alpha1}^{SLL}}
\newcommand{\QbSLLqga}{\tilde{\mathcal{Q}}_{\alpha2}^{SLL}}

\newcommand{\QaSLRqga}{\tilde{\mathcal{Q}}_{\alpha1}^{SLR}}
\newcommand{\QbSLRqga}{\tilde{\mathcal{Q}}_{\alpha2}^{SLR}}
\newcommand{\QaSLLqgb}{\tilde{\mathcal{Q}}_{\beta1}^{SLL}}
\newcommand{\QbSLLqgb}{\tilde{\mathcal{Q}}_{\beta2}^{SLL}}

\newcommand{\QaSLRqgb}{\tilde{\mathcal{Q}}_{\beta1}^{SLR}}
\newcommand{\QbSLRqgb}{\tilde{\mathcal{Q}}_{\beta2}^{SLR}}

\newcommand{\SaSLLq}{\left\langle\tilde{\mathcal{Q}}_1^{SLL}\right\rangle}
\newcommand{\SbSLLq}{\left\langle\tilde{\mathcal{Q}}_2^{SLL}\right\rangle}
\newcommand{\ScSLLq}{\left\langle\tilde{\mathcal{Q}}_3^{SLL}\right\rangle}
\newcommand{\SdSLLq}{\left\langle\tilde{\mathcal{Q}}_4^{SLL}\right\rangle}
\newcommand{\SaSLRq}{\left\langle\tilde{\mathcal{Q}}_1^{SLR}\right\rangle}
\newcommand{\SbSLRq}{\left\langle\tilde{\mathcal{Q}}_2^{SLR}\right\rangle}

\newcommand{\SaSLLqga}{\left\langle
	\tilde{\mathcal{Q}}_{\alpha1}^{SLL}\right\rangle}
\newcommand{\SbSLLqga}{\left\langle
	\tilde{\mathcal{Q}}_{\alpha2}^{SLL}\right\rangle}

\newcommand{\SaSLRqga}{\left\langle
	\tilde{\mathcal{Q}}_{\alpha1}^{SLR}\right\rangle}
\newcommand{\SbSLRqga}{\left\langle
	\tilde{\mathcal{Q}}_{\alpha2}^{SLR}\right\rangle}
\newcommand{\SaSLLqgb}{\left\langle
	\tilde{\mathcal{Q}}_{\beta1}^{SLL}\right\rangle}
\newcommand{\SbSLLqgb}{\left\langle
	\tilde{\mathcal{Q}}_{\beta2}^{SLL}\right\rangle}

\newcommand{\SaSLRqgb}{\left\langle
	\tilde{\mathcal{Q}}_{\beta1}^{SLR}\right\rangle}
\newcommand{\SbSLRqgb}{\left\langle
	\tilde{\mathcal{Q}}_{\beta2}^{SLR}\right\rangle}

\newcommand{\CaSLL}{C_1^{SLL}}
\newcommand{\CbSLL}{C_2^{SLL}}
\newcommand{\CcSLL}{C_3^{SLL}}
\newcommand{\CdSLL}{C_4^{SLL}}
\newcommand{\CaVLL}{C_1^{VLL}}
\newcommand{\CbVLL}{C_2^{VLL}}
\newcommand{\QaSLL}{\mathcal{Q}_1^{SLL}}
\newcommand{\QbSLL}{\mathcal{Q}_2^{SLL}}
\newcommand{\QcSLL}{\mathcal{Q}_3^{SLL}}
\newcommand{\QdSLL}{\mathcal{Q}_4^{SLL}}
\newcommand{\QaVLL}{\mathcal{Q}_1^{VLL}}
\newcommand{\QbVLL}{\mathcal{Q}_2^{VLL}}

\newcommand{\SaSLL}{\left\langle\mathcal{Q}_1^{SLL}\right\rangle}
\newcommand{\SbSLL}{\left\langle\mathcal{Q}_2^{SLL}\right\rangle}
\newcommand{\ScSLL}{\left\langle\mathcal{Q}_3^{SLL}\right\rangle}
\newcommand{\SdSLL}{\left\langle\mathcal{Q}_4^{SLL}\right\rangle}
\newcommand{\SaVLL}{\left\langle\mathcal{Q}_1^{VLL}\right\rangle}
\newcommand{\SbVLL}{\left\langle\mathcal{Q}_2^{VLL}\right\rangle}

\newcommand{\lamPLL}{\Delta^{LL}_d}
\newcommand{\lamPLR}{\Delta^{LR}_d}
\newcommand{\lamPRL}{\Delta^{RL}_d}
\newcommand{\lamPRR}{\Delta^{RR}_d}
\newcommand{\lamKLL}{\Delta^{LL}_s}
\newcommand{\lamKLR}{\Delta^{LR}_s}
\newcommand{\lamKRL}{\Delta^{RL}_s}
\newcommand{\lamKRR}{\Delta^{RR}_s}

\newcommand{\Atree}{\mathcal{A}^\text{full}_\text{tree}}
\newcommand\epsn[1]{\frac{#1}{\varepsilon}}
\newcommand\epsnIR[1]{\frac{#1}{\varepsilon_{IR}}}
\newcommand\lu{L(\mu)}
\newcommand{\IRpart}{\mathrm{IRpart}}

\newcommand{\hc}{\mathrm{h.c.}}

\allowdisplaybreaks[4]

\title{\boldmath Phenomenological study of diquark effects in $\bar{B}_{(s)}^0 \to D_{(s)}^{(*)+} L^-$ decays}

\author[b]{Fang-Min Cai,}
\author[a,1]{Xin-Qiang Li,\note{Corresponding author.}}
\author[a]{Tong-Kai Liu,}
\author[a]{Yan Shi,}
\author[a,b]{and Ya-Dong Yang}

\affiliation[a]{Institute of Particle Physics and Key Laboratory of Quark and Lepton Physics~(MOE), Central China Normal University, Wuhan, Hubei 430079, China}
\affiliation[b]{Institute of Particle and Nuclear Physics, Henan Normal University, Xinxiang 453007, China}

\emailAdd{caifangmin@htu.edu.cn}
\emailAdd{xqli@mail.ccnu.edu.cn}
\emailAdd{liutik@mails.ccnu.edu.cn}
\emailAdd{shiyan@mails.ccnu.edu.cn}
\emailAdd{yangyd@ccnu.edu.cn}

\abstract{Motivated by the deviations between experimental data and Standard Model (SM) predictions for the branching fractions of $\bar{B}_{(s)}^{0} \to D_{(s)}^{(*)+}L^-$ decays with $L=\pi,K^{(*)},\rho$, we investigate whether the type-IV scalar diquark could offer a plausible New Physics (NP) explanation for them. By matching the renormalized amplitudes in the full theory onto those in the low-energy effective theory formulated in the diquark operator basis, we extract the corresponding Wilson coefficients up to next-to-leading order in $\alpha_s$ at the NP scale $\mu_\phi=M_{\phi}$. We then rewrite the obtained operators in the SM basis and apply the renormalization-group running to evolve the Wilson coefficients down to the hadronic scale $\mu_b=m_b$. A coherent effective-field-theory description is therefore constructed for the diquark-mediated $\Delta B=1$ non-leptonic processes. For the hadronic matrix elements of the relevant NP four-quark operators, we employ the QCD factorization approach to evaluate the non-factorizable one-loop vertex corrections, and confront our theoretical predictions to the latest experimental data to constrain the diquark parameter space. In the single-coupling case, we find that the purely left-handed coupling $\Delta_{d(s)}^{LL}$ can reconcile the discrepancies at the $1\sigma$ level, while the mixed right-left coupling $\Delta_{d(s)}^{RL}$ remains compatible with data only at about $2\sigma$. Extending to the two-coupling case, we find that sizable parameter regions remain viable as long as $\Delta_{d(s)}^{LL}\neq 0$. Incorporating the LHCb measurement of the ratio $R_{\pi/\mu\nu_{\mu}} = \mathcal{B}(B_{c}^{-}\to J/\psi \pi^{-})/\mathcal{B}(B_{c}^{-}\to J/\psi \mu^{-}\bar{\nu}_{\mu})$ further reduces the allowed range of $\Delta_{d}^{LL}$, but does not yield a meaningful extra restriction on $\Delta_{d}^{RL}$. Hence, while either $\Delta_{d(s)}^{LL}$ or $\Delta_{d(s)}^{RL}$ can separately provide an acceptable NP interpretation, our phenomenological results favor a scenario dominated by purely left-handed diquark-quark couplings. Finally, within the surviving parameter space, we present updated predictions for the branching fractions and the ratios $R_{(s)L}^{(*)} = \Gamma(\bar{B}^0_{(s)} \to D_{(s)}^{(*)+} L^-)/d\Gamma(\bar{B}^0_{(s)}\to  D_{(s)}^{(*)+} \ell^-\bar{\nu}_{\ell})/dq^2\mid_{q^2=m_L^2}$, which can be probed by forthcoming high-precision measurements from LHCb and Belle II.}

\begin{document}
\maketitle
\flushbottom

\section{Introduction}
\label{sec:intro}

Despite the remarkable success of the Standard Model (SM) in accounting for elementary particles and their interactions, several features of the universe, e.g. neutrino masses, dark matter, matter-antimatter asymmetry, remain unaccounted for by it, motivating searches for New Physics (NP) beyond it. Since direct searches at the Large Hadron Collider (LHC)~\cite{ParticleDataGroup:2026aaa} have not yet uncovered unambiguous NP signatures, indirect probes through high-precision flavor physics have gained increasing importance~\cite{Buchalla:2008jp,Antonelli:2009ws,HeavyFlavorAveragingGroupHFLAV:2024ctg,FlavourLatticeAveragingGroupFLAG:2024oxs}. In this context, non-leptonic $B$-meson decays, which are featured by many interesting observables, provide a particularly powerful laboratory for scrutinizing the SM and exploring NP, yielding distinctive information on the interplay between weak and strong interactions~\cite{BaBar:2014omp,Belle-II:2018jsg,LHCb:2018roe}.

Theoretical predictions for non-leptonic $B$-meson decays are intrinsically challenging because several distinct energy scales enter simultaneously. A standard way to handle this is to use the effective weak Hamiltonian framework~\cite{Buchalla:1995vs,Buras:1998raa}, which separates the decay amplitude into Wilson coefficients that can be computed perturbatively and hadronic matrix elements of local operators. To evaluate these matrix elements, we employ the QCD factorization (QCDF) formalism~\cite{Beneke:1999br,Beneke:2000ry,Beneke:2001ev}. In the heavy-quark limit, the QCDF permits writing the matrix elements as a product of transition form factors and convolutions of light-cone distribution amplitudes (LCDAs) with hard-scattering kernels~\cite{Beneke:1999br,Beneke:2000ry,Beneke:2001ev,Beneke:2003zv,Beneke:2006hg}. Equivalent factorization results can be derived in the soft-collinear effective theory~\cite{Bauer:2000yr,Bauer:2001yt,Beneke:2002ph,Beneke:2002ni,Bauer:2001cu}, the field-theoretic formulation underlying the QCDF, which provides a systematic separation of short- and long-distance dynamics in the heavy-quark limit and thereby enables the inclusion of higher-order perturbative corrections to the hard-scattering kernels~\cite{Li:2005wx,Beneke:2005vv,Li:2006jb,Kivel:2006xc,Beneke:2006mk,Bell:2007tv,Jain:2007dy,Pilipp:2007mg,Bell:2009nk,Beneke:2009ek,Kim:2011jm,Bell:2014zya,Bell:2015koa,Huber:2015bva,Huber:2016xod,Bell:2020qus,Beneke:2020vnb,Beneke:2021jhp}.

In this paper, we focus on the class-I two-body decays $\bar{B}_{(s)}^0 \to D_{(s)}^{(*)+} L^-~(L=\pi,K^{(*)},\rho)$, induced at the quark level by the transitions $b \to c \bar{u} d(s)$. From a theoretical point of view, these channels are particularly clean: at the leading power in $\Lambda_{\text{QCD}}/m_b$ they are governed by the color-allowed tree amplitudes and are not affected by penguin contributions or color-suppressed tree topologies. Nevertheless, there are notable discrepancies between the most recent experimental measurements of their branching ratios~\cite{HeavyFlavorAveragingGroupHFLAV:2024ctg,ParticleDataGroup:2026aaa} and the corresponding SM expectations obtained within the QCDF~\cite{Huber:2016xod,Bordone:2020gao,Cai:2021mlt,Meiser:2024zea}. Such deviations may, in principle, be explained by previously underestimated higher-order power corrections in the SM~\cite{Huber:2016xod,Bordone:2020gao,Meiser:2024zea,Fleischer:2010ca,Endo:2021ifc,Piscopo:2023opf,Davies:2024vmv,Yaouanc:2024gjl} or, more interestingly, could point to contributions from NP~\cite{Bobeth:2014rda,Brod:2014bfa,Lenz:2019lvd,Iguro:2020ndk,Cai:2021mlt,Bordone:2021cca,Fleischer:2021cct,Fleischer:2021cwb,Gershon:2021pnc,Lenz:2022pgw,Panuluh:2024cxc,Atkinson:2024hqp,Meiser:2024zea,Deng:2025znr,Araz:2026zlu,Guo:2026oed}. 

In our previous studies~\cite{Cai:2021mlt,Deng:2025znr}, we found that the observed discrepancies can be accounted for only by model-independent NP four-quark operators with the $(1+\gamma_{5}) \otimes (1-\gamma_{5})$ (SRL) and $(1+\gamma_{5}) \otimes (1+\gamma_{5})$ (SRR) chiral structures. Motivated by this finding, we extend the SM scalar sector by introducing new scalar fields ($\phi$), requiring their induced interactions to respect the SM gauge symmetry of $SU(3)_C \otimes SU(2)_L \otimes U(1)_Y$, and we systematically investigate their implications for these class-I two-body $B$-meson decays. In parallel with the SM Yukawa pattern, the new scalars can couple to quarks ($q$) and leptons ($l$), offering a flexible setup for flavor phenomenology. Many of the possible couplings, however, are severely restricted by existing experimental data~\cite{ParticleDataGroup:2026aaa,HeavyFlavorAveragingGroupHFLAV:2024ctg}. As an illustration, the Higgs-like interactions $\phi \bar{q} q$ and $\phi \bar{l} l$ may generate tree-level $\Delta F=2$ effects and are thus strongly constrained by low-energy precision observables such as the neutral-meson mixings~\cite{Ma:1998pi,Manohar:2006ga,Arnold:2009ay,Chen:2009xjb,Han:2010rf,Grinstein:2011dz,Giudice:2011ak,Li:2013vlx,Cheng:2015lsa,Chang:2015rva,Cho:2017jym,Crivellin:2023saq}. Likewise, the leptoquark-type couplings $\phi \bar{q} l$ and $\phi q l$ have been widely examined in the studies of lepton-flavor universality and leptoquark decays at hadron colliders~\cite{Buchmuller:1986zs,Dorsner:2016wpm}. In addition, the dilepton couplings $\phi l l$ contribute only marginally to the quark-flavor dynamics. Therefore, we restrict our attention to the diquark couplings $\phi q q$~\cite{Ma:1998pi,Manohar:2006ga,Arnold:2009ay,Chen:2009xjb,Han:2010rf,Grinstein:2011dz,Giudice:2011ak}. Because they couple simultaneously to two quarks, the diquarks can naturally avoid stringent leptonic bounds while still producing sizable effects in quark flavor transitions.

\begin{table}[t]
	\tabcolsep 0.10in
	\renewcommand*{\arraystretch}{1.3}
	\begin{center}
		\begin{tabular}{|c c c c c c|} 
			\hline\hline
			Type & $SU(3)_C$ & $SU(2)_L$ & $U(1)_Y$ & Coupling Form & Tree-level $\Delta F=2$? \\
			\hline
			I & $\bar{6}$ & 3 & $-\frac{1}{3}$ & $(q_L q_L)$  & Yes \\
			II & 3 & 3 & $-\frac{1}{3}$ & $[q_L q_L],q_L l_L$  & No \\
			III & $\bar{6}$ & 1 & $-\frac{1}{3}$ & $[q_L q_L],u_R d_R$ & No \\
			IV & 3 & 1 & $-\frac{1}{3}$ & $(q_L q_L),u_R d_R,q_L l_L,u_R e_R $ & No\\
			V & $\bar{6}$ & 1 & $-\frac{4}{3}$ & $(u_R u_R)$  & Yes\\
			VI & 3 & 1 & $-\frac{4}{3}$ &  $[u_R u_R],d_R e_R$  & No\\
			VII & $\bar{6}$ & 1 & $+\frac{2}{3}$ & $(d_R d_R)$ & Yes\\
			VIII & 3 & 1 & $+\frac{2}{3}$ & $[d_R d_R]$ & No\\[1mm]  \hline\hline
		\end{tabular}
	\caption{Scalar diquarks and their representations under the SM gauge group $SU(3)_C\otimes SU(2)_L\otimes U(1)_Y$~\cite{Giudice:2011ak}. The parentheses $(\cdots)$ and square brackets $[\cdots]$ indicate that the corresponding couplings in flavor space are symmetric and antisymmetric, respectively. \label{tab:DQstable}}
	\end{center}
\end{table}

The scalar diquarks can be grouped into eight separate categories according to their gauge quantum numbers and Fermi–Dirac statistics, as compiled in table~\ref{tab:DQstable}~\cite{Giudice:2011ak}. To determine which of these can contribute to $\bar{B}_{(s)}^0 \to D_{(s)}^{(*)+} L^-$ decays, we narrow down the list by applying phenomenological constraints. Since these class-I non-leptonic decays proceed through the quark-level $b \to c \bar{u} d(s)$ transitions, the relevant diquark must couple simultaneously to up- and down-type quarks, thereby eliminating the last four entries of table~\ref{tab:DQstable}. The type-I diquark is likewise not favored, because its tree-level contributions to the $\Delta F=2$ processes are tightly bounded by the neutral-meson mixing data~\cite{Chen:2009xjb,Li:2013vlx,Cheng:2015lsa,Chang:2015rva,Cho:2017jym,Crivellin:2023saq}. In addition, channels like $\bar{B}_{(s)}^0 \to D_{(s)}^{(*)+} K^{(*)-}$ probe second-generation $c$ and $s$ quarks, which requires flavor-diagonal diquark interactions and thus excludes the type-II and type-III diquarks whose flavor couplings are purely antisymmetric. While transforming from the flavor to the physical mass-eigenstate basis can induce flavor-diagonal terms even for these antisymmetric structures, the resulting effects are strongly suppressed by the CKM factors. As a result, the type-IV diquark is the only option left as a viable candidate. In particular, this same diquark setup has also been considered as a possible explanation of the flavor anomalies observed in $b \to s\ell^+\ell^-$ transitions~\cite{Crivellin:2023saq}. Nevertheless, despite the fact that the next-to-leading order (NLO) QCD corrections to diquark production at hadron colliders are well known~\cite{Han:2009ya,Pascual-Dias:2020hxo}, the corresponding $\mathcal{O}(\alpha_s)$ corrections to their contributions in these class-I non-leptonic $B$-meson decays have not yet been provided.

In this work, we fill this gap by carrying out an explicit computation of the NLO QCD corrections to the type-IV diquark contributions to the $\BtoDL$ decays, followed by an extensive phenomenological study of whether these effects can help alleviate the tensions observed between the SM expectations and the experimental measurements for their branching fractions. Making use of the latest data on these branching fractions~\cite{HeavyFlavorAveragingGroupHFLAV:2024ctg,ParticleDataGroup:2026aaa}, we derive constraints on the diquark couplings for both the single- and double-coupling scenarios, corresponding to the presence of one and two non-vanishing effective couplings, respectively. We find that, in the single-coupling case, the purely left-handed coupling $\Delta^{LL}_{d(s)}$ can reproduce the observed anomalies at the $1\sigma$ level, whereas the mixed right-left coupling $\Delta^{RL}_{d(s)}$ remains compatible with the data at $2\sigma$. In addition, the two-coupling fit shows that sizable regions of parameter space are still allowed whenever $\Delta^{LL}_{d(s)} \neq 0$. To sharpen these bounds, we further include the LHCb measurement of the ratio $R_{\pi/\mu\nu_{\mu}}=\mathcal{B}(B_c^- \to J/\psi \pi^-)/\mathcal{B}(B_c^- \to J/\psi \mu^-\bar{\nu}_\mu)$~\cite{LHCb:2014rck} as an independent constraint; this narrows the allowed range of $\Delta^{LL}_d$ while having only a mild impact on $\Delta^{RL}_d$. Finally, we provide updated predictions for the branching fractions and the corresponding ratios $R_{(s)L}^{(*)}$ within the allowed parameter spaces.

This paper is structured as follows. In section~\ref{sec:framework}, we formulate the type-IV diquark Lagrangian and evaluate the tree-level and NLO QCD contributions to the $\Delta B=1$ non-leptonic decays. We then match the full theory onto the effective Hamiltonian in both the diquark and SM operator bases, which yields the corresponding Wilson coefficients at the NP scale $\mu_\phi=M_\phi$. In section~\ref{sec: Phenomenological_analysis}, we carry out an extensive phenomenological study to constrain the NP couplings for both single- and double-coupling scenarios, and to forecast the branching fractions and the ratios $R_{(s)L}^{(*)}$ of $\BtoDL$ decays. Our conclusions are made in section~\ref{sec:conclusion}. For convenience, additional materials on the Fierz transformations involving charge-conjugated fermion fields and the explicit anomalous-dimension matrices (ADMs) for the SM operator basis are collected in appendices \ref{app:FierzCharge} and \ref{app:ano_dim}, respectively.

\section{Wilson coefficients for non-leptonic \texorpdfstring{$\boldsymbol{\Delta B=1}$}{Delta B=1} decays in type-IV diquark}
\label{sec:framework}

\subsection{Feynman rules for the type-IV diquark}
\label{subsec:FR_dq}

We first construct the type-IV diquark Lagrangian and keep only its interactions with gluons and quarks, setting its couplings to leptons to zero. The type-IV diquark is specified through the antisymmetric contraction in color space of two quark fields. Because the antisymmetric component of $3\otimes3$ is $\bar 3$, the color structure is uniquely determined by the Levi-Civita tensor $\epsilon^{\alpha\beta\gamma}$, implying that the diquark transforms as a color triplet. Given its quantum numbers, $\phi\sim(3,1,-1/3)$, it may couple to the right-handed weak-singlet bilinear $\overline{u_R^c}\,d_R$. The left-handed quark doublets must then be contracted with $\epsilon_{ab}$ so that both bilinears couple to the same scalar field. As a result, the renormalizable gauge-basis Lagrangian for the type-IV scalar diquark $\phi$ can be expressed as~\cite{Dorsner:2016wpm}
\begin{equation}
\mathcal{L}=(D_\mu\phi)^\dagger(D^\mu\phi)-M_\phi^2\phi^\dagger\phi
+\bigg\{\Big[\tilde{\lambda}^L_{ij}\,\epsilon^{\alpha\beta\gamma}\epsilon_{ab}\,
\overline{(q^{\alpha a}_{Li})^c}\,q^{\beta b}_{Lj}
+\lambda^R_{ij}\,\epsilon^{\alpha\beta\gamma}\,
\overline{(u^{\alpha}_{Ri})^c}\,d^{\beta}_{Rj}\Big]\phi^\gamma+\mathrm{h.c.}\bigg\},
\end{equation}
where the covariant derivative is given by $D_\mu=\partial_\mu-i g_s t^A G^A_\mu$, with $t^A=\lambda^A/2$ denoting the $SU(3)_C$ generators in the fundamental representation. We label color indices by $\alpha,\beta,\gamma$, weak-isospin indices by $a,b$, and generation indices by $i,j$. Charge conjugation is indicated by the superscript $c$, with $\psi^c=C\bar\psi^{T}$ and $C=i\gamma^2\gamma^0$.

\begin{figure}[t]
	\centering
	\includegraphics[width=0.48\textwidth]{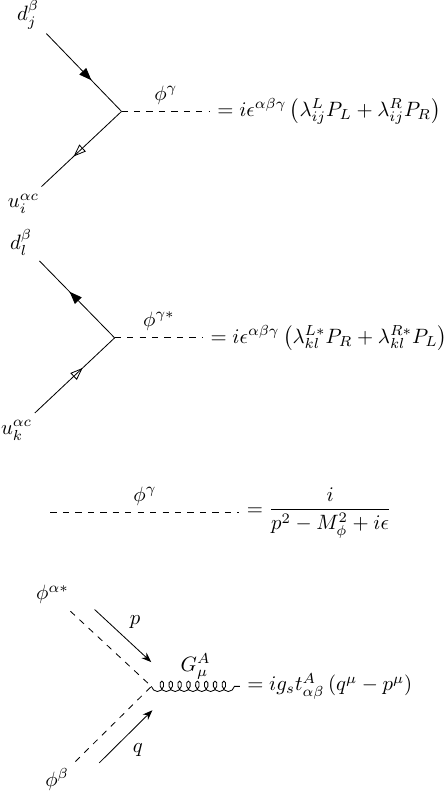}
	\caption{Relevant Feynman rules for the type-IV diquark. \label{fig:dia_DQ_FR} }
\end{figure}

In reducing the doublet contraction, it is useful to note the exchange identity
\begin{equation}
\overline{\psi_1^c}\,P_{L,R}\,\psi_2=\overline{\psi_2^c}\,P_{L,R}\,\psi_1, \label{eq:exchange}
\end{equation}
which follows from $\overline{\psi^c}=-\psi^TC^{-1}$, the Grassmann nature of the fermion fields, and the identities $C^{T}=-C$ and $P_{L,R}^TC^{-1}=C^{-1}P_{L,R}$. Because $\epsilon^{\alpha\beta\gamma}\epsilon_{ab}$ is symmetric under the simultaneous exchange $(\alpha a)\leftrightarrow(\beta b)$, the antisymmetric part of $\tilde{\lambda}^L_{ij}$ does not contribute, and one may take $\tilde{\lambda}^L_{ij}=\tilde{\lambda}^L_{ji}$ without loss of generality. Using eq.~\eqref{eq:exchange} and $(u_L)^c=(u^c)_R$, the Lagrangian becomes
\begin{equation}\label{eq:IV_Lar}
\mathcal{L}=(D_\mu\phi)^\dagger(D^\mu\phi)-M_\phi^2\phi^\dagger\phi
+\bigg\{ \Big[\lambda^L_{ij}\,\epsilon^{\alpha\beta\gamma}\,
\overline{u^{\alpha c}_{i}}P_L d^{\beta}_{j}
+\lambda^R_{ij}\,\epsilon^{\alpha\beta\gamma}\,
\overline{u^{\alpha c}_{i}}P_R d^{\beta}_{j}\Big]\phi^\gamma+\mathrm{h.c.}\bigg\},
\end{equation}
where $P_{L,R}=(1\mp\gamma_5)/2$ are the chiral projection operators, and $\lambda^L_{ik}\equiv (\tilde{\lambda}^L_{ij}+\tilde{\lambda}^L_{ji})V_{jk}=2\tilde{\lambda}^L_{ij}V_{jk}$, with $V_{jk}$ the CKM matrix element. Here $u_i$ and $d_j$ denote mass eigenstates, so that $\lambda^L_{ij}$ and $\lambda^R_{ij}$ are the coupling constants after electroweak symmetry breaking and after rotation to the mass basis; unlike $\tilde{\lambda}^L_{ij}$ in the gauge basis, $\lambda^L_{ij}$ obeys no symmetry constraint. Note that the Hermitian conjugate vertex carries the opposite projector, e.g.\ $\big[\lambda^L\overline{u^c}P_Ld\,\phi\big]^\dagger = \lambda^{L*}\,\bar d\,P_R\,u^c\,\phi^*$. Since the vertices involve charge-conjugated fields, we adopt the fermion-flow conventions of ref.~\cite{Denner:1992vza} throughout this paper. The relevant Feynman rules are collected in figure~\ref{fig:dia_DQ_FR}.

\subsection{Tree-level matching}
\label{subsec:tree_matching}

\begin{figure}[t]
	\centering
	\includegraphics[width=0.45\textwidth]{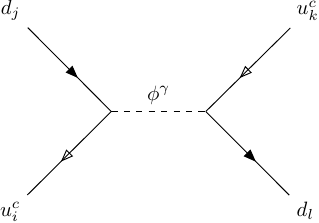}
	\caption{Tree-level Feynman diagram mediated by a type-IV diquark.\label{fig:dia_tree_full}}	
\end{figure}

To assess the phenomenological consequences of a type-IV diquark for class-I two-body decays $\bar{B}_{(s)}^0 \to D_{(s)}^{(*)+} L^-$, we begin by determining the Wilson coefficients through matching the full theory onto the effective Hamiltonian at the diquark mass scale $\mu_\phi=M_{\phi}$. In the full theory, the relevant tree-level Feynman diagram mediated by the diquark is depicted in figure~\ref{fig:dia_tree_full}. Employing the Feynman rules collected in figure~\ref{fig:dia_DQ_FR}, the corresponding amplitude reads
\begin{equation}
\mathcal{A}_{\text{tree}}^{\text{full}} = \frac{-i}{p^2 - M_\phi^2} \, \epsilon^{\alpha\beta\gamma} \epsilon^{\eta\xi\gamma} 
\left[ \overline{u_{i}^{\alpha c}} \left( \lambda_{ij}^L P_L + \lambda_{ij}^R P_R \right) d_{j}^{\beta} \right] 
\left[ \bar{d}_{l}^{\xi} \left( \lambda_{kl}^{L*} P_R + \lambda_{kl}^{R*} P_L \right) u_{k}^{\eta c} \right]\,,
\end{equation}
with $i$, $j$, $k$, and $l$ labeling the quark generations. In the regime $\MPP \gg p^2$ we can drop $p^2$, and using $\epsilon^{\alpha \beta\gamma} \epsilon^{\eta \xi \gamma} = \delta^{\alpha \eta}\delta^{\beta \xi}-\delta^{\alpha \xi}\delta^{\beta \eta}$ the amplitude can be cast into the form
\begin{equation} \label{eq:Afull_tree}
	\mathcal{A}_{\text{tree}}^{\text{full}} = \frac{i}{\MPP} \left[\lamL \lamLx \left(\SaSLRq-\SbSLRq \right) + \lamL \lamRx \left(\SaSLLq-\SbSLLq \right) \right] + \left(L \leftrightarrow R\right)  \,,
\end{equation}
where $\langle \tilde{\mathcal{Q}}_i \rangle = \langle \overline{u^c} d| \tilde{\mathcal{Q}}_i |d \overline{u^c} \rangle$ denotes the tree-level matrix element of the four-quark operator $\tilde{\mathcal{Q}}_i$. In order for the effective theory to reproduce this result, we introduce the following four-quark operators in the diquark basis:
\begin{equation}\label{eq:DQ_operators}
\begin{aligned}
	\tilde{\mathcal{Q}}_1^{SLR} &= \left[\overline{u_i^{\alpha c}} P_L d_j^\beta\right]\otimes\left[\bar{d}_l^\beta P_R u_k^{\alpha c}\right]\, , &
	\tilde{\mathcal{Q}}_2^{SLR} &= \left[\overline{u_i^{\alpha c}} P_L d_j^\beta\right]\otimes\left[\bar{d}_l^\alpha P_R u_k^{\beta c}\right] \, , \\
	\tilde{\mathcal{Q}}_1^{SLL} &= \left[\overline{u_i^{\alpha c}} P_L d_j^\beta\right]\otimes\left[\bar{d}_l^\beta P_L u_k^{\alpha c}\right] \, , &
	\tilde{\mathcal{Q}}_2^{SLL} &= \left[\overline{u_i^{\alpha c}} P_L d_j^\beta\right]\otimes\left[\bar{d}_l^\alpha P_L u_k^{\beta c} \right]\, , \\
	\tilde{\mathcal{Q}}_3^{SLL} &= \left[\overline{u_i^{\alpha c}}\sigma_{\mu\nu} P_L d_j^\beta\right]\otimes\left[\bar{d}_l^\beta \sigma^{\mu\nu} P_L u_k^{\alpha c}\right] \, , &
	\tilde{\mathcal{Q}}_4^{SLL} &= \left[\overline{u_i^{\alpha c}}\sigma_{\mu\nu} P_L d_j^\beta\right]\otimes\left[\bar{d}_l^\alpha \sigma^{\mu\nu} P_L u_k^{\beta c}\right] \,,
\end{aligned}
\end{equation}
with $\sigma_{\mu\nu}=\frac{i}{2}[\gamma_{\mu},\gamma_{\nu}]$. The other operators with flipped chiralities are obtained by the replacement $P_L \leftrightarrow P_R$. The corresponding effective Lagrangian is then given by
\begin{equation} \label{eq:L_DQeff}
	\mathcal{L}^{\text{DQ}}_{\text{eff}} = \sum_{i} \tilde{C}_i \tilde{\mathcal{Q}}_i \,.
\end{equation}
Imposing equality of the tree-level amplitudes in the full and effective descriptions at the scale $\mu_\phi$ yields the leading-order (LO) Wilson coefficients in the diquark basis:
\begin{equation}\label{eq:WC_DQtree}
\begin{aligned}
    \tilde{C}_1^{SLR(0)} &= \frac{\lambda_{ij}^L \lambda_{kl}^{L*}}{M_\phi^2} \, , \qquad &
	\tilde{C}_2^{SLR(0)} &= -\frac{\lambda_{ij}^L \lambda_{kl}^{L*}}{M_\phi^2} \, , \\
	\tilde{C}_1^{SLL(0)} &= \frac{\lambda_{ij}^L \lambda_{kl}^{R*}}{M_\phi^2} \, , \qquad &
	\tilde{C}_2^{SLL(0)} &= -\frac{\lambda_{ij}^L \lambda_{kl}^{R*}}{M_\phi^2} \,, \\
	\tilde{C}_3^{SLL(0)} &= 0 \, , \qquad &
	\tilde{C}_4^{SLL(0)} &= 0 \,,
\end{aligned}
\end{equation}
and the remaining coefficients follow from the exchange $L \leftrightarrow R$.

For the subsequent renormalization-group (RG) running from the high scale $\mu_\phi$ down to the hadronic scale $\mu_b$, it is convenient to translate these results into the standard SM operator basis, for which the one- and two-loop ADMs are known~\cite{Buras:2000if,Buras:2012gm}. In that basis the effective weak Lagrangian is written as
\begin{equation}\label{eq:L_SMeff}    
	\mathcal{L}^{\text{SM}}_{\text{eff}}=\frac{1}{4} \sum_{i}C_i \mathcal{Q}_i \,,
\end{equation}
where the operator fields are ordered as $\left[\bar{q}_1\Gamma_1 q_2\right]\otimes\left[\bar{q}_3\Gamma_2 q_4\right]$ and the relevant four-quark operators read~\cite{Buras:2000if,Buras:2012gm}\footnote{The complete set of four-quark operators relevant for $b \to c \bar{u} d(s)$ transitions in the SM basis also includes the $VLR$ and $SLR$ chiral structures~\cite{Buras:2000if,Buras:2012gm}. Since the type-IV diquark does not generate them, they are not shown here.}
\begin{equation}\label{eq:SM_operators}
\begin{aligned}
	\mathcal{Q}_1^{VLL} &= \left[\bar{c}_\alpha\gamma_{\mu}(1-\gamma_5)b_\beta\right]\otimes\left[\bar{q}_\beta\gamma^{\mu}(1-\gamma_5)u_\alpha\right] \, , \\
	\mathcal{Q}_2^{VLL} &= \left[\bar{c}_\alpha\gamma_{\mu}(1-\gamma_5)b_\alpha\right]\otimes\left[\bar{q}_\beta\gamma^{\mu}(1-\gamma_5)u_\beta\right] \, , \\
	\mathcal{Q}_1^{SLL} &= \left[\bar{c}_\alpha (1-\gamma_5)b_\beta\right]\otimes\left[\bar{q}_\beta(1-\gamma_5)u_\alpha\right] \, , \\
	\mathcal{Q}_2^{SLL} &= \left[\bar{c}_\alpha(1-\gamma_5)b_\alpha\right]\otimes\left[\bar{q}_\beta(1-\gamma_5)u_\beta\right] \, , \\
	\mathcal{Q}_3^{SLL} &= \left[\bar{c}_\alpha\sigma_{\mu\nu}(1-\gamma_5)b_\beta\right]\otimes\left[\bar{q}_\beta \sigma^{\mu\nu}(1-\gamma_5)u_\alpha\right] \, , \\ 
	\mathcal{Q}_4^{SLL} &= \left[\bar{c}_\alpha\sigma_{\mu\nu}(1-\gamma_5)b_\alpha\right]\otimes\left[\bar{q}_\beta\sigma^{\mu\nu}(1-\gamma_5)u_\beta\right] \, ,
\end{aligned}
\end{equation}
where $q=d$ or $s$. Relative to the diquark basis, the SM basis involves different field orderings as well as distinct color contractions. Using the Fierz relations~\cite{Fierz:1939zz,Buras:2020xsm} collected in appendix~\ref{app:FierzCharge}, we connect the two sets of operators and obtain the tree-level Wilson coefficients in the SM basis at the scale $\mu_\phi$:
\begin{equation}\label{eq:WC_SMtree}  
\begin{aligned}
	C_1^{VLL(0)} &= \frac{1}{2} \tilde{C}_1^{SLR(0)} = \frac{\lambda_{ij}^L \lambda_{kl}^{L*}}{2M_\phi^2} \, , \qquad &
	C_2^{VLL(0)} &= \frac{1}{2} \tilde{C}_2^{SLR(0)} = -\frac{\lambda_{ij}^L \lambda_{kl}^{L*}}{2M_\phi^2}  \, ,\\
	C_1^{SLL(0)} &= -\frac{1}{2} \tilde{C}_1^{SLL(0)} = -\frac{\lambda_{ij}^L \lambda_{kl}^{R*}}{2M_\phi^2} \, , \qquad &
	C_2^{SLL(0)} &= -\frac{1}{2} \tilde{C}_2^{SLL(0)} = \frac{\lambda_{ij}^L \lambda_{kl}^{R*}}{2M_\phi^2}  \, ,\\
	C_3^{SLL(0)} &= \frac{1}{8} \tilde{C}_1^{SLL(0)} = \frac{\lambda_{ij}^L \lambda_{kl}^{R*}}{8M_\phi^2} \, , \qquad &
	C_4^{SLL(0)} &= \frac{1}{8} \tilde{C}_2^{SLL(0)} = -\frac{\lambda_{ij}^L \lambda_{kl}^{R*}}{8M_\phi^2}  \, .
\end{aligned}
\end{equation}
The four-quark operators and their corresponding Wilson coefficients not shown above are obtained by applying the chirality interchange $L \leftrightarrow R$.

\subsection{One-loop matching}
\label{subsec:oneloop_matching}

\subsubsection{Calculation in the full theory}
\label{subsubsec:oneloop_full}

\begin{figure}[t]
	\centering
	\includegraphics[width=0.75\textwidth]{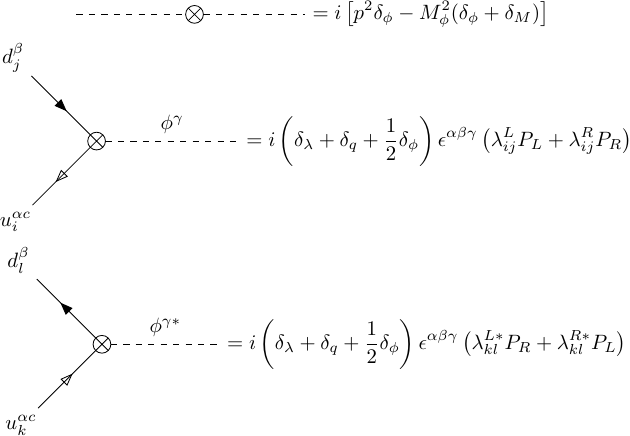}
	\caption{Relevant Feynman rules for the type-IV diquark counterterms. \label{fig:dia_DQ_CT_FR}}
\end{figure}

To compute the one-loop amplitude in the full theory and renormalize it via the counterterm approach, we begin by rewriting the diquark Lagrangian in terms of bare fields and bare parameters. Concretely, in eq.~\eqref{eq:IV_Lar} we make the replacements $\phi \to \phi_0$, $G_\mu^A \to G_{0\mu}^A$, $q \to q_0$, $M_{\phi} \to M_{\phi 0}$, $\lambda^{L,R} \to \lambda_0^{L,R}$, and $g_s \to g_{0s}$. The bare and renormalized quantities are connected by renormalization constants $Z_i$ through
\begin{align} \label{eq:DQRGR}
	&\phi_0 = Z_\phi^{1/2} \phi \,, \qquad
	G_{0\mu}^A = Z_G^{1/2} G_{\mu}^A \, ,\qquad 
	q_0 = Z_q^{1/2} q \,, \nonumber\\[2mm]
	&M_{\phi 0}^2 = Z_M M_{\phi}^2 \, , \qquad
	\lambda_0^{L,R} = Z_\lambda \lambda^{L,R} \, , \qquad
	g_{0s} = Z_g g_s \mu^{\varepsilon} \, ,
\end{align}
where $\mu$ denotes an arbitrary mass scale and $\varepsilon$ appears in dimensional regularization with $D=4-2\varepsilon$ space-time dimensions. We then write each renormalization constant as $Z_i = 1 + \delta_i$, with $\delta_i$ the counterterms introduced to remove the ultraviolet (UV) divergences in the Green functions obtained from calculations based solely on eq.~\eqref{eq:IV_Lar}. Inserting these expressions into the bare Lagrangian leads to
\begin{align} \label{eq:DQphy_Lar}
	\mathcal{L} =& (\partial_\mu \phi)^\dagger (\partial^\mu \phi) + i g_sG^A_\mu[\phi^{\dagger} t^A (\partial^\mu \phi) - (\partial^\mu \phi)^\dagger t^A\phi]- \MPP\phi^{\dagger}\phi\nonumber\\
	&+ \left[\epsabc(\lambda_{ij}^L \overline{u_{i}^{\alpha c}}P_L d_{j}^\beta  +\lambda_{ij}^R \overline{u_{i}^{\alpha c}}P_R d_{j}^\beta )\phi^\gamma +\hc \right] + \delta_\phi (\partial_\mu \phi)^\dagger (\partial^\mu \phi)\nonumber\\
	& + i g_s\left(\delta_\phi+\frac{1}{2}\delta_G\right)G^A_\mu[\phi^{ \dagger} t^A (\partial^\mu \phi) - (\partial^\mu \phi)^\dagger t^A\phi]-(\delta_\phi+\delta_M)\MPP\phi^{\dagger}\phi \nonumber\\
	&
	+\left[\left(\delta_\lambda+\delta_q+\frac{1}{2}\delta_\phi\right) \epsabc \left(\lambda_{ij}^L \overline{u_{i}^{\alpha c}} P_L d_{j}^\beta +
	\lambda_{ij}^R \overline{u_{i}^{\alpha c}} P_R d_{j}^\beta\right) \phi^\gamma  + \hc \right]  \, . 
\end{align}
For simplicity, we do not display those terms as well as their counterterms that do not enter the QCD corrections considered here, such as the diquark–gluon quartic couplings. The first four terms of eq.~\eqref{eq:DQphy_Lar} define the physical Lagrangian, whereas the remaining terms constitute the counterterm Lagrangian. The corresponding Feynman rules are summarized in figures~\ref{fig:dia_DQ_FR} and \ref{fig:dia_DQ_CT_FR}, respectively.

The diquark mass renormalization factor $Z_M$ is fixed in the on-shell scheme, whereas $Z_q$ and $Z_\lambda$ are evaluated in the modified minimal subtraction ($\MSbar$) scheme. Up to NLO in $\alpha_s=\frac{g_s^2}{4\pi}$, they are given, respectively, by
\begin{align} \label{eq:Zi}
	&Z_M = 1 - \asdpi C_F \left(\frac{3}{\varepsilon}+3\lu+7 \right) \, ,\nonumber\\[2mm]
	&Z_q = 1- \asdpi C_F\frac{1}{\varepsilon} \, ,\qquad
	Z_\lambda = 1- \asdpi \frac{4N_c+2}{2N_c}\frac{1}{\varepsilon} \, , 
\end{align}
with $\lu = \ln\frac{\mu^2}{M_\phi^2}$ and $C_F = (N_c^2-1)/(2N_c)$, where $N_c=3$ denotes the number of colors. Because neither the diquark nor the gluon is taken as an external state, explicit expressions for $Z_\phi$ and $Z_G$ are unnecessary. For convenience in the matching computation, we take the external quark masses and momenta to vanish. Moreover, the infrared (IR) divergences are labeled by $\epsnIR{1}$ to clearly separate them from the UV poles.

\begin{figure}[t]
	\centering
	\includegraphics[width=0.70\textwidth]{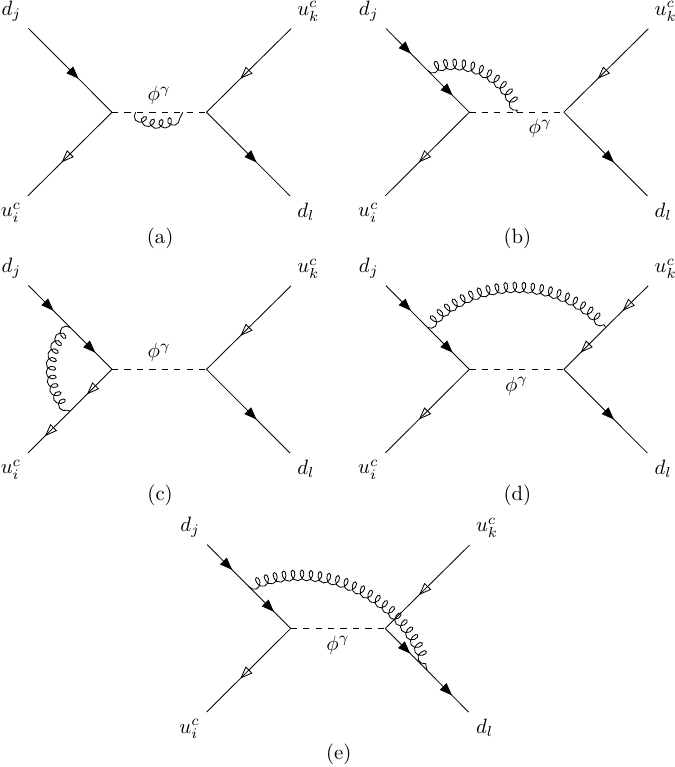}
	\caption{One-loop Feynman diagrams mediated by a type-IV diquark. Contributions from the corresponding symmetric diagrams are also included in our calculation. \label{fig:dia_oneloop_full} }
\end{figure}

\begin{figure}[t]
	\centering
	\includegraphics[width=0.70\textwidth]{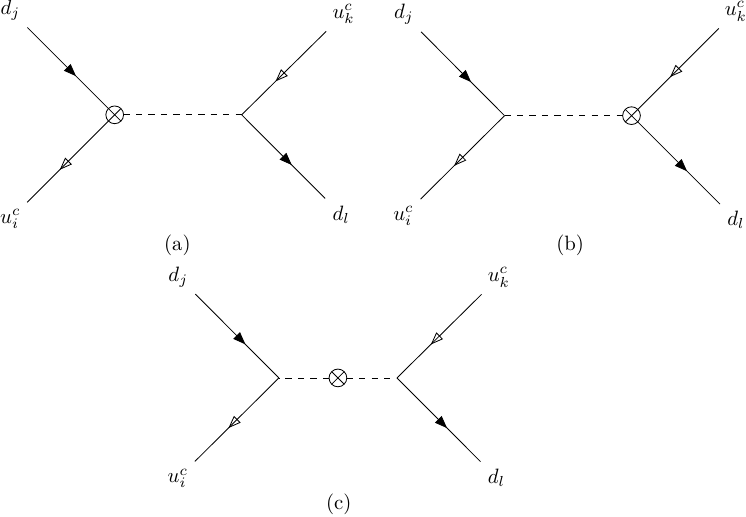}
	\caption{Counterterm Feynman diagrams mediated by a type-IV diquark. \label{fig:dia_counterterm} }
\end{figure}

After fixing the renormalization constants, we move on to compute the one-loop QCD contributions to the tree-level amplitude. The relevant Feynman diagrams are shown in figure~\ref{fig:dia_oneloop_full} and include the diquark self-energy insertions, box-type diagrams, and two classes of vertex corrections: those arising from gluon couplings to a quark–diquark pair and those from standard quark–gluon interactions. In addition, the symmetric counterparts of each topology are understood to be included in the calculation. We also include the counterterm diagrams shown in figure~\ref{fig:dia_counterterm}. For practical implementation, we encode the Lagrangian in \texttt{FeynRules}~\cite{Christensen:2008py,Alloul:2013bka} to produce the model files, then load them into \texttt{FeynArts}~\cite{Kublbeck:1990xc,Hahn:2000kx} to generate the quark-level diagrams and associated amplitudes. The resulting expressions are processed in an automated chain using \texttt{FeynCalc}~\cite{Mertig:1990an,Shtabovenko:2020gxv,Shtabovenko:2023idz,Shtabovenko:2025lxq} for the algebraic steps and \texttt{Package-X}~\cite{Patel:2015tea,Patel:2016fam} for the analytic evaluation of the one-loop integrals.

We next provide the bare (unrenormalized) amplitudes for each one-loop topology separately. For the diquark self-energy graph shown in figure~\ref{fig:dia_oneloop_full}(a), one finds
\begin{align} \label{eq:A_DQsef}
	\mathcal{A}_\Sigma = -\mathcal{A}^\text{full}_\text{tree} \times \asdpi C_F\left(\epsn{1} +  \lu + 1\right)  \, . 
\end{align}
The first set of vertex corrections, displayed in figure~\ref{fig:dia_oneloop_full}(b), in which the gluon couples to a quark–diquark pair, gives
\begin{align} \label{eq:A_DQV1}
	\mathcal{A}_{V1} = 
	-\Atree \times \asdpi \frac{1}{2N_c}\left(\epsn{1}+\lu+1\right)  \, .
\end{align}
Attaching the gluon to any one of the four external quark legs leads to four identical diagrams, so that the total contribution is given by $4 \mathcal{A}_{V1}$. The second set of vertex corrections, shown in figure~\ref{fig:dia_oneloop_full}(c), involves gluon exchange between two quarks; the corresponding amplitude reads
\begin{align} \label{eq:A_DQV2}
	\mathcal{A}_{\Delta1} =& 
	\Atree\times \asdpi 4\frac{1+N_c}{2N_c}\left(\epsn{1}-1\right) + \epsnIR{1}\asdpi\frac{1}{4}\left(-\frac{1+N_c}{2N_c}\right) \frac{i}{\MPP} \nonumber\\
	& \times \biggl[\Bigl\langle 
	\overline{u_{i}^{\alpha c}}\gamma_\mu\gamma_\rho\left(\lamL P_L+\lamR P_R\right)\gamma^\rho\gamma^\mu d_j^\beta \otimes
	\bar{d}_l^\beta \left(\lamLx P_R+\lamRx P_L\right) u_k^{\alpha c}\Bigr\rangle \nonumber\\
	&\;\; - \Bigl\langle\overline{u_{i}^{\alpha c}}\gamma_\mu\gamma_\rho\left(\lamL P_L+\lamR P_R\right)\gamma^\rho\gamma^\mu d_j^\beta \otimes
	\bar{d}_l^\alpha \left(\lamLx P_R+\lamRx P_L\right) u_k^{\beta c} \Bigr\rangle \biggr]\, .
\end{align}
Together with its mirrored diagram, where the gluon is attached to the opposite pair of quarks, this topology acquires an overall factor of two.

For the box diagram shown in figure~\ref{fig:dia_oneloop_full}(d), one obtains its amplitude as
\begin{align} \label{eq:A_Box1}
	\mathcal{A}_{\Delta2} =& \asdpi \frac{i}{\MPP}\left(\lu+\frac{3}{2}\right)\left[\lamL\lamLx\left(\frac{1}{2N_c}\SaSLRq+\frac{N^2_c-N_c-1}{2N_c}\SbSLRq\right) \nonumber \right.\\ 
	&+\lamL\lamRx\left(\frac{1}{2N_c}\SaSLLq+\frac{N_c^2-N_c-1}{2N_c}\SbSLLq\right) \nonumber \\
	&-\left.\lamL\lamRx\frac{1}{4}\left(\frac{1}{2N_c}\ScSLLq+\frac{N_c^2-N_c-1}{2N_c}\SdSLLq\right)\right] \nonumber \\
	+&\asdpi \frac{i}{\MPP} \epsnIR{1} \frac{1}{4} \left[\lamL\lamLx\left(\frac{1}{2N_c}\SaSLRqga+\frac{N_c^2-N_c-1}{2N_c}\SbSLRqga\right)\right. \nonumber\\
	&\left.+\lamL\lamRx\left(\frac{1}{2N_c}\SaSLLqga+\frac{N_c^2-N_c-1}{2N_c}\SbSLLqga\right)\right] + (L \leftrightarrow R) \, ,
\end{align}
where the additional operators appearing here are introduced as
\begin{align} \label{eq:OP2ga1}
	\QaSLLqga &=\left[\overline{u_{i}^{\alpha c}}\gamma^\rho\gamma^\nu P_L d_j^\beta\right]\otimes\left[\bar{d}_l^\beta \gamma_\rho\gamma_\nu P_L u_k^{\alpha c}\right]\, , \nonumber\\
	\QbSLLqga &=\left[\overline{u_{i}^{\alpha c}}\gamma^\rho\gamma^\nu P_L d_j^\beta\right]\otimes\left[\bar{d}_l^\alpha \gamma_\rho\gamma_\nu P_L u_k^{\beta  c}\right]\, , \nonumber\\
	\QaSLRqga &=\left[\overline{u_{i}^{\alpha c}}\gamma^\rho\gamma^\nu P_L d_j^\beta\right]\otimes\left[\bar{d}_l^\beta \gamma_\rho\gamma_\nu P_R u_k^{\alpha c}\right]\, , \nonumber\\
	\QbSLRqga &=\left[\overline{u_{i}^{\alpha c}}\gamma^\rho\gamma^\nu P_L d_j^\beta\right]\otimes\left[\bar{d}_l^\alpha \gamma_\rho\gamma_\nu P_R u_k^{\beta  c}\right]\, .
\end{align}
In order to enable a straightforward comparison of the IR singularities in the full and the effective theory, the operators multiplying the IR-divergent pieces are left in their original form. By contrast, the structures accompanying the UV-divergent contributions and the finite parts are reduced using the so-called Greek method~\cite{Tracas:1982gp,Buras:1998raa}. 

Along the same lines, for the second box topology in figure~\ref{fig:dia_oneloop_full}(e), the corresponding amplitude can be written as
\begin{align} \label{eq:A_Box2}
	\mathcal{A}_{\Delta3} =& -\asdpi \frac{i}{\MPP} \left(\lu+\frac{3}{2}\right) \bigg[ 
	\lamL\lamLx \left( \frac{1}{2N_c}\SbSLRq + \frac{N_c^2-N_c-1}{2N_c}\SaSLRq \right) \nonumber\\[0.1cm]
	& \quad + \lamL\lamRx \left( \frac{1}{2N_c}\SbSLLq + \frac{N_c^2-N_c-1}{2N_c}\SaSLLq \right) \nonumber\\[0.1cm]
	& \quad + \lamL\lamRx \frac{1}{4} \left( \frac{1}{2N_c}\SdSLLq + \frac{N_c^2-N_c-1}{2N_c}\ScSLLq \right) \bigg] \nonumber\\[0.2cm]
	  &- \asdpi \frac{i}{\MPP} \epsnIR{1} \frac{1}{4} \bigg[ 
	\lamL\lamLx \left( \frac{1}{2N_c}\SbSLRqgb + \frac{N_c^2-N_c-1}{2N_c}\SaSLRqgb \right) \nonumber\\[0.1cm]
	& \quad + \lamL\lamRx \left( \frac{1}{2N_c}\SbSLLqgb + \frac{N_c^2-N_c-1}{2N_c}\SaSLLqgb \right) \bigg] + (L \leftrightarrow R) \, ,
\end{align}
and the corresponding new operators are defined by
\begin{align} \label{eq:OP2ga2}
	\QaSLLqgb &=\left[\overline{u_{i}^{\alpha c}}\gamma^\rho\gamma^\nu P_L d_j^\beta\right]\otimes\left[\bar{d}_l^\beta \gamma_\nu\gamma_\rho P_L u_k^{\alpha c}\right]\, , \nonumber\\
	\QbSLLqgb &=\left[\overline{u_{i}^{\alpha c}}\gamma^\rho\gamma^\nu P_L d_j^\beta\right]\otimes\left[\bar{d}_l^\alpha \gamma_\nu\gamma_\rho P_L u_k^{\beta  c}\right]\, , \nonumber\\
	\QaSLRqgb &=\left[\overline{u_{i}^{\alpha c}}\gamma^\rho\gamma^\nu P_L d_j^\beta\right]\otimes\left[\bar{d}_l^\beta \gamma_\nu\gamma_\rho P_R u_k^{\alpha c}\right]\, , \nonumber\\
	\QbSLRqgb &=\left[\overline{u_{i}^{\alpha c}}\gamma^\rho\gamma^\nu P_L d_j^\beta\right]\otimes\left[\bar{d}_l^\alpha \gamma_\nu\gamma_\rho P_R u_k^{\beta  c}\right]\, .
\end{align}
Finally, both box configurations discussed above possess a symmetric partner diagram of the same size, which together imply an overall multiplicative factor of two.

By adding the tree-level and one-loop contributions and including the counterterm-diagram contributions shown in figure~\ref{fig:dia_counterterm}, we obtain the fully renormalized total amplitude in the full theory up to NLO in $\alpha_s$:
\begin{align} \label{eq:A_Full}
	\mathcal{A}_{\text{total}}^{\text{full}} =& \left[ 1 + \asdpi \left( \frac{N_c-1}{N_c}\lu + \frac{3N_c^2-5N_c-12}{2N_c} \right) \right] \frac{i}{\MPP} \bigg[ \lamL\lamLx \left( \SaSLRq - \SbSLRq \right) \nonumber\\[0.1cm]
	& \quad + \lamL\lamRx \left( \SaSLLq - \SbSLLq \right) \bigg] - \asdpi \frac{N_c-1}{4} \left( \lu + \frac{3}{2} \right) \frac{i}{\MPP} \lamL\lamRx \nonumber\\[0.2cm]
    & \times \left( \ScSLLq + \SdSLLq \right)  + \IRpart + (L \leftrightarrow R) \, ,
\end{align}
where we label the remaining IR-divergent pieces by ``IRpart''; these terms will be canceled by the corresponding contributions in the effective theory.

\subsubsection{Calculation in the effective theory formulated in the diquark basis}
\label{subsubsec:oneloop_eft_diquarkbasis}

\begin{figure}[t]
	\centering
	\includegraphics[width=0.70\textwidth]{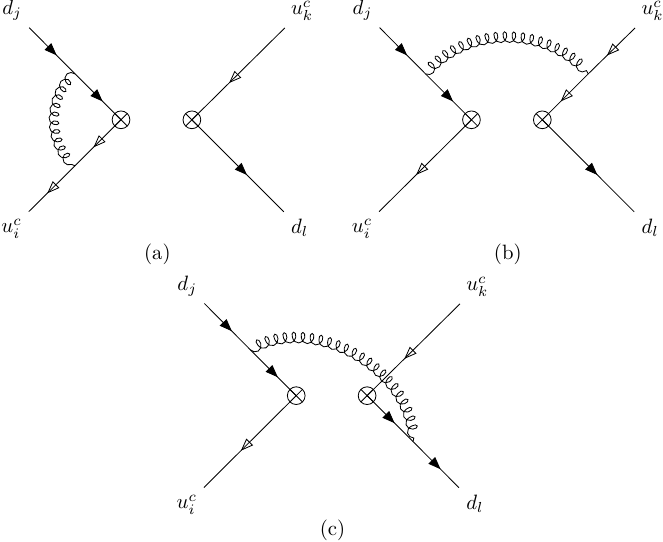}
	\caption{One-loop Feynman diagrams in the effective theory formulated in the diquark basis, where ``$\otimes$'' indicates the insertion of the effective four-quark operators defined in eq.~\eqref{eq:DQ_operators}. The contributions of the associated symmetric diagrams are understood to be included implicitly.\label{fig:dia_oneloop_DQ} }
\end{figure}

Starting with the effective Lagrangian in eq.~\eqref{eq:L_DQeff}, we calculate the NLO QCD corrections to the matrix elements of the pertinent four-quark operators expressed in the diquark basis (see eq.~\eqref{eq:DQ_operators}). The corresponding one-loop Feynman diagrams are shown in figure~\ref{fig:dia_oneloop_DQ}, with the symmetric counterparts implicitly included. We evaluate the NLO QCD corrections only for those operator matrix elements whose Wilson coefficients are non-zero at tree level, while the matrix elements of the remaining operators are taken at the LO in $\alpha_s$.

Computing the vertex corrections shown in figure~\ref{fig:dia_oneloop_DQ}(a), including the insertion of the operators defined in eq.~\eqref{eq:DQ_operators}, one obtains the amplitudes
\begin{align} \label{eq:A_DQeff1}
	&\mathcal{A}_{\QaSLRq}^{\Delta1} = i\asdpi\left(\epsn{1}-1\right) 4\left(\frac{1}{2N_c}\SaSLRq-\frac{1}{2}\SbSLRq\right) \nonumber\\[1mm]
	&-i\epsnIR{1}\asdpi\frac{1}{4}\left(\frac{1}{2N_c}\left\langle\overline{u_{i}^{\alpha c}}\gamma_\mu\gamma_\rho P_L\gamma^\rho\gamma^\mu d_j^\beta \otimes
	\bar{d}_l^\beta P_R u_k^{\alpha c}\right\rangle  -\frac{1}{2}\left\langle\overline{u_{i}^{\alpha c}}\gamma_\mu\gamma_\rho P_L\gamma^\rho\gamma^\mu d_j^\beta \otimes
	\bar{d}_l^\alpha P_R u_k^{\beta c}\right\rangle\right), \nonumber\\[0.2cm]
	&\mathcal{A}_{\QbSLRq}^{\Delta1} =i\asdpi\left(\epsn{1}-1\right) 4\left(\frac{1}{2N_c}\SbSLRq-\frac{1}{2}\SaSLRq\right) \nonumber\\[1mm]
	&-i\epsnIR{1}\asdpi\frac{1}{4}\left(\frac{1}{2N_c}\left\langle\overline{u_{i}^{\alpha c}}\gamma_\mu\gamma_\rho P_L\gamma^\rho\gamma^\mu d_j^\beta \otimes
	\bar{d}_l^\alpha P_R u_k^{\beta c}\right\rangle -\frac{1}{2}\left\langle\overline{u_{i}^{\alpha c}}\gamma_\mu\gamma_\rho P_L\gamma^\rho\gamma^\mu d_j^\beta \otimes
	\bar{d}_l^\beta P_R u_k^{\alpha c}\right\rangle\right), \nonumber\\[0.2cm]
	&\mathcal{A}_{\QaSLLq}^{\Delta1} =i\asdpi\left(\epsn{1}-1\right) 4\left(\frac{1}{2N_c}\SaSLLq-\frac{1}{2}\SbSLLq\right) \nonumber\\[1mm]
	&-i\epsnIR{1}\asdpi\frac{1}{4}\left(\frac{1}{2N_c}\left\langle\overline{u_{i}^{\alpha c}}\gamma_\mu\gamma_\rho P_L\gamma^\rho\gamma^\mu d_j^\beta \otimes
	\bar{d}_l^\beta P_L u_k^{\alpha c}\right\rangle -\frac{1}{2}\left\langle\overline{u_{i}^{\alpha c}}\gamma_\mu\gamma_\rho P_L\gamma^\rho\gamma^\mu d_j^\beta \otimes
	\bar{d}_l^\alpha P_L u_k^{\beta c}\right\rangle\right), \nonumber\\[0.2cm]
	&\mathcal{A}_{\QbSLLq}^{\Delta1} =i\asdpi\left(\epsn{1}-1\right) 4\left(\frac{1}{2N_c}\SbSLLq-\frac{1}{2}\SaSLLq\right) \nonumber\\[1mm]
	&-i\epsnIR{1}\asdpi\frac{1}{4}\left(\frac{1}{2N_c}\left\langle\overline{u_{i}^{\alpha c}}\gamma_\mu\gamma_\rho P_L\gamma^\rho\gamma^\mu d_j^\beta \otimes
	\bar{d}_l^\alpha P_L u_k^{\beta c}\right\rangle -\frac{1}{2}\left\langle\overline{u_{i}^{\alpha c}}\gamma_\mu\gamma_\rho P_L\gamma^\rho\gamma^\mu d_j^\beta \otimes
	\bar{d}_l^\beta P_L u_k^{\alpha c}\right\rangle\right).
\end{align}
In the same way, for the diagram topologies displayed in figure~\ref{fig:dia_oneloop_DQ}(b) one finds
\begin{align} \label{eq:A_DQeff2}
	\mathcal{A}_{\QaSLRq}^{\Delta2} &=i\asdpi\left(\epsn{1}+1\right)\left(-\frac{1}{2N_c}\SaSLRq+\frac{1}{2}\SbSLRq\right) \nonumber\\[1mm]
	&-i\epsnIR{1}\asdpi\frac{1}{4}\left(-\frac{1}{2N_c} \SaSLRqga+\frac{1}{2} \SbSLRqga\right)\, , \nonumber\\[2mm]
	\mathcal{A}_{\QbSLRq}^{\Delta2} &=i\asdpi\left(\epsn{1}+1\right) C_F \SbSLRq -i\epsnIR{1}\asdpi \frac{1}{4} C_F \SbSLRqga \, ,\nonumber\\[2mm]
	\mathcal{A}_{\QaSLLq}^{\Delta2} 
	&=i\asdpi\left[\frac{1}{4\varepsilon}\left(\frac{1}{2N_c}\ScSLLq-\frac{1}{2}\SdSLLq\right) -\left(\epsn{1}-\frac{1}{2}\right)\left(\frac{1}{2N_c}\SaSLLq-\frac{1}{2}\SbSLLq\right)\right]\nonumber\\[1mm]
	&-i\epsnIR{1}\asdpi\frac{1}{4}\left(-\frac{1}{2N_c}\SaSLLqga+\frac{1}{2}\SbSLLqga\right) \, , \nonumber\\[2mm]
	\mathcal{A}_{\QbSLLq}^{\Delta2}
	&=i\asdpi C_F\left[\left(\epsn{1}-\frac{1}{2}\right)\SbSLLq-\frac{1}{4\varepsilon}\SdSLLq\right] -i\epsnIR{1}\asdpi \frac{1}{4}C_F \SbSLLqga  \, .
\end{align}
Finally, evaluating the diagrams of figure~\ref{fig:dia_oneloop_DQ}(c) leads to
\begin{align} \label{eq:A_DQeff3}
	\mathcal{A}_{\QaSLRq}^{\Delta3} &=i\asdpi\left(\epsn{1}-2\right)C_F\SaSLRq-i\epsnIR{1}\asdpi\frac{1}{4}C_F \SaSLRqgb \,, \nonumber\\[2mm]
	\mathcal{A}_{\QbSLRq}^{\Delta3} &=i\asdpi\left(\epsn{1}-2\right) \left(-\frac{1}{2N_c}\SbSLRq+\frac{1}{2}\SaSLRq\right) \nonumber\\[1mm]
	&-i\epsnIR{1}\asdpi \frac{1}{4} \left(-\frac{1}{2N_c}\SbSLRqgb+\frac{1}{2}\SaSLRqgb\right)\,,\nonumber\\[2mm]
	\mathcal{A}_{\QaSLLq}^{\Delta3} 
	&= i\asdpi C_F\left[\left(\epsn{1}-\frac{1}{2}\right)\SaSLLq+\frac{1}{4\varepsilon}\ScSLLq\right] -i\epsnIR{1} \asdpi \frac{1}{4}C_F \SaSLLqgb \,, \nonumber\\[2mm]
	\mathcal{A}_{\QbSLLq}^{\Delta3}
	&= i\asdpi \left[\left(\epsn{1}-\frac{1}{2}\right)\left(\frac{1}{2}\SaSLLq-\frac{1}{2N_c}\SbSLLq\right) +\frac{1}{4\varepsilon}\left(\frac{1}{2}\ScSLLq-\frac{1}{2N_c}\SdSLLq\right)\right]\nonumber\\[1mm]
	&-i\epsnIR{1}\asdpi\frac{1}{4}\left(-\frac{1}{2N_c}\SbSLLqgb+\frac{1}{2}\SaSLLqgb\right)
	\, .
\end{align}
To get the full one-loop contributions, we must multiply the amplitudes given above by an overall factor of two to include contributions of the symmetric counterparts, whereas the chirality-flipped pieces are obtained by exchanging $L \leftrightarrow R$. Adding these one-loop QCD corrections to the tree-level expressions then gives the complete unrenormalized amplitude in the effective theory formulated in the diquark basis:
\begin{align} \label{eq:A_DQunren}
	\mathcal{A}_{\text{unren}}^{\text{DQ,eff}} &= 
	i \bigg\{ \tilde{C}_1^{SLR(0)} \left\langle \tilde{\mathcal{Q}}_1^{SLR} \right\rangle + \tilde{C}_2^{SLR(0)} \left\langle \tilde{\mathcal{Q}}_2^{SLR} \right\rangle + \tilde{C}_1^{SLL(0)} \left\langle \tilde{\mathcal{Q}}_1^{SLL} \right\rangle + \tilde{C}_2^{SLL(0)} \left\langle \tilde{\mathcal{Q}}_2^{SLL} \right\rangle \bigg\} \nonumber\\[0.3cm]
	&+ i \frac{\alpha_s}{4\pi} \Bigg\{ \left[ \tilde{C}_1^{SLR(1)} + \tilde{C}_1^{SLR(0)} \left( \frac{N_c^2+2}{N_c} \frac{1}{\varepsilon} - \frac{2N_c^2+3}{N_c} \right) + \tilde{C}_2^{SLR(0)} \left(2-\frac{3}{\varepsilon}\right) \right] \left\langle \tilde{\mathcal{Q}}_1^{SLR} \right\rangle \nonumber\\[0.2cm]
	&\hspace{1.05cm} + \left[ \tilde{C}_2^{SLR(1)} + \tilde{C}_2^{SLR(0)} \left( \frac{N_c^2+2}{N_c} \frac{1}{\varepsilon} + \frac{N_c^2-3}{N_c} \right) + \tilde{C}_1^{SLR(0)} \left(5-\frac{3}{\varepsilon}\right) \right] \left\langle \tilde{\mathcal{Q}}_2^{SLR} \right\rangle \nonumber\\[0.2cm]
	&\hspace{1.05cm} + \left[ \tilde{C}_1^{SLL(1)} + \tilde{C}_1^{SLL(0)} \left( \frac{N_c^2+2}{N_c} \frac{1}{\varepsilon} - \frac{N_c^2+6}{2N_c} \right) + \tilde{C}_2^{SLL(0)} \left( \frac{7}{2}-\frac{3}{\varepsilon} \right) \right] \left\langle \tilde{\mathcal{Q}}_1^{SLL} \right\rangle \nonumber\\[0.2cm]
	&\hspace{1.05cm} + \left[ \tilde{C}_2^{SLL(1)} + \tilde{C}_2^{SLL(0)} \left( \frac{N_c^2+2}{N_c} \frac{1}{\varepsilon} - \frac{N_c^2+6}{2N_c} \right) + \tilde{C}_1^{SLL(0)} \left( \frac{7}{2}-\frac{3}{\varepsilon} \right) \right] \left\langle \tilde{\mathcal{Q}}_2^{SLL} \right\rangle \nonumber\\[0.2cm]
	&\hspace{1.05cm} + \left[ \tilde{C}_3^{SLL(1)} + \frac{1}{4} \left( \tilde{C}_1^{SLL(0)} \frac{N_c}{\varepsilon} + \tilde{C}_2^{SLL(0)} \frac{1}{\varepsilon} \right) \right] \left\langle \tilde{\mathcal{Q}}_3^{SLL} \right\rangle \nonumber\\[0.2cm]
	&\hspace{1.05cm} + \left[ \tilde{C}_4^{SLL(1)} - \frac{1}{4} \left( \tilde{C}_1^{SLL(0)} \frac{1}{\varepsilon} + \tilde{C}_2^{SLL(0)} \frac{N_c}{\varepsilon} \right) \right] \left\langle \tilde{\mathcal{Q}}_4^{SLL} \right\rangle \Bigg\} \nonumber\\[0.2cm]
	&+ \text{IRpart} + (L \leftrightarrow R) \,,
\end{align}
where the Wilson coefficients have been expanded up to $\mathcal{O}(\alpha_s)$, $\tilde{C}_i^X = \tilde{C}_i^{X(0)} + \frac{\alpha_s}{4\pi} \tilde{C}_i^{X(1)}$, for $X = SLR, SLL$, with the tree-level coefficients $\tilde{C}_{1,2}^{SLR(0)}$ and $\tilde{C}_{1,2}^{SLL(0)}$ given by eq.~\eqref{eq:WC_DQtree}.

The unrenormalized amplitude in eq.~\eqref{eq:A_DQunren} is made finite by renormalizing the quark fields and the effective operators according to
\begin{align}\label{eq:DQren} 
	q_0 = \sqrt{Z_q}\, q \,, \qquad
	\tilde{Q}^{\text{bare}}_i = Z^{\text{DQ}}_{ij}\, \tilde{Q}_j \, ,
\end{align}
where the quark wave-function renormalization factor $Z_q$ has already been specified in eq.~\eqref{eq:Zi}. The corresponding operator basis is collected in the vector
\begin{align}
	\tilde{\mathbf{Q}}^T=(\QaSLRq,\QbSLRq,\QaSLLq,\QbSLLq,\QcSLLq,\QdSLLq)\, . 
\end{align}
Working in the $\MSbar$ scheme, the operator renormalization constant matrix $\hat{Z}^{\text{DQ}}$, expanded up to NLO in $\alpha_s$, reads
\begin{align} \label{DQZ}
	\hat{Z}^{\text{DQ}} = 1 + \asdpi \epsn{1}
	\left(\begin{array}{cccccc}
		\frac{3}{N_c} & -3 & 0 & 0 & 0 & 0 \\[0.2cm]
		-3 & \frac{3}{N_c} & 0 & 0 & 0 & 0 \\[0.2cm]
		0 & 0 & \frac{3}{N_c} & -3 & \frac{N_c}{4} & -\frac{1}{4} \\[0.2cm]
		0 & 0 & -3 & \frac{3}{N_c} & \frac{1}{4} & -\frac{N_c}{4} \\[0.2cm]
		0 & 0 & 12N_c & -12 & \frac{2N_c^2-5}{N_c} & 3 \\[0.2cm]
		0 & 0 & 12 & -12N_c & 3 & \frac{2N_c^2-5}{N_c}
	\end{array}\right) \, ,
\end{align}
which implies the associated one-loop ADM
\begin{align} \label{eq:DQ_Gamma}
	\hat{\gamma}^{\text{DQ}(0)} = 
	\left(\begin{array}{cccccc}
		-\frac{6}{N_c} & 6 & 0 & 0 & 0 & 0 \\[0.2cm]
		6 & -\frac{6}{N_c} & 0 & 0 & 0 & 0 \\[0.2cm]
		0 & 0 & -\frac{6}{N_c} & 6 & -\frac{N_c}{2} & \frac{1}{2} \\[0.2cm]
		0 & 0 & 6 & -\frac{6}{N_c} & -\frac{1}{2} & \frac{N_c}{2} \\[0.2cm]
		0 & 0 & -24N_c & 24 & \frac{10-4N_c^2}{N_c} & -6 \\[0.2cm]
		0 & 0 & -24 & 24N_c & -6 & \frac{10-4N_c^2}{N_c}
	\end{array}\right) \, .
\end{align}
Applying eq.~\eqref{eq:DQren} to eq.~\eqref{eq:A_DQunren} then yields the renormalized one-loop effective amplitude
\begin{align} \label{eq:A_DQeff_ren}
	\mathcal{A}_{\text{ren}}^{\text{DQ,eff}} &= 
	i \bigg\{ \tilde{C}_1^{SLR(0)} \left\langle \tilde{\mathcal{Q}}_1^{SLR} \right\rangle + \tilde{C}_2^{SLR(0)} \left\langle \tilde{\mathcal{Q}}_2^{SLR} \right\rangle + \tilde{C}_1^{SLL(0)} \left\langle \tilde{\mathcal{Q}}_1^{SLL} \right\rangle + \tilde{C}_2^{SLL(0)} \left\langle \tilde{\mathcal{Q}}_2^{SLL} \right\rangle \bigg\} \nonumber\\[0.3cm]
	&+ i \frac{\alpha_s}{4\pi} \Bigg\{ \left[ \tilde{C}_1^{SLR(1)} - \tilde{C}_1^{SLR(0)} \left( \frac{2N_c^2+3}{N_c} \right) + 2 \tilde{C}_2^{SLR(0)} \right] \left\langle \tilde{\mathcal{Q}}_1^{SLR} \right\rangle \nonumber\\[0.2cm]
	&\hspace{1.05cm} + \left[ \tilde{C}_2^{SLR(1)} + \tilde{C}_2^{SLR(0)} \left( \frac{N_c^2-3}{N_c} \right) + 5 \tilde{C}_1^{SLR(0)} \right] \left\langle \tilde{\mathcal{Q}}_2^{SLR} \right\rangle \nonumber\\[0.2cm]
	&\hspace{1.05cm} + \left[ \tilde{C}_1^{SLL(1)} - \tilde{C}_1^{SLL(0)} \left( \frac{N_c^2+6}{2N_c} \right) + \frac{7}{2} \tilde{C}_2^{SLL(0)} \right] \left\langle \tilde{\mathcal{Q}}_1^{SLL} \right\rangle \nonumber\\[0.2cm]
	&\hspace{1.05cm} + \left[ \tilde{C}_2^{SLL(1)} - \tilde{C}_2^{SLL(0)} \left( \frac{N_c^2+6}{2N_c} \right) + \frac{7}{2} \tilde{C}_1^{SLL(0)} \right] \left\langle \tilde{\mathcal{Q}}_2^{SLL} \right\rangle \nonumber\\[0.2cm]
	&\hspace{1.05cm}+ \tilde{C}_3^{SLL(1)} \left\langle \tilde{\mathcal{Q}}_3^{SLL} \right\rangle + \tilde{C}_4^{SLL(1)} \left\langle \tilde{\mathcal{Q}}_4^{SLL} \right\rangle \Bigg\} \nonumber\\[0.2cm]
	& + \text{IRpart} + (L \leftrightarrow R) \,.
\end{align}

By equating the full-theory amplitude in eq.~\eqref{eq:A_Full} to the corresponding renormalized effective-theory result in eq.~\eqref{eq:A_DQeff_ren}, we determine the one-loop Wilson coefficients at the scale $\mu_\phi$ to be
\begin{align} \label{eq:WC_DQ_oneloop}
	\tilde{C}_1^{SLR(1)} &= \frac{1}{\MPP} \lamL\lamLx \left( \frac{N_c-1}{N_c} \lu + \frac{7N_c^2-N_c-6}{2N_c} \right) \, , \nonumber\\[2mm]
	\tilde{C}_2^{SLR(1)} &= - \frac{1}{\MPP} \lamL\lamLx \left( \frac{N_c-1}{N_c} \lu + \frac{N_c^2+5N_c-6}{2N_c} \right) \, , \nonumber\\[2mm]
	\tilde{C}_1^{SLL(1)} &= \frac{1}{\MPP} \lamL\lamRx \left( \frac{N_c-1}{N_c} \lu + \frac{2N_c^2+N_c-3}{N_c} \right) \, , \nonumber\\[2mm]
	\tilde{C}_2^{SLL(1)} &= - \frac{1}{\MPP} \lamL\lamRx \left( \frac{N_c-1}{N_c} \lu + \frac{2N_c^2+N_c-3}{N_c} \right) \, , \nonumber\\[2mm]
	\tilde{C}_3^{SLL(1)} &= - \frac{1}{\MPP} \lamL\lamRx \frac{N_c-1}{4} \left( \lu + \frac{3}{2} \right) \, , \nonumber\\[2mm]
	\tilde{C}_4^{SLL(1)} &= - \frac{1}{\MPP} \lamL\lamRx \frac{N_c-1}{4} \left( \lu + \frac{3}{2} \right) \, .
\end{align}

A well-defined effective field theory must accurately capture the IR behavior of the underlying full theory. By comparing the IR-divergent contributions of eqs.~\eqref{eq:A_DQV2}, \eqref{eq:A_Box1}, and \eqref{eq:A_Box2} in the full theory with the corresponding effective-theory results in eqs.~\eqref{eq:A_DQeff1}, \eqref{eq:A_DQeff2}, and \eqref{eq:A_DQeff3}, we verify that they cancel exactly, providing an important consistency check of our computation.

\subsubsection{Calculation in the effective theory formulated in the SM basis}
\label{subsubsec:oneloop_eft_SMbasis}

\begin{figure}[t]
	\centering
	\includegraphics[width=0.70\textwidth]{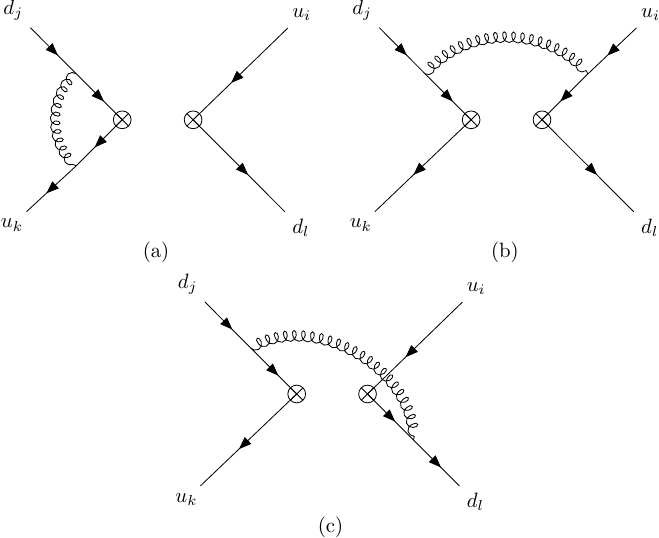}
	\caption{One-loop Feynman diagrams in the effective theory formulated in the SM basis. The other captions are the same as in figure~\ref{fig:dia_oneloop_DQ}. \label{fig:dia_oneloop_SM} }
\end{figure}

Since the Fierz identities~\cite{Fierz:1939zz,Buras:2020xsm} summarized in appendix~\ref{app:FierzCharge} are valid only in four space-time dimensions, they are not adequate at one loop. For this reason, we convert between the diquark and SM operator bases by explicitly matching the corresponding one-loop effective amplitudes~\cite{Aebischer:2018acj,Aebischer:2022aze}. The effective weak Lagrangian in the SM basis is given in eq.~\eqref{eq:L_SMeff}, and the associated one-loop diagrams are shown in figure~\ref{fig:dia_oneloop_SM}. Proceeding analogously to the diquark-basis calculation, we obtain the bare NLO operator matrix elements
\begin{align} \label{QSM}
	\SaVLL_\mathrm{NLO}^{\text{bare}}=&\left[1+\asdpi\left(\frac{N_c^2+2}{N_c}\epsn{1}+\frac{-2N_c^2+3}{N_c}\right)\right]\SaVLL  
	-\asdpi\left(\epsn{3}+1\right)\SbVLL  \, , \nonumber \\[2mm]
	\SbVLL_\mathrm{NLO}^{\text{bare}}=&\left[1+\asdpi\left(\frac{N_c^2+2}{N_c}\epsn{1}+\frac{-2N_c^2+3}{N_c}\right)\right]\SbVLL
	-\asdpi\left(\epsn{3}+1\right)\SaVLL  \, , \nonumber \\[2mm]
	\SaSLL_\mathrm{NLO}^{\text{bare}}=&\left[1+\asdpi\left(\frac{N_c^2-4}{N_c}\epsn{1}+\frac{-N_c^2+8}{2N_c}\right)\right]\SaSLL
	+\asdpi\left(\epsn{3}-\frac{7}{2}\right)\SbSLL \nonumber \\[1mm]
	&+\asdpi\frac{1}{4}\left(\frac{N_c^2-2}{N_c}\epsn{1}\ScSLL+\epsn{1}\SdSLL \right) \, , \nonumber \\[2mm]
	\SbSLL_\mathrm{NLO}^{\text{bare}}=&\left[1+\asdpi 8C_F \left(\epsn{1}-1\right) \right]\SbSLL
	+\asdpi\frac{1}{4\varepsilon}\left(-\frac{2}{N_c}\SdSLL+2\ScSLL \right) \, , \nonumber \\[2mm]
	\ScSLL_\mathrm{NLO}^{\text{bare}}=&\left[1+\asdpi \left(3N_c\epsn{1}-\frac{7N_c^2-4}{2N_c}\right) \right]\ScSLL
	-\asdpi\left(\epsn{3}-\frac{3}{2}\right)\SdSLL \nonumber \\[1mm] 
    &+\asdpi\left(\epsn{12}-20\right)\left(\frac{N_c^2-2}{N_c}\SaSLL+\SbSLL \right) \, , \nonumber \\[2mm]
	\SdSLL_\mathrm{NLO}^{\text{bare}}=&\left(1+\asdpi\frac{2}{N_c}\right) \langle \QdSLL \rangle -\asdpi 2 \langle \QcSLL \rangle \nonumber\\[1mm] 
    &+\asdpi\left(\epsn{12}-20\right)\left(-\frac{2}{N_c}\SbSLL+2\SaSLL \right) \, ,
\end{align}
with $\langle \mathcal{Q}_i \rangle = \langle \bar{u}d| \mathcal{Q}_i |d\bar{u}\rangle$ on the right-hand side representing the tree-level matrix element of the operator $\mathcal{Q}_i$. We do not display the IR-divergent terms, since they cancel automatically in the matching. The field renormalization coincides with eq.~\eqref{eq:DQren}, whereas the operator renormalization is defined by
\begin{align}\label{SMrenOp} 
	Q^{\text{bare}}_i = Z^{\text{SM}}_{ij} Q_j \, ,
\end{align}
and we collect the four-quark operators into the vector
\begin{align}
	\mathbf{Q}^T=(\QaVLL,\QbVLL,\QaSLL,\QbSLL,\QcSLL,\QdSLL)\, .
\end{align}
Within the $\MSbar$ scheme, the renormalization constant matrix $\hat{Z}^{\text{SM}}$ is given by \cite{Buras:2000if,Buras:2012gm}
\begin{align} \label{SMZ}
	\hat{Z}^{\text{SM}}=1-2\asdpi\epsn{1} \hat{\gamma}^{(0)}\, , 
\end{align}
where $\hat{\gamma}^{(0)}$ denotes the one-loop ADM listed in appendix~\ref{app:ano_dim}. Consequently, the bare NLO matrix elements $\left\langle\mathcal{Q}_i\right\rangle^{\text{bare}}_\mathrm{NLO}$ and the renormalized ones $\left\langle\mathcal{Q}_i\right\rangle_\mathrm{NLO}$ are related through
\begin{align}\label{eq:SMren}
\left\langle\mathcal{Q}_i\right\rangle^{\text{bare}}_\mathrm{NLO}=Z_q^{-2} Z^{\text{SM}}_{ij} \left\langle\mathcal{Q}_j\right\rangle_\mathrm{NLO} \, . 
\end{align}

The one-loop ADMs in the two bases are connected by the matrix $\hat{R}$, defined as
\begin{align} \label{Fierz6}
	\hat{R}=\left(\begin{array}{cccccc}
		\frac{1}{2} & 0 & 0 & 0 & 0 & 0\\[0.2cm]
		0&\frac{1}{2}& 0 & 0 & 0 & 0\\[0.2cm]
		0&0&-\frac{1}{2}&0&\frac{1}{8}&0\\[0.2cm]
		0&0&0&-\frac{1}{2}&0&\frac{1}{8}\\[0.2cm]
		0&0&6&0&\frac{1}{2}&0\\[0.2cm]
		0&0&0&6&0&\frac{1}{2}
	\end{array}\right)\,,
\end{align}
and the corresponding operators satisfy
\begin{align} \label{DQtoSMQ}
	\tilde{Q}_i =  \frac{1}{4}R_{ij} Q_j \,.
\end{align}
The prefactor $1/4$ appears because the diquark-basis operators $\tilde{Q}_i$ are written using the projectors $P_{L,R}=(1\mp\gamma_5)/2$, whereas the SM-basis operators $Q_j$ are defined with $1\pm\gamma_5$. Requiring the amplitudes to agree in the two bases yields
\begin{align} \label{DQeqSMZ}
	\tilde{C}_i Z^{\text{DQ}}_{ij}\langle \tilde{Q}_j\rangle_\mathrm{NLO}= C_i Z^{\text{SM}}_{ij} \frac{1}{4} \langle Q_j\rangle_\mathrm{NLO} \,.
\end{align}
At tree level, this matching condition simplifies to
\begin{align} \label{DQtoSMC}
	\tilde{C}^{(0)}_j = C^{(0)}_iR^{-1}_{ij} \,.
\end{align}
Inserting eqs.~\eqref{DQtoSMQ} and~\eqref{DQtoSMC} into eq.~\eqref{DQeqSMZ} leads to the relation between the renormalization constant matrices in the two operator bases,
\begin{align} \label{DQeqSMRZ}
\hat{R}^{-1} \hat{Z}^{\text{DQ}} \hat{R} = \hat{Z}^{\text{SM}} \,.
\end{align}
Consequently, the ADMs obey the corresponding similarity transformation,
\begin{align} \label{eq:DQeqSMgammaZ}
	\hat{R}^{-1} \hat{\gamma}^{\text{DQ}(0)} \hat{R} = \hat{\gamma}^{(0)} \,.
\end{align}
Employing the explicit results in eqs.~\eqref{eq:DQ_Gamma} and~\eqref{eq:gamma0_SM} verifies that the effective description remains consistent when switching between the two bases.

Using the renormalization prescription in eq.~\eqref{eq:SMren}, the renormalized amplitude up to NLO in $\alpha_s$ can be written as
\begin{align} \label{eq:A_SMeff_ren}
	\mathcal{A}^{\text{SM,eff}}_{\text{ren}} &= 
	i \frac{1}{4} \bigg\{ C_1^{VLL(0)} \left\langle \mathcal{Q}_1^{VLL} \right\rangle + C_2^{VLL(0)} \left\langle \mathcal{Q}_2^{VLL} \right\rangle + C_1^{SLL(0)} \left\langle \mathcal{Q}_1^{SLL} \right\rangle + C_2^{SLL(0)} \left\langle \mathcal{Q}_2^{SLL} \right\rangle \nonumber\\[0.1cm]
	&\qquad + C_3^{SLL(0)} \left\langle \mathcal{Q}_3^{SLL} \right\rangle + C_4^{SLL(0)} \left\langle \mathcal{Q}_4^{SLL} \right\rangle \bigg\} \nonumber\\[0.3cm]
	&+ i \frac{1}{4} \frac{\alpha_s}{4\pi} \Bigg\{ \left[ C_1^{VLL(1)} + C_1^{VLL(0)} \left( \frac{3-2N_c^2}{N_c} \right) - C_2^{VLL(0)} \right] \left\langle \mathcal{Q}_1^{VLL} \right\rangle \nonumber\\[0.2cm]
	&\quad + \left[ C_2^{VLL(1)} + C_2^{VLL(0)} \left( \frac{3-2N_c^2}{N_c} \right) - C_1^{VLL(0)} \right] \left\langle \mathcal{Q}_2^{VLL} \right\rangle \nonumber\\[0.2cm]
	&\quad + \left[ C_1^{SLL(1)} + C_1^{SLL(0)} \left( \frac{8-N_c^2}{2N_c} \right) - 20 C_3^{SLL(0)} \left( \frac{N_c^2-2}{N_c} \right) - 40 C_4^{SLL(0)} \right] \left\langle \mathcal{Q}_1^{SLL} \right\rangle \nonumber\\[0.2cm]
	&\quad + \left[ C_2^{SLL(1)} - 8 C_F C_2^{SLL(0)} - \frac{7}{2} C_1^{SLL(0)} - 20 C_3^{SLL(0)} + \frac{40}{N_c} C_4^{SLL(0)} \right] \left\langle \mathcal{Q}_2^{SLL} \right\rangle \nonumber\\[0.2cm]
	&\quad + \left[ C_3^{SLL(1)} - C_3^{SLL(0)} \left( \frac{7N_c^2-4}{2N_c} \right) - 2 C_4^{SLL(0)} \right] \left\langle \mathcal{Q}_3^{SLL} \right\rangle \nonumber\\[0.2cm]
	&\quad + \left[ C_4^{SLL(1)} + C_4^{SLL(0)} \left( \frac{2}{N_c} \right) + \frac{3}{2} C_3^{SLL(0)} \right] \left\langle \mathcal{Q}_4^{SLL} \right\rangle \Bigg\} \nonumber\\[0.2cm] 
	& + (L \leftrightarrow R) \,,
\end{align}
with the expansion $C_i^X = C_i^{X(0)} + \frac{\alpha_s}{4\pi} C_i^{X(1)}$ for $X = VLL, SLL$. The tree-level coefficients $C_i^{X(0)}$ have already been given in eq.~\eqref{eq:WC_SMtree}. By equating the renormalized expressions in eqs.~\eqref{eq:A_DQeff_ren} and \eqref{eq:A_SMeff_ren}, one obtains the one-loop conversion formulae relating the Wilson coefficients in the two operator bases:
\begin{align} \label{eq:WC_SMDQtran}
	C_1^{VLL(1)} &= \frac{1}{2} \left(\tilde{C}_1^{SLR(1)} - \frac{6}{N_c} \tilde{C}_1^{SLR(0)} + 3 \tilde{C}_2^{SLR(0)} \right) , \nonumber\\[0.2cm]
	C_2^{VLL(1)} &= \frac{1}{2} \left(\tilde{C}_2^{SLR(1)} + 6 \tilde{C}_1^{SLR(0)} + \frac{3N_c^2-6}{N_c} \tilde{C}_2^{SLR(0)}\right) , \nonumber\\[0.2cm]
	C_1^{SLL(1)} &= -\frac{1}{2} \tilde{C}_1^{SLL(1)} + 6 \tilde{C}_3^{SLL(1)} + \frac{5N_c^2-3}{2N_c} \tilde{C}_1^{SLL(0)} + \frac{13}{4} \tilde{C}_2^{SLL(0)} , \nonumber\\[0.2cm]
	C_2^{SLL(1)} &= -\frac{1}{2} \tilde{C}_2^{SLL(1)} + 6 \tilde{C}_4^{SLL(1)} - \frac{7N_c^2+6}{4N_c} \tilde{C}_2^{SLL(0)} - \tilde{C}_1^{SLL(0)} , \nonumber\\[0.2cm]
	C_3^{SLL(1)} &= \frac{1}{8} \tilde{C}_1^{SLL(1)} + \frac{1}{2} \tilde{C}_3^{SLL(1)} + \frac{3N_c^2-5}{8N_c} \tilde{C}_1^{SLL(0)} + \frac{11}{16} \tilde{C}_2^{SLL(0)} , \nonumber\\[0.2cm]
	C_4^{SLL(1)} &= \frac{1}{8} \tilde{C}_2^{SLL(1)} + \frac{1}{2} \tilde{C}_4^{SLL(1)} - \frac{N_c^2+10}{16N_c} \tilde{C}_2^{SLL(0)} + \frac{1}{4} \tilde{C}_1^{SLL(0)} .
\end{align}
Combining the tree-level pieces with the one-loop QCD corrections, the Wilson coefficients in the SM basis up to $\mathcal{O}(\alpha_s)$ at the matching scale $\mu_\phi$ become
\begin{align} \label{eq:WC_SM_oneloop}
	\CaVLL &= \frac{\lamL\lamLx}{2\MPP} \left[ 1 + \asdpi \left( \frac{N_c-1}{N_c} \lu + \frac{7N_c^2-7N_c-18}{2N_c} \right) \right] \, , \nonumber \\[2mm]
	\CbVLL &= -\frac{\lamL\lamLx}{2\MPP} \left[ 1 + \asdpi \left( \frac{N_c-1}{N_c} \lu + \frac{7N_c^2-7N_c-18}{2N_c} \right) \right] \, , \nonumber \\[2mm]
	\CaSLL &= -\frac{\lamL\lamRx}{2\MPP} \left[ 1 + \asdpi \left( \frac{3N_c^2-2N_c-1}{N_c} \lu + \frac{3(N_c+2)}{2} \right) \right] \, , \nonumber \\[2mm]
	\CbSLL &= \frac{\lamL\lamRx}{2\MPP} \left[ 1 + \asdpi \left( \frac{-3N_c^2+4N_c-1}{N_c} \lu + \frac{7}{2} + N_c \right) \right] \, , \nonumber \\[2mm]
	\CcSLL &= \frac{\lamL\lamRx}{8\MPP} \left[ 1 - \asdpi \left( \frac{(N_c-1)^2}{N_c} \lu + \frac{-7N_c^2+6N_c+16}{2N_c} \right) \right] \, , \nonumber \\[2mm]
	\CdSLL &= -\frac{\lamL\lamRx}{8\MPP} \left[ 1 + \asdpi \left( \frac{N_c^2-1}{N_c} \lu + \frac{6N_c^2-5N_c-16}{2N_c} \right) \right] \, .
\end{align}
The other six coefficients follow straightforwardly from the replacement $L \leftrightarrow R$.

\begin{figure}[t] 
	\centering
	\includegraphics[width=0.98\textwidth]{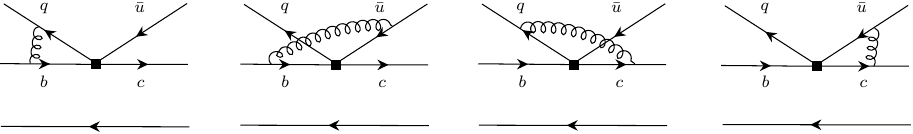}
	\caption{Non-factorizable one-loop vertex corrections to the matrix elements of the four-quark operators in eq.~\eqref{eq:SM_operators} within the QCDF approach~\cite{Beneke:1999br,Beneke:2000ry,Beneke:2001ev}. \label{fig:vertexcorrection} }
\end{figure} 

Following the tree-level and one-loop matching at the NP scale $\mu_\phi$, we evolve the Wilson coefficients down to the hadronic scale $\mu_b$ relevant for non-leptonic $\Delta B=1$ decays, using the ADMs given in appendix~\ref{app:ano_dim}. We then evaluate the hadronic matrix elements of the four-quark operators in eq.~\eqref{eq:SM_operators} within the QCDF framework~\cite{Beneke:1999br,Beneke:2000ry,Beneke:2001ev}, taking into account the non-factorizable one-loop vertex corrections shown in figure~\ref{fig:vertexcorrection}. The complete formulae for these matrix elements are provided in refs.~\cite{Cai:2021mlt,Meiser:2024zea} and are not reproduced here. Note that the hadronic matrix elements of the SM current-current operators~--~sharing the same structures as $\mathcal{Q}_1^{VLL}$ and $\mathcal{Q}_2^{VLL}$ in eq.~\eqref{eq:SM_operators}~--~have been computed in QCDF up to the next-next-to-leading order (NNLO) in $\alpha_s$~\cite{Huber:2016xod,Huber:2015bva}. We will employ these results in the phenomenological study of the type-IV diquark contributions to $\bar{B}^0_{(s)} \to D_{(s)}^{(*)+} L^-$ decays, which will be presented in the next section.

\section{Phenomenological analysis of diquark effects in \texorpdfstring{$\boldsymbol{\bar{B}^0_{(s)} \to D_{(s)}^{(*)+} L^-}$}{bar{B}0(s) to D(*)+(s) L-} decays}
\label{sec: Phenomenological_analysis}

\subsection{Input parameters}
\label{subsec:Input parameters}

\begin{table}[t]
	\newcolumntype{Y}{>{\centering\arraybackslash}X}
	\small
	\tabcolsep 0.18in
	\renewcommand*{\arraystretch}{1.3}
	\begin{center}
		\renewcommand{\tabcolsep}{3pt}
		\begin{tabularx}{1\textwidth}{|Y Y Y Y|}
			\hline\hline
			\multicolumn{4}{|c|}{QCD and electroweak parameters~\cite{ParticleDataGroup:2026aaa}}  \\
			\hline
			$G_F\left[ 10^{-5}\,\text{GeV}^{-2}\right]$ & $\alpha_s(m_{Z})$& $m_Z\left[\text{GeV}\right]$ & $m_W\left[ \text{GeV}\right]$  \\
			$1.1663785$ & $0.1180\pm0.0009$ & $91.1879\pm0.0020$ & $80.3625\pm0.0077$ \\
			\hline
		\end{tabularx}
		
		\begin{tabularx}{1\textwidth}{|Y Y Y Y Y|}
			\multicolumn{5}{|c|}{Quark masses $\left[\text{GeV}\right]$~\cite{ParticleDataGroup:2026aaa,ATLAS:2014wva}}      \\
			\hline
			$m_t^{\text{pole}}$ & $\bar{m}_b(\bar{m}_b)$ & $\bar{m}_c(\bar{m}_c)$ & $\bar{m}_s(2~\text{GeV})$ & $2\bar{m}_s/(\bar{m}_u+\bar{m}_d)$\\
			$172.60\pm0.27$ & $4.186\pm0.006$ & $1.2729\pm0.0045$ &  $0.0929\pm0.0007$ & $27.33^{+0.18}_{-0.14}$\\
			\hline
		\end{tabularx}
		
		\begin{tabularx}{1\textwidth}{|Y Y Y|}
			\multicolumn{3}{|c|}{CKM matrix elements~\cite{Charles:2004jd,ckm2025}}\\
			\hline
			$\left|V_{ud}\right|$ & $\left|V_{us}\right|$ & $\left|V_{cb}\right|\left[ 10^{-3}\right] $  \\
			$0.974343^{+0.000049}_{-0.000048}$ & $0.22504^{+0.00020}_{-0.00022}$ & $41.59^{+0.23}_{-0.84}$ \\
			\hline
		\end{tabularx}
		
		\begin{tabularx}{1\textwidth}{|Y Y Y Y|}
			\multicolumn{4}{|c|}{Lifetimes and masses of $\bar{B}^0_{(s)}$, $D_{(s)}^{(*)+}$ mesons~\cite{ParticleDataGroup:2026aaa,HeavyFlavorAveragingGroupHFLAV:2024ctg}}\\
			\hline
			$\tau_{B^0}$[ps] & $m_{B^0}$[MeV] & $m_{D^+}$[MeV] & $m_{D^{*+}}$[MeV]\\
			$1.511\pm0.003$& $5279.72\pm0.08$& $1869.66\pm0.05$ & $2010.27\pm0.04$ \\
			$\tau_{B_s^0}$[ps]  & $m_{B_s^0}$[MeV] & $m_{D_s^+}$[MeV] & $m_{D_s^{*+}}$[MeV]\\
			$1.515\pm0.006$ & $5366.93\pm0.10$ & $ 1968.35\pm0.07$ & $2112.2\pm0.4$ \\
			\hline
		\end{tabularx}
	
		\begin{tabularx}{1\textwidth}{|Y Y Y|}
		   \multicolumn{3}{|c|}{Lifetimes and masses of $B_c$, $J/\psi$ mesons~\cite{ParticleDataGroup:2026aaa,HeavyFlavorAveragingGroupHFLAV:2024ctg}}\\
	       \hline
		   $\tau_{B_c}$[ps] & $m_{B_c}$[MeV] & $m_{J/\psi}$[MeV]\\
	       $0.510\pm0.009$& $6274.47\pm0.32$& $3096.900\pm0.006$  \\
	 	   \hline
  	    \end{tabularx}
		
		\begin{tabularx}{1\textwidth}{|Y | Y Y Y Y | Y|}
			\multicolumn{6}{|c|}{Masses, decay constants and Gegenbauer moments of light mesons}\\
			\hline
			& $\pi^-$ & $K^-$ & $\rho^-$ & $K^{*-}$ & References\\
			\hline
			$m_L$[MeV] & $139.57$ & $493.68$ & $775.11 $ & $891.88 $ & \cite{ParticleDataGroup:2026aaa} \\
			\hline
			$f_L$[MeV] & $130.2\pm1.7$ & $155.6\pm0.4$ & $216\pm6$ & $211\pm7$ & \\
			$f_L^\perp$[MeV] & -- & -- & $160\pm11$ & $163\pm8$ &\cite{Rosner:2015wva,Bharucha:2015bzk,Dimou:2012un} \\
			\hline
			$\alpha_1^L$ & -- & $-0.0525_{-0.0031}^{+0.0033}$ & -- & $0.06\pm0.04$ & \\
			$\alpha_2^L$ & $0.116_{-0.020}^{+0.019}$ & $0.106_{-0.016}^{+0.015}$ & $0.17\pm0.07$ & $0.16\pm0.09$ &\cite{Bharucha:2015bzk,Dimou:2012un,Arthur:2010xf,RQCD:2019osh} \\
			\hline\hline
		\end{tabularx}
        \caption{Summary of the theoretical input parameters employed in this work. The transverse decay constants $f_L^\perp$ are determined at the scale $\mu=1~\text{GeV}$, while the Gegenbauer moments of light pseudoscalar and vector mesons are evaluated at $\mu=2~\text{GeV}$ and $\mu=1~\text{GeV}$, respectively. \label{table:input} }
        \end{center}
\end{table}

Within the QCDF framework, the predicted observables for $\bar{B}^0_{(s)} \to D_{(s)}^{(*)+} L^-$ decays depend on a number of input parameters, such as QCD and electroweak parameters, quark masses, CKM matrix elements, $B$-meson lifetimes, light-meson decay constants, and the Gegenbauer moments of light mesons. The numerical inputs used in this work, along with the corresponding references, are collected in table~\ref{table:input}. Moreover, because the decay $B_{c}^{-}\to J/\psi \pi^{-}$ is driven by the same quark-level $b\to c\bar{u}d$ transition, we also use the ratio of non-leptonic to semi-leptonic branching fractions $R_{\pi/\mu\nu_{\mu}}$ to constrain the diquark couplings; the associated input parameters are provided in the same table. We use the two-loop relation between the pole and $\MSbar$ masses~\cite{Chetyrkin:2000yt} to convert the top-quark pole mass $m_t^{\text{pole}}$ into the scale-invariant mass $\overline{m}_t(\overline{m}_t)$. The strong coupling $\alpha_s$ is evaluated at three-loop order~\cite{Czakon:2004bu}. To assess theoretical uncertainties for a given observable, we vary the relevant inputs within their $1\sigma$ intervals and combine the resulting individual errors in quadrature. We also include the uncertainty associated with varying the renormalization scale over $\mu_b \in [m_b/2, 2m_b]$.

\begin{table}[t]
	\tabcolsep 0.16in
	\newcolumntype{Y}{>{\centering\arraybackslash}X}
	\renewcommand*{\arraystretch}{1.3}
	\begin{center}
		\renewcommand{\tabcolsep}{3pt}
		\begin{tabularx}{1\textwidth}{|Y Y Y Y|}
			\hline\hline
			$F_0^{B\to D}(m^2_\pi)$ & $A_0^{B\to D^*}(m^2_\pi)$ & $F_0^{B\to D}(m^2_K)$ & $A_0^{B\to D^*}(m^2_K)$\\
			$0.644\pm0.025$& $0.582\pm0.038$& $0.648\pm0.024$ & $0.589\pm0.037$ \\
			\hline
			$F_0^{B\to D}(m^2_\rho)$ & $F_+^{B\to D}(m^2_\rho)$ & $F_0^{B\to D}(m^2_{K^*})$ & $F_+^{B\to D}(m^2_{K^*})$ \\
			$0.654\pm0.023$& $0.662\pm0.023$& $0.657\pm0.023$ & $0.667\pm0.023$ \\
			\hline
			$F_0^{B_s\to D_s}(m^2_\pi)$ & $A_0^{B_s\to D_s^*}(m^2_\pi)$ & $F_0^{B_s\to D_s}(m^2_K)$ & $A_0^{B_s\to D_s^*}(m^2_K)$\\
			$0.664\pm0.012$& $0.619\pm0.033$& $0.667\pm0.012$ & $0.625\pm0.033$ \\
			\hline
			$F_T^{B\to D}(m^2_\rho)$ & $F_T^{B\to D}(m^2_{K^*})$ & & \\
			$0.823\pm0.027$& $0.827\pm0.027$ & & \\
			\hline
			$V^{B_c\to J/\psi}(m^2_\pi)$ & $A_0^{B_c\to J/\psi}(m^2_\pi)$ & $A_1^{B_c\to J/\psi}(m^2_\pi)$ & $A_2^{B_c\to J/\psi}(m^2_\pi)$\\
			$0.752\pm0.058$& $0.466\pm0.025$& $0.456\pm0.019$ & $0.438\pm0.071$ \\
			\hline\hline
		\end{tabularx}
		\caption{Numerical values of the relevant transition form factors $F_0^{B_{(s)}\to D^{}_{(s)}}(q^2)$, $A_0^{B_{(s)}\to D^{*}_{(s)}}(q^2)$, $F_+^{B\to D}(q^2)$, $F_T^{B\to D}(q^2)$, as well as $V^{B_c\to J/\psi}(q^2)$ and $A_{0,1,2}^{B_c\to J/\psi}(q^2)$ evaluated at $q^2=m_L^2$. \label{table:formfactor} }
	\end{center}
\end{table}

Building upon our earlier analysis in ref.~\cite{Cai:2021mlt}, this work incorporates the most recent experimental measurements of the branching fractions~\cite{HeavyFlavorAveragingGroupHFLAV:2024ctg,ParticleDataGroup:2026aaa}, together with updated fits of the $B_{(s)}\to D_{(s)}^{(*)}$ transition form factors. Their $q^2$ dependence is described using the Boyd-Grinstein-Lebed (BGL) parameterization~\cite{Boyd:1994tt,Boyd:1995sq,Boyd:1997kz}. In particular, for the form factors $F_0^{B_{(s)}\to D^{}_{(s)}}$, $A_0^{B_{(s)}\to D^{*}_{(s)}}$, and $F_+^{{B}\to D}$, we use the fit results obtained from the combined ``Lattice $\oplus$ LCSR'' strategy of ref.~\cite{Cui:2023jiw}. For $F_T^{B\to D}$, we take the results derived in refs.~\cite{Bernlochner:2017jka,Bernlochner:2022ywh} within the heavy-quark effective theory, employing the ``$\text{L}_{w\geq1}$'' scheme. For the $B_c \to J/\psi$ transition form factors, we adopt the lattice QCD determinations from ref.~\cite{Harrison:2025kxm}, which are parameterized in the BGL framework across the full kinematic region. The numerical inputs for all these form factors are collected in table~\ref{table:formfactor}. In addition, based on the combined ``$\text{Lattice} \oplus \text{LCSR} \oplus \text{Exp.}$'' fits for the $B_{(s)}\to D_{(s)}^{(*)}$ form factors~\cite{Cui:2023jiw}, we extract the differential decay rates for the semi-leptonic decays $B_{(s)}^0\to D_{(s)}^{(*)+} \ell^- \bar{\nu}_{\ell}$ at $q^2=m_L^2$, which we then use to update the experimental values of the ratios $R_{(s)L}^{(*)}$ (see eq.~\eqref{eq:nonlep2Ratio} below).

\subsection{SM predictions of the observables}
\label{subsec:SM_predictions}

In the QCDF approach, the decay widths for the class-I $\bar{B}^0_{(s)} \to D_{(s)}^{(*)+} L^-$ decays can be written in terms of the light-meson decay constants $f_L$, the $B_{(s)}\to D_{(s)}^{(*)}$ hadronic form factors, and the effective parameters $a_1(D_{(s)}^{(*)+} L^-)$, namely
\begin{align}\label{eq:decayrate_B2DL}
	 \Gamma(\bar{B}^0_{(s)} \to D_{(s)}^+ P^-) 
	&= \frac{G_F^2 (m_{B_{(s)}}^2-m_{D_{(s)}}^2)^2|\vec{q}|}{16\pi m_{B_{(s)}}^2}
	|V_{cb}V^*_{uq}|^2 |a_1(D_{(s)}^{+} P^-)|^2 f_{P}^2
	\big[F_0^{B_{(s)}\to D_{(s)}}(m_{P}^2)\big]^2 \, , \nonumber \\[0.2cm] 
	\Gamma(\bar{B}^0_{(s)} \to D_{(s)}^{*+} P^-) 
	&= \frac{G_F^2|\vec{q}|^3}{4\pi}
	|V_{cb}V^*_{uq}|^2 |a_1(D_{(s)}^{*+} P^-)|^2 f_{P}^2
	\big[A_0^{B_{(s)}\to D_{(s)}^*}(m_{P}^2)\big]^2 \, , \nonumber \\[0.2cm] 
	\Gamma(\bar{B}^0_{(s)} \to D_{(s)}^+ V^-) 
	&= \frac{G_F^2|\vec{q}|^3}{4\pi}
	|V_{cb}V^*_{uq}|^2 |a_1(D_{(s)}^{+} V^-)|^2 f_{V}^2
	\big[F_+^{B_{(s)}\to D_{(s)}}(m_{V}^2)\big]^2 \,,
\end{align}
with $G_F$ denoting the Fermi constant, $V_{cb}$ and $V_{uq}$ the relevant CKM elements, and $|\vec{q}|$ the magnitude of the three-momentum of the final state in the $B_{(s)}$ rest frame. It is given by
\begin{align}
	|\vec{q}|=\frac{1}{2m_{B_{(s)}}}\sqrt{\left(m_{B_{(s)}}^2-m_{D_{(s)}^{(*)}}^2-m_L^2\right)^2-4m_{D_{(s)}^{(*)}}^2m_L^2} \, .
\end{align}
By multiplying the decay widths with the $B_{(s)}$-meson lifetimes $\tau_{B_{(s)}}$, one obtains the corresponding branching fractions
\begin{equation}
	\mathcal{B}_{(s)}^{(*)}(L) = \mathcal{B}(\bar{B}^0_{(s)} \to D_{(s)}^{(*)+} L^-) = \tau_{B_{(s)}} \, 
    \Gamma(\bar{B}^0_{(s)} \to D_{(s)}^{(*)+} L^-) \, .
\end{equation}

\begin{table}[t] 
	\tabcolsep 0.13in
	\renewcommand*{\arraystretch}{1.3}
	\begin{center}
		\begin{tabular}{|cccccc|}
			\hline
			\hline
			Decay channel & LO & NLO & NNLO & Exp.~\cite{HeavyFlavorAveragingGroupHFLAV:2024ctg,ParticleDataGroup:2026aaa} & Deviation $(\sigma)$ \\ \hline
			$\Bdecay{D^+}{\pi^-}$ & $3.73$ & $3.96^{+0.36}_{-0.38}$ & $4.07^{+0.34}_{-0.37}$ & $2.48\pm0.12$ & $4.0$ \\ 
			$\Bdecay{D^{*+}}{\pi^-}$ & $2.83$ & $3.01^{+0.42}_{-0.42}$ & $3.10^{+0.42}_{-0.42}$ & $2.652\pm0.064$ & $1.0$ \\  
			$\Bdecay{D^+}{\rho^-}$ & $9.83$ & $10.43^{+1.03}_{-1.08}$ & $10.72^{+1.00}_{-1.06}$ & $7.6\pm1.2$ & $1.9$ \\ 
			\hline
			$\Bdecay{D^+}{K^-}$ & $2.84$ & $3.02^{+0.25}_{-0.28}$ & $3.11^{+0.24}_{-0.27}$ & $2.03\pm0.12$ & $3.7$ \\
			$\Bdecay{D^{*+}}{K^-}$ & $2.13$ & $2.26^{+0.30}_{-0.30}$ & $2.34^{+0.31}_{-0.30}$ & $2.146\pm0.079$ & $0.6$ \\
			$\Bdecay{D^{+}}{K^{*-}}$ & $4.93$ & $5.23^{+0.54}_{-0.56}$ & $5.35^{+0.52}_{-0.55}$ & $4.60\pm0.78$ & $0.8$ \\  
			\hline
			$\Bdecay[\bar{B}^0_s]{D^+_s}{\pi^-}$ & $4.05$ & $4.30^{+0.24}_{-0.30}$ & $4.42^{+0.21}_{-0.28}$ & $2.83^{+0.18}_{-0.19}$ & $4.8$ \\
			$\Bdecay[\bar{B}^0_s]{D^{*+}_s}{\pi^-}$ & $3.27$ & $3.47^{+0.41}_{-0.42}$ & $3.58^{+0.41}_{-0.42}$ & $2.19^{+0.62}_{-0.53}$ & $1.9$ \\  
			\hline
			$\Bdecay[\bar{B}^0_s]{D_s^+}{K^-}$ & $3.09$ & $3.28^{+0.17}_{-0.21}$ & $3.38^{+0.14}_{-0.19}$ & $2.12\pm0.14$ & $5.3$ \\
			$\Bdecay[\bar{B}^0_s]{D_s^{*+}}{K^-}$ & $2.45$ & $2.61^{+0.30}_{-0.30}$ & $2.70^{+0.30}_{-0.30}$ & $1.49^{+0.45}_{-0.37}$ & $2.2$ \\
			\hline\hline
		\end{tabular}
	\caption{Comparison of our theoretical predictions at various orders in $\alpha_s$ with the experimental measurements of the branching ratios for $\bar{B}^0_{(s)} \to D_{(s)}^{(*)+} L^-$ decays. Results for the $b \to c\bar{u}d$ and $b \to c\bar{u}s$ transitions are quoted in units of $10^{-3}$ and $10^{-4}$, respectively. The last column reports the difference between the NNLO predictions and the measured values. \label{tab:SMBr} }
	\end{center}
\end{table}

To more transparently assess the present mismatches in the $\bar{B}^0_{(s)} \to D_{(s)}^{(*)+} L^-$ decays, we confront in table~\ref{tab:SMBr} the SM predictions computed at different orders in $\alpha_s$ with the most recent experimental results~\cite{HeavyFlavorAveragingGroupHFLAV:2024ctg,ParticleDataGroup:2026aaa}. From table~\ref{tab:SMBr}, it is evident that several theoretical expectations based on QCDF differ from the measurements at the level of roughly $4\!-\!6\sigma$. Such pronounced tensions prompt us to explore whether higher-order power corrections within the SM have been underestimated~\cite{Huber:2016xod,Bordone:2020gao,Meiser:2024zea,Fleischer:2010ca,Endo:2021ifc,Piscopo:2023opf,Davies:2024vmv,Yaouanc:2024gjl} or whether, more intriguingly, they hint at NP effects beyond the SM~\cite{Bobeth:2014rda,Brod:2014bfa,Lenz:2019lvd,Iguro:2020ndk,Cai:2021mlt,Bordone:2021cca,Fleischer:2021cct,Fleischer:2021cwb,Gershon:2021pnc,Lenz:2022pgw,Panuluh:2024cxc,Atkinson:2024hqp,Meiser:2024zea,Deng:2025znr,Araz:2026zlu,Guo:2026oed}. In this work, we study the type-IV diquark model as a plausible NP explanation for the observed discrepancies.

\begin{table}[t]
	\centering
	\tabcolsep 0.21in
	\renewcommand{\arraystretch}{1.3} 
	\begin{tabular}{|l c c c c c|}
		\hline \hline
		$R^{(*)}_{(s)L}$ & LO & NLO & NNLO & Exp. & Deviation ($\sigma$) \\
		\hline
		$R_\pi$ & $1.01$ & $1.07^{+0.04}_{-0.05}$ & $1.10^{+0.03}_{-0.03}$ & $0.68 \pm 0.04$ & $7.7$ \\
		$R^*_\pi$ & $1.00$ & $1.06^{+0.04}_{-0.05}$ & $1.09^{+0.03}_{-0.04}$ & $0.83 \pm 0.04$ & $5.0$ \\
		$R_\rho$ & $2.77$ & $2.94^{+0.19}_{-0.19}$ & $3.02^{+0.17}_{-0.18}$ & $2.17 \pm 0.35$ & $2.1$ \\
		\hline
		$R_K$ & $0.78$ & $0.83^{+0.03}_{-0.03}$ & $0.85^{+0.01}_{-0.02}$ & $0.56 \pm 0.04$ & $6.6$ \\
		$R^*_K$ & $0.71$ & $0.75^{+0.03}_{-0.03}$ & $0.78^{+0.01}_{-0.02}$ & $0.64 \pm 0.04$ & $3.3$ \\
		$R_{K^*}$ & $1.41$ & $1.49^{+0.11}_{-0.11}$ & $1.53^{+0.10}_{-0.10}$ & $1.34 \pm 0.23$ & $0.8$ \\
		\hline
		$R_{s\pi}$ & $1.01$ & $1.07^{+0.04}_{-0.05}$ & $1.10^{+0.03}_{-0.03}$ & $0.76 \pm 0.05$ & $5.2$ \\
		$R^*_{s\pi}$ & $1.00$ & $1.06^{+0.04}_{-0.05}$ & $1.09^{+0.03}_{-0.04}$ & $0.66 \pm 0.18$ & $2.4$ \\
		\hline
		$R_{sK}$ & $0.78$ & $0.83^{+0.03}_{-0.03}$ & $0.85^{+0.01}_{-0.02}$ & $0.58 \pm 0.04$ & $6.0$ \\
		$R^*_{sK}$ & $0.71$ & $0.75^{+0.03}_{-0.03}$ & $0.78^{+0.01}_{-0.02}$ & $0.43 \pm 0.12$ & $2.9$ \\
		\hline \hline
	\end{tabular}
	\caption{Comparison between theoretical predictions at different orders in $\alpha_s$ and experimental measurements for the ratios $R^{(*)}_{(s)L}$. The values for the $b \to c\bar{u}d$ and $b \to c\bar{u}s$ transitions are given in units of $\text{GeV}^2$ and $10^{-1}\,\text{GeV}^2$, respectively. The last column shows the deviation between the NNLO predictions and experimental data.}
	\label{tab:SMRatio}
\end{table}

For $\bar{B}^0_{(s)} \to D_{(s)}^{(*)+} L^-$ decays, a very informative observable for scrutinizing the factorization hypothesis is the ratio of the non-leptonic decay width to the corresponding differential semi-leptonic decay width evaluated at the kinematic point $q^2=m_L^2$~\cite{Bjorken:1988kk,Neubert:1997uc,Beneke:2000ry}:
\begin{align} \label{eq:nonlep2Ratio}
	R_{(s)L}^{(*)} \equiv \frac{\Gamma(\bar{B}^0_{(s)} \to D_{(s)}^{(*)+} L^-)}{d\Gamma(\bar{B}^0_{(s)}\to  D_{(s)}^{(*)+} \ell^-\bar{\nu}_{\ell})/dq^2\mid_{q^2=m_L^2}} = 6\pi^2\,|V_{uq}|^2\,f_L^2\,|a_1(D_{(s)}^{(*)+} L^-)|^2\, X_L^{(*)} \, .
\end{align}
If the light-lepton masses are neglected, one obtains $X_L=X_L^*=1$ when $L^-$ is a vector meson~\cite{Neubert:1997uc,Beneke:2000ry}. In contrast, for a light pseudoscalar meson $L^-$, the factor $X_L$ is determined by a combination of form-factor ratios and kinematic prefactors (see eq.~(68) of ref.~\cite{Neubert:1997uc} for explicit formulae). To check whether the anomalies mentioned above remain visible in these theoretically cleaner observables, we list in table~\ref{tab:SMRatio} the SM predictions for $R_{(s)L}^{(*)}$ at various orders in $\alpha_s$ and confront them with the current experimental measurements. Notably, for several channels the discrepancies between theory and experiment in $R_{(s)L}^{(*)}$ are significantly larger than those found for the corresponding branching fractions. The reason is that, in the SM, these ratios depend only weakly on the CKM element $|V_{cb}|$ and exhibit substantially reduced form-factor uncertainties relative to the branching fractions. 

Once the type-IV diquark effects are included, however, the picture becomes more nuanced. Even though the ratios still profit from smaller form-factor uncertainties, the dependence on $V_{cb}$ returns through the semi-leptonic denominator. Consequently, the overall theory uncertainties for these ratios are only marginally smaller than, or can even be comparable to, those for the corresponding branching fractions. On the experimental side, the branching fractions are measured directly, whereas the ratios are obtained by combining non-leptonic results with semi-leptonic inputs, which introduces additional uncertainties from $|V_{cb}|$ and $B_{(s)}\to D_{(s)}^{(*)}$ transition form factors. Since this experimental limitation offsets the theoretical advantage, we therefore rely on the branching fractions rather than these ratios when constraining the NP parameter space.

\subsection{Constraints on the type-IV diquark couplings}
\label{subsec:constraint_DQ}

We now examine the phenomenological implications of the type-IV diquark scenario for the $\BtoDL$ decays. At the quark level, these processes are mediated by the $b\to c\bar{u}d(s)$ transitions, represented by the Feynman diagrams in figures~\ref{fig:dia_tree_full} and~\ref{fig:dia_oneloop_full} for the particular choice of flavor indices $i=1$, $j=3$, $l=1(2)$, and $k=2$. Expressed in terms of the diquark coupling constants, the corresponding branching fractions are given by
\begin{align}\label{eq:BR_NP}
	\mathcal{B}(\Bdecay{D^+}{\pi^-}) =& \left| (2.01+0.09i)-(0.96+0.09i)\lamPLL-(3.44+0.28i)\lamPLR \right. \nonumber \\
	&\left. + (3.44+0.28i)\lamPRL+(0.96+0.09i)\lamPRR \right|^2 \times 10^{-3} \, , \nonumber \\[2mm]
	\mathcal{B}(\Bdecay{D^{*+}}{\pi^-}) =& \left| (1.76+0.06i)-(0.83+0.05i)\lamPLL+(1.83+0.26i)\lamPLR \right .\nonumber \\  
	&\left. + (1.83+0.26i)\lamPRL-(0.83+0.05i)\lamPRR \right|^2 \times 10^{-3} \, , \nonumber \\[2mm]
	\mathcal{B}(\Bdecay{D^+}{\rho^-}) =& \left| (3.27+0.15i)-(1.55+0.14i)\lamPLL+(0.06+0.02i)\lamPLR \right. \nonumber \\  
	&\left. + (0.06+0.02i)\lamPRL-(1.55+0.14i)\lamPRR \right|^2 \times 10^{-3} \, ,  \nonumber \\[2mm]
	\mathcal{B}(\Bdecay{D^+}{K^-}) =& \left| (1.76+0.08i)-(3.62+0.31i)\lamKLL -(11.48+0.92i)\lamKLR \right. \nonumber \\  
	&\left. + (11.48+0.92i)\lamKRL+(3.62+0.31i)\lamKRR \right|^2 \times 10^{-4} \, ,  \nonumber \\[2mm]
	\mathcal{B}(\Bdecay{D^{*+}}{K^-}) =& \left| (1.53+0.05i)-(3.14+0.19i)\lamKLL +(6.05+0.85i)\lamKLR \right. \nonumber \\  
	&\left. + (6.05+0.85i)\lamKRL-(3.14+0.19i)\lamKRR \right|^2 \times 10^{-4} \, ,  \nonumber \\[2mm]
	\mathcal{B}(\Bdecay{D^{+}}{K^{*-}}) =& \left| (2.31+0.11i)-(4.74+0.46i)\lamKLL +(0.24+0.06i)\lamKLR \right. \nonumber \\  
	&\left. + (0.24+0.06i)\lamKRL-(4.74+0.46i)\lamKRR \right|^2 \times 10^{-4} \, ,  \nonumber \\[2mm]
	\mathcal{B}(\Bdecay[\bar{B}^0_s]{D^+_s}{\pi^-}) =& \left| (2.10+0.10i)-(1.00+0.09i)\lamPLL-(3.59+0.29i)\lamPLR  \right. \nonumber \\  
	&\left. + (3.59+0.29i)\lamPRL+(1.00+0.09i)\lamPRR \right|^2 \times 10^{-3} \, ,  \nonumber \\[2mm]
	\mathcal{B}(\Bdecay[\bar{B}^0_s]{D^{*+}_s}{\pi^-}) =& \left| (1.89+0.07i)-(0.90+0.06i)\lamPLL+(1.96+0.27i)\lamPLR \right. \nonumber \\  
	&\left. + (1.96+0.27i)\lamPRL-(0.90+0.06i)\lamPRR \right|^2 \times 10^{-3} \, ,  \nonumber \\[2mm]
	\mathcal{B}(\Bdecay[\bar{B}^0_s]{D_s^+}{K^-}) =& \left| (1.84+0.08i)-(3.77+0.33i)\lamKLL-(11.97+0.96i)\lamKLR \right. \nonumber \\  
	&\left. + (11.97+0.96i)\lamKRL+(3.77+0.33i)\lamKRR \right|^2 \times 10^{-4} \, ,  \nonumber \\[2mm]
	\mathcal{B}(\Bdecay[\bar{B}^0_s]{D_s^{*+}}{K^-}) =& \left| (1.64+0.06i)-(3.37+0.20i)\lamKLL +(6.50+0.91i)\lamKLR \right. \nonumber \\
	&\left. +(6.50+0.91i)\lamKRL-(3.37+0.20i)\lamKRR \right|^2 \times 10^{-4} \, ,
\end{align}
where we have regrouped the products of the coupling constants $\lambda^{L,R}_{ij}$ in eq.~\eqref{eq:WC_SM_oneloop} into the effective couplings $\Delta_d^{XY}$ and $\Delta_s^{XY}$ (with $X,Y \in \{L,R\}$), defined by
\begin{align} \label{eq:lamsimd}
	\Delta_{d(s)}^{LL} &= \lambda^L_{13}\lambda^{L*}_{21(2)} \, , &
	\Delta_{d(s)}^{LR} &= \lambda^L_{13}\lambda^{R*}_{21(2)} \, , \nonumber \\[2mm]
	\Delta_{d(s)}^{RL} &= \lambda^R_{13}\lambda^{L*}_{21(2)} \, , &
	\Delta_{d(s)}^{RR} &= \lambda^R_{13}\lambda^{R*}_{21(2)} \, .
\end{align}

In what follows, we assume that all coupling constants $\lambda^{L,R}_{ij}$, and hence $\Delta_{d,s}^{XY}$, are real. Although the theoretical uncertainties are not shown explicitly in eq.~\eqref{eq:BR_NP}, they are included in our numerical analysis. The decay modes with $L=\pi,\rho$ are sensitive only to $\Delta_d^{XY}$, while those with $L=K^{(*)}$ depend solely on $\Delta_s^{XY}$. This separation enables us to place constraints on the two flavor sectors independently. Since the branching fractions probe each of the four coupling combinations $\Delta_{d(s)}^{XY}$ (see eq.~\eqref{eq:BR_NP}), we consider two benchmark setups: (i) only one effective coupling different from zero (case~I) and (ii) two effective couplings different from zero (case~II).

\subsubsection{Case I: single non-zero effective coupling}
\label{subsubsection:CaseI}

\begin{figure}[t]
	\centering
    \includegraphics[width=0.82\textwidth]{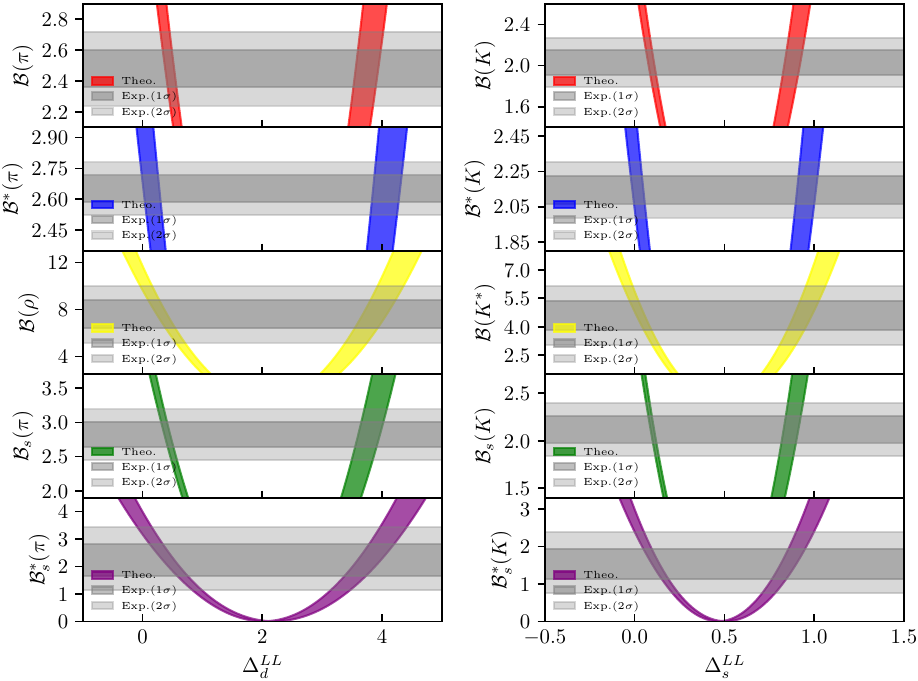}
	\caption{Bounds on the effective couplings $\Delta_{d}^{LL}$ (left) and $\Delta_{s}^{LL}$ (right) derived from the measured branching fractions of $\BtoDL$ decays with $L=\pi,\rho$ and $L=K^{(*)}$, respectively. The colored bands represent the theory expectations, while the horizontal shaded areas denote the experimental confidence ranges at $1\sigma$ (darker) and $2\sigma$ (lighter) for each decay mode. \label{figure:LLSpace}}
\end{figure}

\begin{figure}[t]
	\centering
    \includegraphics[width=0.82\textwidth]{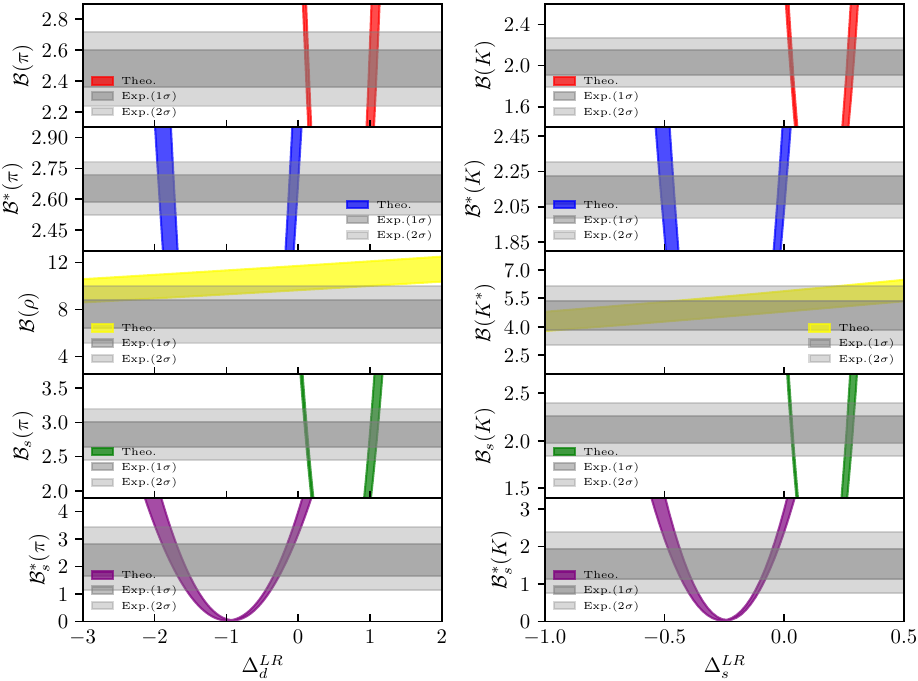}
	\caption{Bounds on the effective couplings $\lamPLR$ (left) and $\lamKLR$ (right) derived from the measured branching fractions of $\BtoDL$ decays with $L=\pi,\rho$ and $L=K^{(*)}$, respectively. The other captions are the same as in figure~\ref{figure:LLSpace}. \label{figure:LRSpace}}
\end{figure}

\begin{figure}[htbp]
	\centering
    \includegraphics[width=0.82\textwidth]{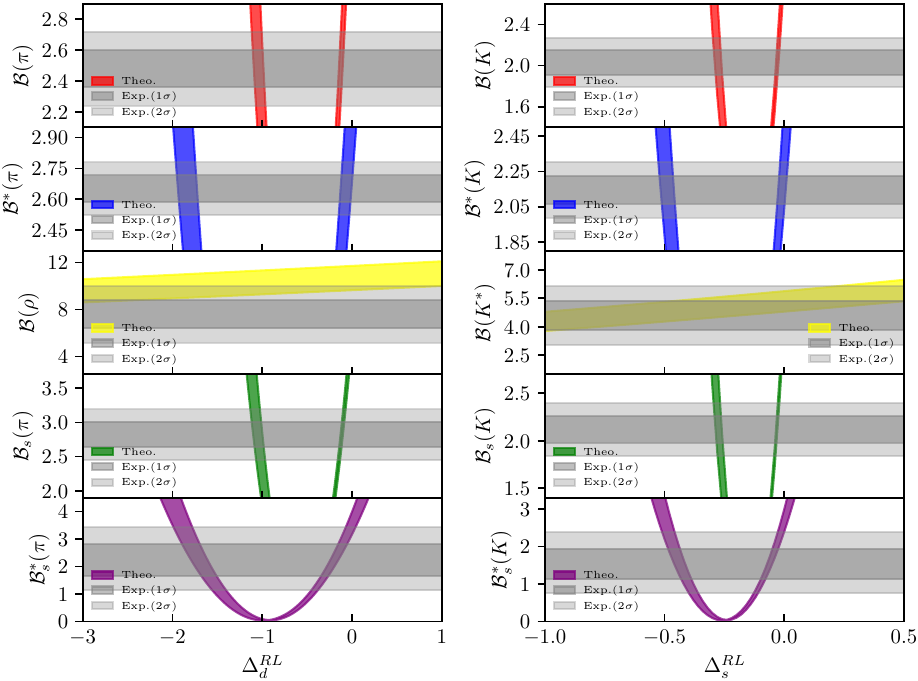}
	\caption{Bounds on the effective couplings $\lamPRL$ (left) and $\lamKRL$ (right) derived from the measured branching fractions of $\BtoDL$ decays with $L=\pi,\rho$ and $L=K^{(*)}$, respectively. The other captions are the same as in figure~\ref{figure:LLSpace}. \label{figure:RLSpace}}
\end{figure}

\begin{figure}[htbp]
	\centering
    \includegraphics[width=0.82\textwidth]{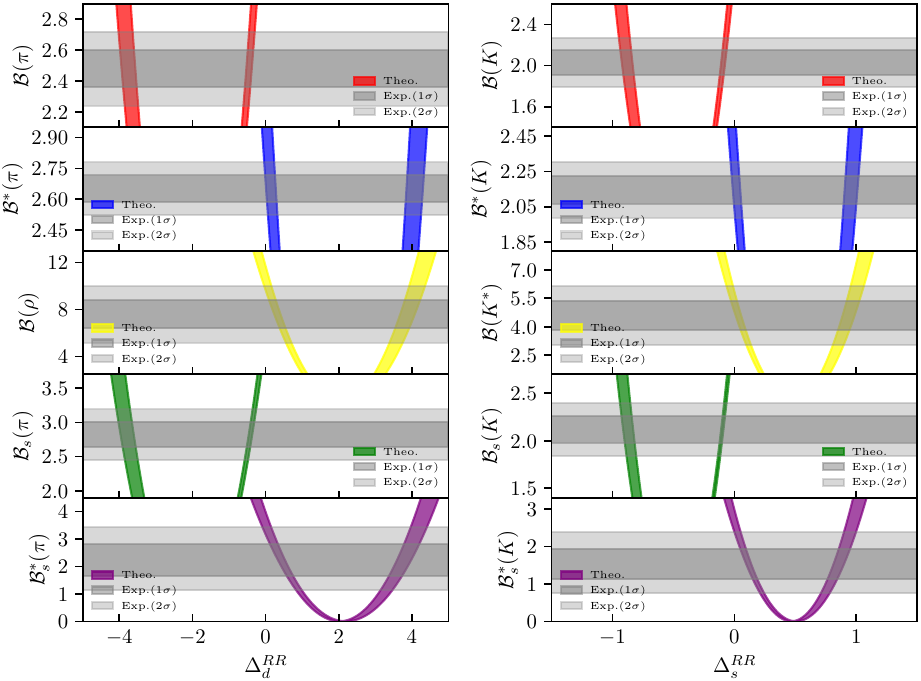}
	\caption{Bounds on the effective couplings $\lamPRR$ (left) and $\lamKRR$ (right) derived from the measured branching fractions of $\BtoDL$ decays with $L=\pi,\rho$ and $L=K^{(*)}$, respectively. The other captions are the same as in figure~\ref{figure:LLSpace}. \label{figure:RRSpace}}
\end{figure}

For case~I, we study the four chiral configurations with $XY=LL$, $LR$, $RL$, and $RR$ one by one. For each non-zero effective coupling $\Delta_{d(s)}^{XY}$, eq.~\eqref{eq:lamsimd} implies the presence of two non-vanishing coupling constants, $\lambda_{ij}^{L,R}$.
\begin{enumerate}
\item[\textbullet] \textbf{S1:} 
We first study the case in which the only non-vanishing contribution comes from the product of two left–left couplings, $\Delta_{d(s)}^{LL}=\lambda^L_{13}\lambda^{L*}_{21(2)}$. Under the constraints of the experimental branching fractions listed in table~\ref{tab:SMBr}, we derive the allowed parameter regions displayed in figure~\ref{figure:LLSpace}. One can see that $\lamPLL$ and $\lamKLL$ individually can account for the observed anomalies in $\bar{B}^0_{(s)} \to D_{(s)}^{(*)+} L^-$ decays within $1\sigma$. The resulting $1\sigma$ ranges read
\begin{align} \label{lamLL1S}
	\lamPLL \in [3.830, 3.981]\,, \qquad \lamKLL \in [0.897, 0.926]\,.
\end{align}
When extending to $2\sigma$, the allowed intervals broaden to
\begin{align} \label{lamLL2S}
	\lamPLL \in [0.297, 0.324] \cup [3.808, 4.022]\,, \quad
	\lamKLL \in [0.064, 0.065] \cup [0.889, 0.939]\,.
\end{align} 

\item[\textbullet] \textbf{S2:} 
We next examine the case in which only the product of left–right couplings, $\Delta_{d(s)}^{LR}=\lambda^L_{13}\lambda^{R*}_{21(2)}$, is non-zero. The resulting bounds are displayed in figure~\ref{figure:LRSpace}. Evidently, neither $\lamPLR$ nor $\lamKLR$ can explain the observed anomalies, leaving no parameter space allowed, even at the $2\sigma$ level.

\item[\textbullet] \textbf{S3:} 
In the third scenario, we take only the product of right–left couplings, $\Delta_{d(s)}^{RL}=\lambda^R_{13}\lambda^{L*}_{21(2)}$, to be non-vanishing. The corresponding parameter regions are displayed in figure~\ref{figure:RLSpace}. At the $1\sigma$ level, $\lamPRL$ does not explain the discrepancies in the $L=\pi,\rho$ modes, while $\lamKRL$ can accommodate the anomalies observed in the $L=K^{(*)}$ channels. For $\lamKRL$, the allowed $1\sigma$ range is given by 
\begin{align} \label{lamRL1S}
	 \lamKRL \in [-0.029, -0.025]\,.
\end{align}
At the $2\sigma$ confidence level, however, both effective couplings are able to describe the observed deviations, leading to the allowed intervals
\begin{align} \label{lamRL2S}
	\lamPRL \in [-0.148, -0.082]\,, \qquad \lamKRL \in [-0.034, -0.020]\,.
\end{align}

\item[\textbullet] \textbf{S4:} 
Lastly, we consider the case in which only the product of two right–right couplings, $\Delta_{d(s)}^{RR}=\lambda^R_{13}\lambda^{R*}_{21(2)}$, is taken to be non-zero. As shown in figure~\ref{figure:RRSpace}, neither option can account for the discrepancies between theoretical predictions and experimental results for the $\BtoDL$ decays, leaving no viable parameter space even at the $2\sigma$ level.
\end{enumerate}

\subsubsection{Case II: two non-zero effective couplings}
\label{subsubsec:CaseII}

\begin{figure}[t]
	\centering
    \includegraphics[width=0.99\textwidth]{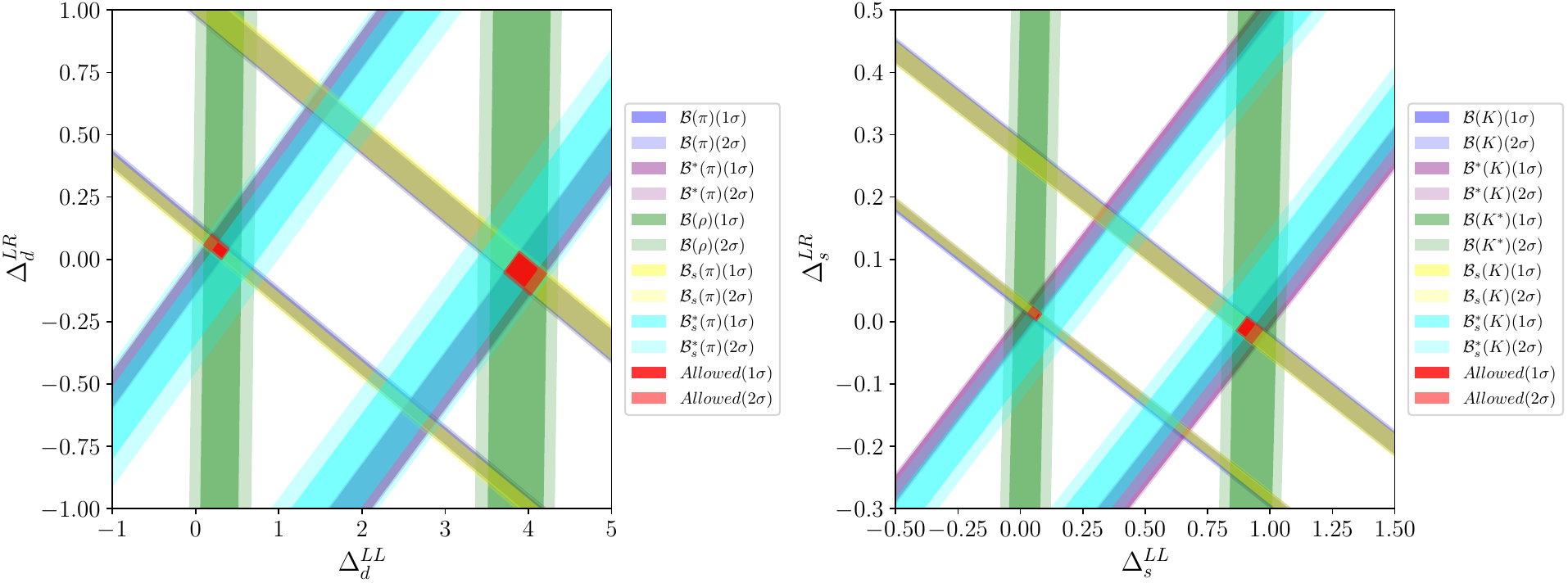}
	\caption{Constraints on the effective couplings $(\Delta_d^{LL}, \Delta_d^{LR})$ (left) and $(\Delta_s^{LL}, \Delta_s^{LR})$ (right) from the measured branching fractions of $\bar{B}^0_{(s)} \to D_{(s)}^{(*)+} L^-$ decays with $L = \pi, \rho$ and $L = K^{(*)}$, respectively. The shaded colored bands represent the $1\sigma$ (darker) and $2\sigma$ (lighter) confidence ranges for each decay channel, while the red regions indicate the allowed parameter spaces at the $1\sigma$ (dark red) and $2\sigma$ (light red) levels. \label{figure:LLLRSpace} }
\end{figure}

\begin{figure}[t]
	\centering
    \includegraphics[width=0.99\textwidth]{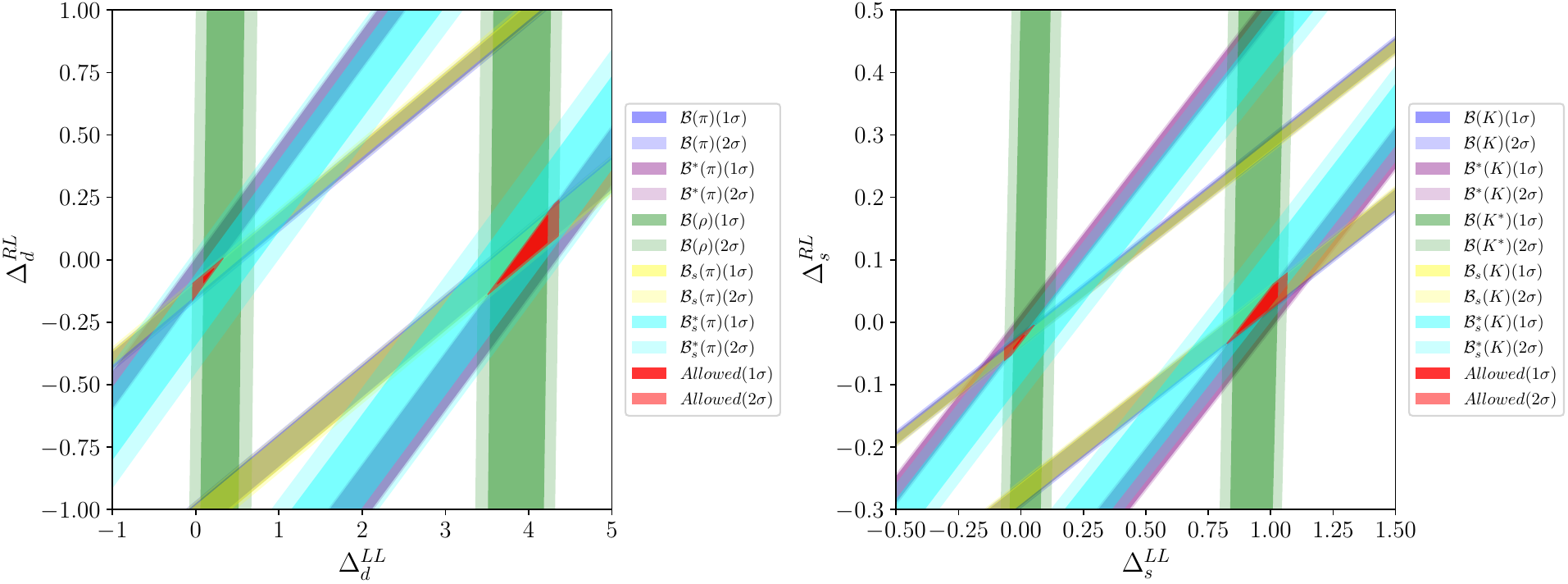}
	\caption{Constraints on the effective couplings $(\Delta_d^{LL}, \Delta_d^{RL})$ (left) and $(\Delta_s^{LL}, \Delta_s^{RL})$ (right) from the measured branching fractions of $\bar{B}^0_{(s)} \to D_{(s)}^{(*)+} L^-$ decays with $L = \pi, \rho$ and $L = K^{(*)}$, respectively. The other captions are the same as in figure~\ref{figure:LLLRSpace}. \label{figure:LLRLSpace} }
\end{figure}

\begin{figure}[t]
	\centering
    \includegraphics[width=0.99\textwidth]{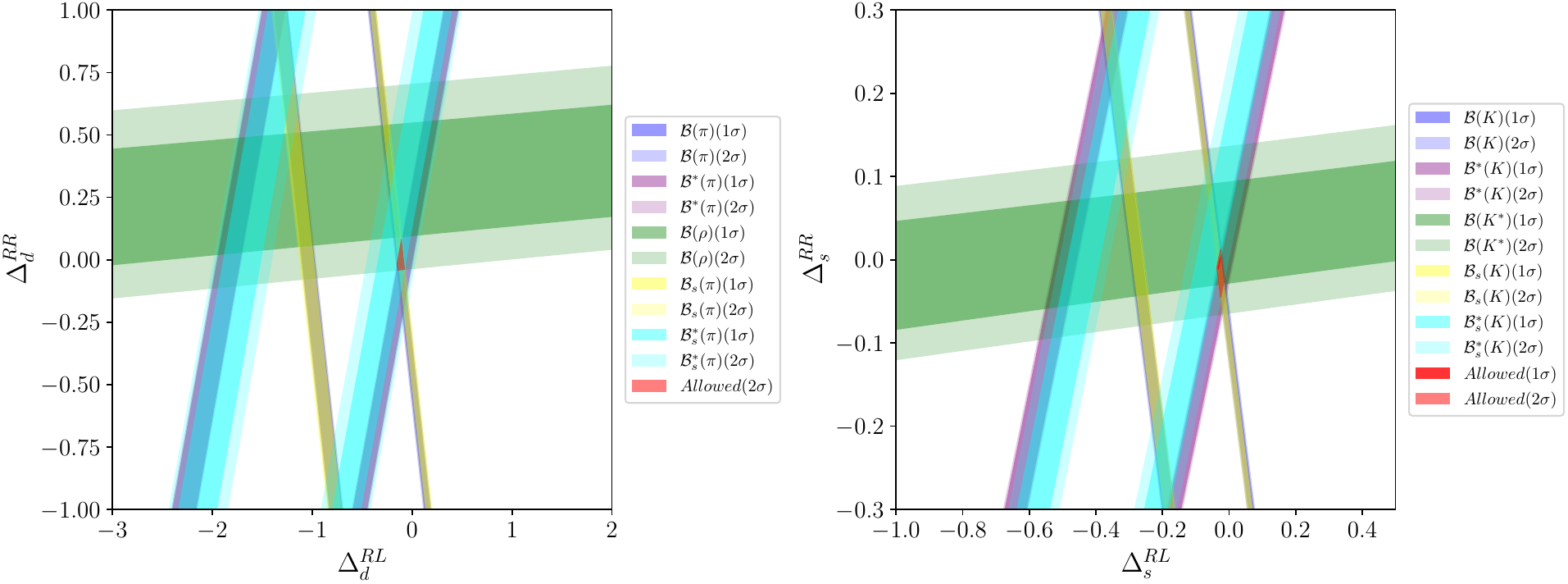}
	\caption{Constraints on the effective couplings $(\Delta_d^{RL}, \Delta_d^{RR})$ (left) and $(\Delta_s^{RL}, \Delta_s^{RR})$ (right) from the measured branching fractions of $\bar{B}^0_{(s)} \to D_{(s)}^{(*)+} L^-$ decays with $L = \pi, \rho$ and $L = K^{(*)}$, respectively. The other captions are the same as in figure~\ref{figure:LLLRSpace}. \label{figure:RLRRSpace} }
\end{figure}

\begin{figure}[t]
	\centering
    \includegraphics[width=0.99\textwidth]{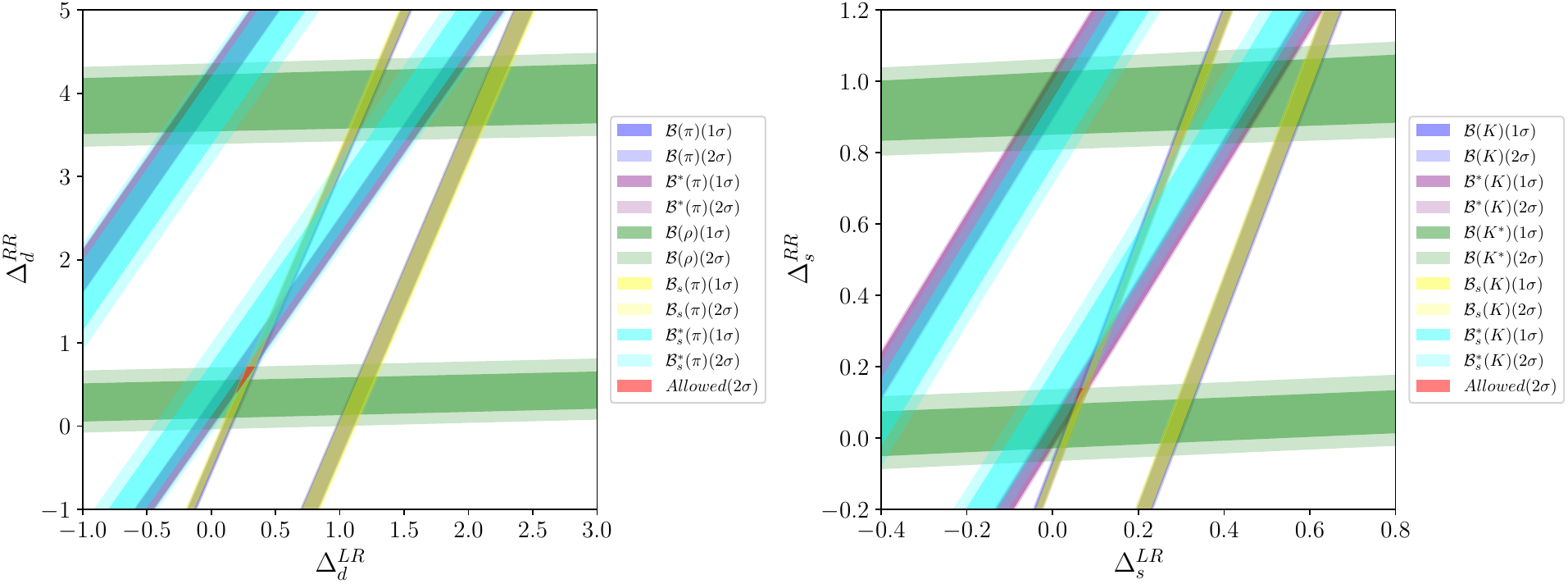}
	\caption{Constraints on the effective couplings $(\Delta_d^{LR}, \Delta_d^{RR})$ (left) and $(\Delta_s^{LR}, \Delta_s^{RR})$ (right) from the measured branching fractions of $\bar{B}^0_{(s)} \to D_{(s)}^{(*)+} L^-$ decays with $L = \pi, \rho$ and $L = K^{(*)}$, respectively. The other captions are the same as in figure~\ref{figure:LLLRSpace}. \label{figure:LRRRSpace} }
\end{figure}

We now turn to case~II, in which only two of the four effective couplings $\Delta_{d(s)}^{XY}$ are non-zero at the same time, corresponding to setting one coupling constant $\lambda_{ij}^L$ or $\lambda_{ij}^R$ to zero. Within this setup, we investigate the following four possible scenarios and show the allowed regions of parameter space for each.
\begin{enumerate}
\item[\textbullet] \textbf{S1$^{\prime}$:} 
We begin by setting $\lambda_{13}^R=0$, ensuring that $\Delta_{d(s)}^{LL}$ and $\Delta_{d(s)}^{LR}$ are the only effective couplings that remain non-vanishing. The viable parameter region for this case is displayed in figure~\ref{figure:LLLRSpace}. We observe that a relatively sizable portion of the parameter space can accommodate the theory–experiment deviations in the relevant branching fractions at the $1\sigma$ level.

\item[\textbullet] \textbf{S2$^{\prime}$:} 
Second, we impose $\lambda_{21(22)}^R=0$, so that $\Delta_{d(s)}^{LL}$ and $\Delta_{d(s)}^{RL}$ remain as the relevant effective couplings. The regions of parameter space consistent with the current experimental data are shown in figure~\ref{figure:LLRLSpace}. As in S1$^{\prime}$, this setup can likewise account for the observed deviations at the $1\sigma$ level.

\item[\textbullet] \textbf{S3$^{\prime}$:}
Next, we take $\lambda_{13}^L=0$, which leads to non-vanishing $\Delta_{d(s)}^{RL}$ and $\Delta_{d(s)}^{RR}$. The phenomenologically viable parameter region is shown in figure~\ref{figure:RLRRSpace}. In contrast to the first two cases, this setup yields almost no allowed parameter space at the $1\sigma$ level and, even at the $2\sigma$ level, the remaining allowed region is highly constrained.

\item[\textbullet] \textbf{S4$^{\prime}$:}
Lastly, taking $\lambda_{21(22)}^L=0$ implies that $\Delta_{d(s)}^{LR}$ and $\Delta_{d(s)}^{RR}$ are the only effective couplings that remain non-vanishing. As shown in figure~\ref{figure:LRRRSpace}, and in analogy with S3$^{\prime}$, this case results in only a marginal region of parameter space at the $2\sigma$ level.
\end{enumerate}

\subsection{Complementary constraints on \texorpdfstring{$\Delta_d^{LL}$}{Delta-d-LL} and \texorpdfstring{$\Delta_d^{RL}$}{Delta-d-RL} from \texorpdfstring{$R_{\pi/\mu\nu_{\mu}}$}{R-pi-mu-nu}}
\label{subsec:R_pimunu}

To further narrow down the viable parameter space of the NP couplings, we now incorporate constraints from the non-leptonic decay $B_c^- \to J/\psi \pi^-$, which proceeds through the same quark-level $b \to c \bar{u} d$ transition as $\bar{B}^0_{(s)} \to D_{(s)}^{(*)+} L^-$ with $L=\pi,\rho$. To reduce hadronic uncertainties associated with the $B_c^- \to J/\psi$ transition form factors~\cite{Harrison:2025kxm}, we consider the ratio of the branching fraction of $B_c^- \to J/\psi \pi^-$ to that of the semi-leptonic channel $B_c^- \to J/\psi \mu^- \bar{\nu}_\mu$. This ratio, denoted by $R_{\pi/\mu\nu_\mu}$, provides an additional and complementary constraint on the diquark couplings. Concretely, it is defined as
\begin{equation}
	R_{\pi/\mu\nu_{\mu}} = \frac{\mathcal{B}(B_{c}^{-}\to J/\psi \pi^{-})}{\mathcal{B}(B_{c}^{-}\to J/\psi \mu^{-}\bar{\nu}_{\mu})}\,,
\end{equation}
with the semi-leptonic branching fraction evaluated by integrating over the entire kinematic range in $q^2$,
\begin{equation}
	\mathcal{B}(B_{c}^{-}\to J/\psi \mu^{-}\bar{\nu}_{\mu}) = \int_{m_{\mu}^{2}}^{(m_{B_{c}}-m_{J/\psi})^2}\frac{d\mathcal{B}(B_{c}^{-}\to J/\psi \mu^{-}\bar{\nu}_{\mu})}{dq^2}dq^2\,.
\end{equation}

\begin{figure}[t]
	\centering
	\includegraphics[width=0.48\textwidth]{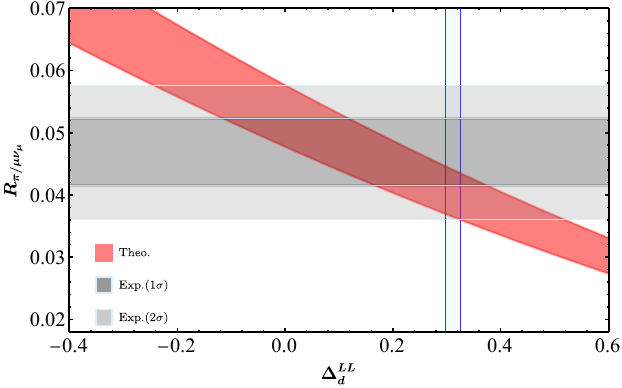}\quad 
	\includegraphics[width=0.48\textwidth]{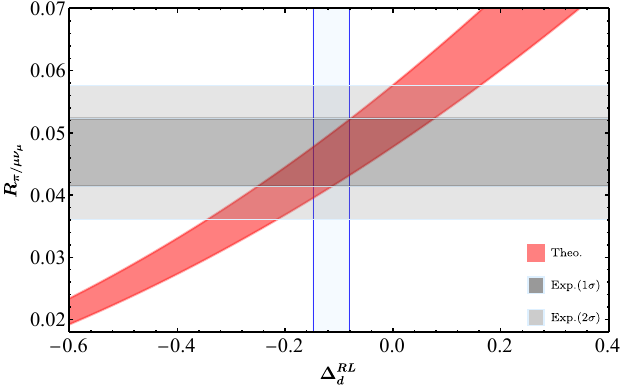}
	\caption{Constraints on the effective couplings $\Delta_d^{LL}$ (left) and $\Delta_d^{RL}$ (right) from the measured ratio $R_{\pi/\mu\nu_\mu}$. The red bands represent the theoretical predictions, while the horizontal shaded regions indicate the $1\sigma$ (dark gray) and $2\sigma$ (light gray) experimental confidence intervals for the ratio~\cite{LHCb:2014rck}. The vertical light blue bands illustrate the $2\sigma$ allowed regions derived from the combined constraints of the five $\bar{B}^0_{(s)} \to D_{(s)}^{(*)+} L^-$ decays with $L = \pi, \rho$. \label{fig:deltafromRpimunu}}
\end{figure}

On the experimental side, the LHCb collaboration has determined this ratio to be~\cite{LHCb:2014rck}
\begin{equation} \label{eq:LHCb-Rpimunu}
	R_{\pi/\mu\nu_{\mu}}^\mathrm{exp} = 0.0469 \pm 0.0028~(\text{stat}) \pm 0.0046~(\text{syst})\,.
\end{equation}
Building on our earlier analysis~\cite{Deng:2025znr} and updating the input parameters as listed in table~\ref{table:input}, we obtain the SM prediction within the QCDF approach at NNLO in $\alpha_s$ as
\begin{align}
	R_{\pi/\mu\nu_{\mu}}^\mathrm{NNLO} = 0.0527 \pm 0.0050\,.
\end{align}
The preceding analysis of $\bar{B}^0_{(s)} \to D_{(s)}^{(*)+} L^-$ ($L=\pi,\rho$) decays has already ruled out two of the single-coupling options in case~I, so that only $\Delta_{d}^{LL}$ and $\Delta_{d}^{RL}$ remain as viable solutions, with their allowed intervals specified in eqs.~\eqref{lamLL1S}, \eqref{lamLL2S} and \eqref{lamRL2S}. Confronting our predictions (including the diquark contributions) with the corresponding LHCb measurement yields the constraints displayed in figure~\ref{fig:deltafromRpimunu}. While $R_{\pi/\mu\nu_{\mu}}$ significantly narrows the allowed parameter space for $\Delta_{d}^{LL}$ (in particular, eliminating the $1\sigma$ region as well as the second solution compatible at $2\sigma$ with $\bar{B}^0_{(s)} \to D_{(s)}^{(*)+} L^-$ ($L=\pi,\rho$) decays), it does not impose any appreciable constraint on $\Delta_{d}^{RL}$. Hence, by combining these complementary bounds, we obtain the preferred $2\sigma$ ranges
\begin{align}\label{eq:deltaLLRL_range}
	\Delta_{d}^{LL} &\in [0.297, 0.324] \,,\qquad
	\Delta_{d}^{RL} \in [-0.148, -0.082]\,.
\end{align}

\subsection{Predictions of observables in different NP scenarios}
\label{subsec:NP_predictions}

\begin{table}[t]
	\centering
	\tabcolsep 0.40in
	\renewcommand{\arraystretch}{1.3} 
	\begin{tabular}{|ccc|}
		\hline \hline
		Couplings & Allowed ranges & Scenarios \\
		\hline
		$\Delta_{d}^{LL}$ & $0.3105 \pm 0.0138$ & S1 \\
		$\Delta_{d}^{RL}$ & $-0.1151 \pm 0.0327$ & S3 \\
        \hline
		$\Delta_{s}^{LL}$ & $0.0644\pm 0.0006 \cup 0.9136 \pm 0.0250$ & S1 \\
		$\Delta_{s}^{RL}$ & $-0.0269 \pm 0.0068$ & S3 \\
		\hline \hline
	\end{tabular}
	\caption{Allowed ranges of the effective couplings $\Delta_{d(s)}^{LL}$ and $\Delta_{d(s)}^{RL}$ at the $2\sigma$ confidence level under the combined constraints from the $\bar{B}^0_{(s)} \to D_{(s)}^{(*)+} L^-$ branching fractions and the ratio $R_{\pi/\mu\nu_{\mu}}$ for the S1 and S3 scenarios. \label{tab:NP_couplings_ranges}}
\end{table}

Under the single-coupling assumption of case I, we assess the impact of the constrained couplings $\Delta_{d(s)}^{LL}$ and $\Delta_{d(s)}^{RL}$ on the $\bar{B}^0_{(s)} \to D_{(s)}^{(*)+} L^-$ branching fractions and the ratios $R_{(s)L}^{(*)}$. As listed in table~\ref{tab:NP_couplings_ranges}, these two scenarios correspond to S1 (only $\Delta_{d(s)}^{LL}$ non-zero) and S3 (only $\Delta_{d(s)}^{RL}$ non-zero), respectively. For each coupling, we take the midpoint of its $2\sigma$ allowed interval as the central value and half its width as the associated uncertainty. The resulting theoretical predictions are collected in table~\ref{tab:predictions_S1_S3}. 

For the $b\to c\bar{u}s$ transitions ($L=K^{(*)}$ channels) in scenario S1, the two sets of predictions in table~\ref{tab:predictions_S1_S3} reflect the two disjoint allowed intervals of $\Delta_s^{LL}$ identified in table~\ref{tab:NP_couplings_ranges}. The first interval is tightly constrained ($\Delta_s^{LL} = 0.0644\pm 0.0006$) and consequently gives predictions with substantially smaller uncertainties.

\begin{table}[t]
	\centering
	\tabcolsep 0.12in
	\renewcommand{\arraystretch}{1.3} 
	\begin{tabular}{|lccccc|}
		\hline \hline
		Observables & \multicolumn{2}{c}{S1} & S3 & SM & Exp. \\
		\hline
		$\mathcal{B}(\bar{B}^0 \to D^+ \pi^-)$ & \multicolumn{2}{c}{$2.95^{+0.25}_{-0.28}$} & $2.62^{+0.44}_{-0.43}$ & $4.07^{+0.34}_{-0.37}$ & $2.48 \pm 0.12$ \\
		$\mathcal{B}(\bar{B}^0 \to D^{*+} \pi^-)$ & \multicolumn{2}{c}{$2.26^{+0.31}_{-0.31}$} & $2.40^{+0.38}_{-0.37}$ & $3.10^{+0.42}_{-0.42}$ & $2.652 \pm 0.064$ \\
		$\mathcal{B}(\bar{B}^0 \to D^+ \rho^-)$ & \multicolumn{2}{c}{$7.79^{+0.74}_{-0.80}$} & $10.67^{+1.00}_{-1.06}$ & $10.72^{+1.00}_{-1.06}$ & $7.6 \pm 1.2$ \\
		$\mathcal{B}(\bar{B}^0 \to D^+ K^-)$ & $2.34^{+0.18}_{-0.20}$ & $2.43^{+0.61}_{-0.56}$ & $2.11^{+0.29}_{-0.29}$ & $3.11^{+0.24}_{-0.27}$ & $2.03 \pm 0.12$ \\
		$\mathcal{B}(\bar{B}^0 \to D^{*+} K^-)$ & $1.76^{+0.23}_{-0.23}$ & $1.81^{+0.48}_{-0.45}$ & $1.87^{+0.27}_{-0.27}$ & $2.34^{+0.31}_{-0.30}$ & $2.146 \pm 0.079$ \\
		$\mathcal{B}(\bar{B}^0 \to D^+ K^{*-})$ & $4.03^{+0.40}_{-0.42}$ & $4.19^{+1.10}_{-1.00}$ & $5.32^{+0.52}_{-0.55}$ & $5.35^{+0.52}_{-0.55}$ & $4.60 \pm 0.78$ \\
		$\mathcal{B}(\bar{B}_s^0 \to D_s^+ \pi^-)$ & \multicolumn{2}{c}{$3.21^{+0.16}_{-0.21}$} & $2.85^{+0.43}_{-0.43}$ & $4.42^{+0.21}_{-0.28}$ & $2.83^{+0.18}_{-0.19}$ \\
		$\mathcal{B}(\bar{B}_s^0 \to D_s^{*+} \pi^-)$ & \multicolumn{2}{c}{$2.60^{+0.30}_{-0.31}$} & $2.78^{+0.38}_{-0.39}$ & $3.58^{+0.41}_{-0.42}$ & $2.19^{+0.62}_{-0.53}$ \\
		$\mathcal{B}(\bar{B}_s^0 \to D_s^+ K^-)$ & $2.54^{+0.10}_{-0.15}$ & $2.64^{+0.64}_{-0.59}$ & $2.30^{+0.27}_{-0.28}$ & $3.38^{+0.14}_{-0.19}$ & $2.12 \pm 0.14$ \\
		$\mathcal{B}(\bar{B}_s^0 \to D_s^{*+} K^-)$ & $2.03^{+0.22}_{-0.23}$ & $2.08^{+0.53}_{-0.50}$ & $2.15^{+0.27}_{-0.28}$ & $2.70^{+0.30}_{-0.30}$ & $1.49^{+0.45}_{-0.37}$ \\
		\hline
		$R_\pi$ & \multicolumn{2}{c}{$0.80^{+0.02}_{-0.03}$} & $0.71^{+0.10}_{-0.10}$ & $1.10^{+0.03}_{-0.03}$ & $0.68 \pm 0.04$ \\
		$R_\pi^*$ & \multicolumn{2}{c}{$0.80^{+0.02}_{-0.03}$} & $0.85^{+0.07}_{-0.07}$ & $1.09^{+0.03}_{-0.04}$ & $0.83 \pm 0.04$ \\
		$R_\rho$ & \multicolumn{2}{c}{$2.19^{+0.13}_{-0.13}$} & $3.01^{+0.17}_{-0.18}$ & $3.02^{+0.17}_{-0.18}$ & $2.17 \pm 0.35$ \\
		$R_K$ & $0.64^{+0.01}_{-0.01}$ & $0.67^{+0.17}_{-0.15}$ & $0.58^{+0.06}_{-0.06}$ & $0.85^{+0.01}_{-0.02}$ & $0.56 \pm 0.04$ \\
		$R_K^*$ & $0.58^{+0.01}_{-0.01}$ & $0.60^{+0.15}_{-0.13}$ & $0.62^{+0.04}_{-0.04}$ & $0.78^{+0.01}_{-0.02}$ & $0.64 \pm 0.04$ \\
		$R_{K^*}$ & $1.15^{+0.08}_{-0.08}$ & $1.20^{+0.32}_{-0.28}$ & $1.52^{+0.10}_{-0.10}$ & $1.53^{+0.10}_{-0.10}$ & $1.34 \pm 0.23$ \\
		$R_{s\pi}$ & \multicolumn{2}{c}{$0.80^{+0.02}_{-0.03}$} & $0.71^{+0.10}_{-0.10}$ & $1.10^{+0.03}_{-0.03}$ & $0.76 \pm 0.05$ \\
		$R_{s\pi}^*$ & \multicolumn{2}{c}{$0.80^{+0.02}_{-0.03}$} & $0.85^{+0.07}_{-0.07}$ & $1.09^{+0.03}_{-0.04}$ & $0.66 \pm 0.18$ \\
		$R_{sK}$ & $0.64^{+0.01}_{-0.01}$ & $0.67^{+0.17}_{-0.15}$ & $0.58^{+0.06}_{-0.06}$ & $0.85^{+0.01}_{-0.02}$ & $0.58 \pm 0.04$ \\
		$R_{sK}^*$ & $0.59^{+0.01}_{-0.01}$ & $0.60^{+0.15}_{-0.13}$ & $0.62^{+0.04}_{-0.04}$ & $0.78^{+0.01}_{-0.02}$ & $0.43 \pm 0.12$ \\
		\hline \hline
	\end{tabular}
	\caption{Predicted branching fractions (in units of $10^{-3}$ for $b \to c\bar{u}d$ and $10^{-4}$ for $b \to c\bar{u}s$ transitions) of $\bar{B}^0_{(s)} \to D^{(*)+}_{(s)} L^-$ decays and the ratios $R_{(s)L}^{(*)}$ (in units of $\text{GeV}^2$ for $b \to c\bar{u}d$ and $10^{-1}\text{ GeV}^2$ for $b \to c\bar{u}s$ transitions), in the S1 and S3 scenarios defined in table~\ref{tab:NP_couplings_ranges}. For the S1 scenario, the two sub-columns correspond to the two allowed ranges of the $\Delta_s^{LL}$ coupling shown in table~\ref{tab:NP_couplings_ranges}. The columns labeled by ``SM'' and ``Exp.'' represent the NNLO SM predictions and the experimental measurements in tables~\ref{tab:SMBr} and \ref{tab:SMRatio}, respectively.}
	\label{tab:predictions_S1_S3}
\end{table}

As can be seen from table~\ref{tab:predictions_S1_S3}, the two scenarios exhibit markedly different phenomenological signatures. Compared to the SM predictions labelled by the column ``SM'', the scenario S1 produces a uniform suppression of all branching fractions and ratios, bringing the theoretical predictions into closer agreement with the experimental measurements. The scenario S3, by contrast, suppresses the pseudoscalar channels ($L=\pi, K$) while leaving the vector channels ($L=\rho, K^*$) essentially unaffected. This striking dichotomy originates from the distinct Lorentz structures of the underlying four-quark operators. The effective couplings $\Delta^{LL}_{d(s)}$ associate with the $(V-A)\otimes(V-A)$ operators, which interfere coherently with the SM vector and axial-vector currents in both pseudoscalar and vector final states. The effective couplings $\Delta^{RL}_{d(s)}$, on the other hand, correspond to scalar and tensor operators. Within the QCDF framework, the scalar operator generates a chirally enhanced contribution to pseudoscalar emission, whereas its hadronic matrix element for vector meson emission vanishes identically, and the tensor contribution remains numerically suppressed. Consequently, the ability of scenario S1 to simultaneously resolve the observed discrepancies across all channels points to a left-handed $(V-A)$ Lorentz structure as the preferred NP contribution, with $\Delta^{LL}_{d(s)}$ dominant over $\Delta^{RL}_{d(s)}$~--~a pattern consistent with a purely left-handed diquark coupling to quarks.

\section{Conclusion}
\label{sec:conclusion}

In this paper, motivated by the deviations observed between the SM predictions and the experimental measurements for the branching fractions of $\bar{B}_{(s)}^{0}\to D_{(s)}^{(*)+}L^{-}$ decays, we have performed a systematic study of the type-IV scalar diquark as a possible NP resolution. By matching the renormalized amplitudes in the full theory to those in the corresponding effective theory formulated in the diquark operator basis, we determine the relevant Wilson coefficients at the NP scale $\mu_\phi=M_{\phi}$, including contributions up to NLO in $\alpha_s$. We subsequently translate the resulting operators into the conventional SM basis and apply the RG evolution to run the Wilson coefficients down to the hadronic scale $\mu_b=m_b$. This procedure yields a consistent effective-field-theory framework for the type-IV diquark-generated $\Delta B=1$ non-leptonic transitions.

For the hadronic matrix elements of the NP four-quark operators involved, we have consistently included the non-factorizable vertex corrections up to NLO in $\alpha_s$, while for the SM pieces the NNLO corrections have also been taken into account. Comparing our updated SM predictions with the most recent experimental data, we observe that the tension still persists at approximately the level of $4\!-\!6\sigma$. To quantify the impact of the type-IV scalar diquark on these class-I non-leptonic $B$-meson decays, we have derived the general expressions for the branching fractions in terms of the effective diquark couplings, and then identified the viable regions of the NP parameter space in two representative setups. In the single-coupling scenario, the purely left-handed coupling $\Delta_{d(s)}^{LL}$ can account for the observed discrepancies within $1\sigma$, whereas the mixed right-left coupling $\Delta_{d(s)}^{RL}$ remains consistent only at the $2\sigma$ level. In the two-coupling scenario, on the other hand, there exist substantially allowed parameter regions as long as $\Delta_{d(s)}^{LL}\neq 0$. To further narrow down the allowed parameter space, we have also incorporated the LHCb measurement of the ratio $R_{\pi/\mu\nu_{\mu}}$. We find that $R_{\pi/\mu\nu_{\mu}}$ can further reduce the allowed range of $\Delta_{d}^{LL}$, while providing no significant additional restriction on $\Delta_{d}^{RL}$. Thus, although both $\Delta_{d(s)}^{LL}$ and $\Delta_{d(s)}^{RL}$ can individually serve as viable NP explanations, our phenomenological results indicate that a scenario with purely left-handed diquark–quark couplings is favored by the current data. Finally, within the remaining allowed parameter space, we have provided updated predictions for the $\bar{B}_{(s)}^{0}\to D_{(s)}^{(*)+}L^{-}$ branching fractions and the ratios $R_{(s)L}^{(*)}$, which can be tested with the forthcoming high-precision data from LHCb and Belle II.

\acknowledgments

This work is supported by the National Natural Science Foundation of China under Grant Nos.~12635005, 12475094 and 12135006, as well as the Science and Technology Innovation Leading Talent Support Program of Henan Province under Grant No.~254000510039. F.~M. Cai is supported by the China Postdoctoral Foundation under Grant No.~GZB20230195 and the Natural Science Foundation of Henan Province under Grant No.~262300421308.

\appendix

\section{Fierz transformations with charge-conjugated fermion fields}
\label{app:FierzCharge}

Fierz identities~\cite{Fierz:1939zz}, which are also often referred to as Fierz relations or Fierz transformations, make it possible to rewrite a given product of Dirac-spinor bilinears into an equivalent one. This rearrangement can render the resulting matrix elements between external states simpler to compute than those of the original expression. In this work, in order to map the diquark basis onto the SM basis, we employ the following Fierz identities involving two charge-conjugated fermion fields~\cite{Buras:2020xsm}:
\begin{align}
    (\overline{\psi}_1 P_A \psi_2^c)\otimes(\overline{\psi^c_3} P_A \psi_4) &= \frac{1}{8} (\overline{\psi}_1 \sigma_{\mu\nu} P_A \psi_3)\otimes(\overline{\psi}_2 \sigma^{\mu\nu} P_A \psi_4)-\frac{1}{2} (\overline{\psi}_1 P_A \psi_3)\otimes(\overline{\psi}_2 P_A \psi_4),\\
	(\overline{\psi}_1 P_A \psi_2^c)\otimes(\overline{\psi^c_3} P_B \psi_4) &= \frac{1}{2} (\overline{\psi}_1 \gamma_\mu P_B \psi_3)\otimes(\overline{\psi}_2 \gamma^\mu P_B \psi_4), \\[0.2cm]
	(\overline{\psi}_1 \gamma_\mu P_A \psi_2^c)\otimes(\overline{\psi^c_3} \gamma^\mu P_A \psi_4) &= 2(\overline{\psi}_1 P_B \psi_3)\otimes(\overline{\psi}_2 P_A \psi_4), \\[0.2cm]
	(\overline{\psi}_1 \gamma_\mu P_A \psi_2^c)\otimes(\overline{\psi^c_3} \gamma^\mu P_B \psi_4) &= (\overline{\psi}_1 P_A \psi_3)\otimes(\overline{\psi}_2 P_B \psi_4), \\[0.2cm]
	(\overline{\psi}_1 \sigma_{\mu\nu} P_A \psi_2^c)\otimes(\overline{\psi^c_3} \sigma^{\mu\nu} P_A \psi_4) &= 6(\overline{\psi}_1 P_A \psi_3)\otimes(\overline{\psi}_2 P_A \psi_4) + \frac{1}{2} (\overline{\psi}_1 \sigma_{\mu\nu} P_A \psi_3)\otimes(\overline{\psi}_2 \sigma^{\mu\nu} P_A \psi_4).
\end{align}  
In these expressions, the chirality projectors $P_{A,B}=(1 \pm \gamma_5)/2$ are to be taken with $P_A\neq P_B$ within any given identity. Concretely, choosing $P_A=P_L$ implies $P_B=P_R$, and vice versa.

\section{Anomalous-dimension matrices in the SM operator basis}
\label{app:ano_dim}

In this appendix, we provide the one- and two-loop ADMs~\cite{Buras:2012gm,Buras:2000if} for the SM operator basis introduced in eq.~\eqref{eq:SM_operators}. Our definition of the antisymmetric tensor, $\sigma^{\mu\nu}=\frac{i}{2}[\gamma^\mu,\gamma^\nu]$, differs from the convention used in ref.~\cite{Buras:2012gm} by a factor of $i$, which implies an overall sign flip in the mixing-matrix entries connecting tensor and scalar operators. In addition, there is no mixing between operators belonging to different sectors, and the ADMs for the chirality-flipped operators ($1 \pm \gamma_5 \to 1 \mp \gamma_5$) coincide with those of the basis in eq.~\eqref{eq:SM_operators}.

First, the one-loop ADMs for the $VLL$ and $SLL$ sectors are given, respectively, by~\cite{Buras:2012gm,Buras:2000if}
\begin{align} \label{eq:gamma0_SM}
	\hat{\gamma}^{(0)VLL} &= \begin{pmatrix}
		-\frac{6}{N_c} & 6 \\[1mm]
		6 & -\frac{6}{N_c}
	\end{pmatrix} \, ,\nonumber\\[2mm]
	\hat{\gamma}^{(0)SLL} &= \begin{pmatrix}
		\frac{6}{N_c} & -6 & -\frac{N_c}{2} + \frac{1}{N_c} &  -\frac{1}{2} \\[1mm] 
		0 & -6N_c + \frac{6}{N_c} & -1 & \frac{1}{N_c} \\[1mm] 
		\frac{48}{N_c} - 24N_c & -24 & -\frac{2}{N_c} - 4N_c & 6 \\[1mm] 
		-48 & \frac{48}{N_c} & 0 & 2N_c - \frac{2}{N_c}
	\end{pmatrix} \, ,
\end{align}
where $N_c$ represents the number of colors. The two-loop ADMs for these two sectors in the $\MSbar$ scheme are given by~\cite{Buras:2012gm,Buras:2000if}
\begin{align}\label{eq:gamma1_SM}
	\hat{\gamma}^{(1)VLL} &= \begin{pmatrix}
		-\frac{22}{3} - \frac{57}{2N_c^{2}} - \frac{2}{3N_c}N_f & -\frac{19}{6}N_c + \frac{39}{N_c} + \frac{2}{3}N_f 
        \\[2mm] 
		-\frac{19}{6}N_c + \frac{39}{N_c} + \frac{2}{3}N_f & -\frac{22}{3} - \frac{57}{2N_c^{2}} - \frac{2}{3N_c}N_f
	\end{pmatrix} \, , \nonumber\\[2mm]
	\gamma_{11}^{(1)SLL} &= -\frac{N_c^{2}}{2} + \frac{148}{3} - \frac{107}{2N_c^{2}} - 2N_cN_f - \frac{10}{3N_c}N_f, \nonumber \\[1mm]
	\gamma_{12}^{(1)SLL} &= -\frac{178}{3}N_c + \frac{64}{N_c} + \frac{16}{3}N_f \, , \nonumber \\[1mm]
	\gamma_{13}^{(1)SLL} &= -\frac{107}{36}N_c^{2} + \frac{71}{18} + \frac{4}{N_c^{2}} + \frac{1}{18}N_cN_f - \frac{N_f}{9N_c} \, , \nonumber \\[2mm]
	\gamma_{14}^{(1)SLL} &= \frac{109}{36}N_c - \frac{8}{N_c} + \frac{N_f}{18} \, , \nonumber \\[1mm]
	\gamma_{21}^{(1)SLL} &= -26N_c + \frac{104}{N_c}, \nonumber \\[2mm]
	\gamma_{22}^{(1)SLL} &= -\frac{203}{6}N_c^{2} + \frac{28}{3} - \frac{107}{2N_c^{2}} + \frac{10}{3}N_cN_f - \frac{10}{3N_c}N_f \, , \nonumber \\[2mm]
	\gamma_{23}^{(1)SLL} &= -\frac{89}{18}N_c - \frac{2}{N_c} + \frac{1}{9}N_f \, , \nonumber \\[1mm]
	\gamma_{24}^{(1)SLL} &= \frac{53}{18} + \frac{4}{N_c^{2}} - \frac{1}{9N_c}N_f \, , \nonumber \\[1mm]
	\gamma_{31}^{(1)SLL} &= -\frac{676}{3}N_c^{2} + \frac{1880}{3} + \frac{320}{N_c^{2}} + \frac{88}{3}N_cN_f - \frac{176}{3N_c}N_f \, , \nonumber\\[2mm]
	\gamma_{32}^{(1)SLL} &= -\frac{820}{3}N_c - \frac{448}{N_c} + \frac{88}{3}N_f \, , \nonumber \\[1mm]
	\gamma_{33}^{(1)SLL} &= -\frac{257}{18}N_c^{2} - \frac{116}{9} + \frac{21}{2N_c^{2}} + \frac{22}{9}N_cN_f + \frac{2}{9N_c}N_f \, , \nonumber \\[2mm]
	\gamma_{34}^{(1)SLL} &= \frac{50}{3}N_c - \frac{8}{3}N_f, \nonumber \\[2mm]
	\gamma_{41}^{(1)SLL} &= -\frac{488}{3}N_c - \frac{416}{N_c} + \frac{176}{3}N_f \, , \nonumber \\[1mm]
	\gamma_{42}^{(1)SLL} &= \frac{776}{3} + \frac{320}{N_c^{2}} - \frac{176}{3N_c}N_f \, , \nonumber \\[1mm]
	\gamma_{43}^{(1)SLL} &= \frac{22}{3}N_c - \frac{40}{N_c} + \frac{8}{3}N_f \, , \nonumber \\[1mm]
	\gamma_{44}^{(1)SLL} &= \frac{343}{18}N_c^{2} + \frac{28}{9} + \frac{21}{2N_c^{2}} - \frac{26}{9}N_cN_f + \frac{2}{9N_c}N_f \, ,
\end{align}
where $N_f$ denotes the number of active quark flavors.

\bibliographystyle{JHEP}
\bibliography{ref}

\providecommand{\href}[2]{#2}\begingroup\raggedright\begin{thebibliography}{100}

\bibitem{ParticleDataGroup:2026aaa}
{\bf Particle Data Group} Collaboration, F.~Takahashi et~al., {\it {Review of Particle Physics}},  {\it Int. J. Mod. Phys. A} {\bf 41} (2026) 2630011.

\bibitem{Buchalla:2008jp}
G.~Buchalla et~al., {\it {$B$, $D$ and $K$ decays}},  {\it Eur. Phys. J. C} {\bf 57} (2008) 309--492, [\href{http://arxiv.org/abs/0801.1833}{{\tt arXiv:0801.1833}}].

\bibitem{Antonelli:2009ws}
M.~Antonelli et~al., {\it {Flavor Physics in the Quark Sector}},  {\it Phys. Rept.} {\bf 494} (2010) 197--414, [\href{http://arxiv.org/abs/0907.5386}{{\tt arXiv:0907.5386}}].

\bibitem{HeavyFlavorAveragingGroupHFLAV:2024ctg}
{\bf Heavy Flavor Averaging Group (HFLAV)} Collaboration, S.~Banerjee et~al., {\it {Averages of b-hadron, c-hadron, and {\ensuremath{\tau}}-lepton properties as of 2023}},  {\it Phys. Rev. D} {\bf 113} (2026), no.~1 012008, [\href{http://arxiv.org/abs/2411.18639}{{\tt arXiv:2411.18639}}].

\bibitem{FlavourLatticeAveragingGroupFLAG:2024oxs}
{\bf Flavour Lattice Averaging Group (FLAG)} Collaboration, Y.~Aoki et~al., {\it {FLAG review 2024}},  {\it Phys. Rev. D} {\bf 113} (2026), no.~1 014508, [\href{http://arxiv.org/abs/2411.04268}{{\tt arXiv:2411.04268}}].

\bibitem{BaBar:2014omp}
{\bf BaBar, Belle} Collaboration, A.~J. Bevan et~al., {\it {The Physics of the B Factories}},  {\it Eur. Phys. J. C} {\bf 74} (2014) 3026, [\href{http://arxiv.org/abs/1406.6311}{{\tt arXiv:1406.6311}}].

\bibitem{Belle-II:2018jsg}
{\bf Belle-II} Collaboration, W.~Altmannshofer et~al., {\it {The Belle II Physics Book}},  {\it PTEP} {\bf 2019} (2019), no.~12 123C01, [\href{http://arxiv.org/abs/1808.10567}{{\tt arXiv:1808.10567}}]. [Erratum: PTEP 2020, 029201 (2020)].

\bibitem{LHCb:2018roe}
{\bf LHCb} Collaboration, R.~Aaij et~al., {\it {Physics case for an LHCb Upgrade II - Opportunities in flavour physics, and beyond, in the HL-LHC era}},  \href{http://arxiv.org/abs/1808.08865}{{\tt arXiv:1808.08865}}.

\bibitem{Buchalla:1995vs}
G.~Buchalla, A.~J. Buras, and M.~E. Lautenbacher, {\it {Weak decays beyond leading logarithms}},  {\it Rev. Mod. Phys.} {\bf 68} (1996) 1125--1144, [\href{http://arxiv.org/abs/hep-ph/9512380}{{\tt hep-ph/9512380}}].

\bibitem{Buras:1998raa}
A.~J. Buras, {\it {Weak Hamiltonian, CP violation and rare decays}}.
\newblock 6, 1998.

\bibitem{Beneke:1999br}
M.~Beneke, G.~Buchalla, M.~Neubert, and C.~T. Sachrajda, {\it {QCD factorization for $B \rightarrow \pi \pi$ decays: Strong phases and CP violation in the heavy quark limit}},  {\it Phys. Rev. Lett.} {\bf 83} (1999) 1914--1917, [\href{http://arxiv.org/abs/hep-ph/9905312}{{\tt hep-ph/9905312}}].

\bibitem{Beneke:2000ry}
M.~Beneke, G.~Buchalla, M.~Neubert, and C.~T. Sachrajda, {\it {QCD factorization for exclusive, nonleptonic $B$ meson decays: General arguments and the case of heavy light final states}},  {\it Nucl. Phys. B} {\bf 591} (2000) 313--418, [\href{http://arxiv.org/abs/hep-ph/0006124}{{\tt hep-ph/0006124}}].

\bibitem{Beneke:2001ev}
M.~Beneke, G.~Buchalla, M.~Neubert, and C.~T. Sachrajda, {\it {QCD factorization in $B \to \pi K, \pi \pi $ decays and extraction of Wolfenstein parameters}},  {\it Nucl. Phys. B} {\bf 606} (2001) 245--321, [\href{http://arxiv.org/abs/hep-ph/0104110}{{\tt hep-ph/0104110}}].

\bibitem{Beneke:2003zv}
M.~Beneke and M.~Neubert, {\it {QCD factorization for $B \rightarrow PP$ and $B \to PV$ decays}},  {\it Nucl. Phys. B} {\bf 675} (2003) 333--415, [\href{http://arxiv.org/abs/hep-ph/0308039}{{\tt hep-ph/0308039}}].

\bibitem{Beneke:2006hg}
M.~Beneke, J.~Rohrer, and D.~Yang, {\it {Branching fractions, polarisation and asymmetries of $B \to VV$ decays}},  {\it Nucl. Phys.} {\bf B774} (2007) 64--101, [\href{http://arxiv.org/abs/hep-ph/0612290}{{\tt hep-ph/0612290}}].

\bibitem{Bauer:2000yr}
C.~W. Bauer, S.~Fleming, D.~Pirjol, and I.~W. Stewart, {\it {An Effective field theory for collinear and soft gluons: Heavy to light decays}},  {\it Phys. Rev. D} {\bf 63} (2001) 114020, [\href{http://arxiv.org/abs/hep-ph/0011336}{{\tt hep-ph/0011336}}].

\bibitem{Bauer:2001yt}
C.~W. Bauer, D.~Pirjol, and I.~W. Stewart, {\it {Soft collinear factorization in effective field theory}},  {\it Phys. Rev. D} {\bf 65} (2002) 054022, [\href{http://arxiv.org/abs/hep-ph/0109045}{{\tt hep-ph/0109045}}].

\bibitem{Beneke:2002ph}
M.~Beneke, A.~P. Chapovsky, M.~Diehl, and T.~Feldmann, {\it {Soft collinear effective theory and heavy to light currents beyond leading power}},  {\it Nucl. Phys. B} {\bf 643} (2002) 431--476, [\href{http://arxiv.org/abs/hep-ph/0206152}{{\tt hep-ph/0206152}}].

\bibitem{Beneke:2002ni}
M.~Beneke and T.~Feldmann, {\it {Multipole expanded soft collinear effective theory with nonAbelian gauge symmetry}},  {\it Phys. Lett. B} {\bf 553} (2003) 267--276, [\href{http://arxiv.org/abs/hep-ph/0211358}{{\tt hep-ph/0211358}}].

\bibitem{Bauer:2001cu}
C.~W. Bauer, D.~Pirjol, and I.~W. Stewart, {\it {A Proof of factorization for $B \to D \pi$}},  {\it Phys. Rev. Lett.} {\bf 87} (2001) 201806, [\href{http://arxiv.org/abs/hep-ph/0107002}{{\tt hep-ph/0107002}}].

\bibitem{Li:2005wx}
X.-q. Li and Y.-d. Yang, {\it {Revisiting $B \to \pi \pi$, $\pi K$ decays in QCD factorization approach}},  {\it Phys. Rev. D} {\bf 72} (2005) 074007, [\href{http://arxiv.org/abs/hep-ph/0508079}{{\tt hep-ph/0508079}}].

\bibitem{Beneke:2005vv}
M.~Beneke and S.~J{\"a}ger, {\it {Spectator scattering at NLO in non-leptonic b decays: Tree amplitudes}},  {\it Nucl. Phys.} {\bf B751} (2006) 160--185, [\href{http://arxiv.org/abs/hep-ph/0512351}{{\tt hep-ph/0512351}}].

\bibitem{Li:2006jb}
X.-q. Li and Y.-d. Yang, {\it {Reexamining charmless $B \to PV$ decays in QCD factorization approach}},  {\it Phys. Rev. D} {\bf 73} (2006) 114027, [\href{http://arxiv.org/abs/hep-ph/0602224}{{\tt hep-ph/0602224}}].

\bibitem{Kivel:2006xc}
N.~Kivel, {\it {Radiative corrections to hard spectator scattering in $B \to \pi \pi$ decays}},  {\it JHEP} {\bf 05} (2007) 019, [\href{http://arxiv.org/abs/hep-ph/0608291}{{\tt hep-ph/0608291}}].

\bibitem{Beneke:2006mk}
M.~Beneke and S.~J{\"a}ger, {\it {Spectator scattering at NLO in non-leptonic B decays: Leading penguin amplitudes}},  {\it Nucl. Phys.} {\bf B768} (2007) 51--84, [\href{http://arxiv.org/abs/hep-ph/0610322}{{\tt hep-ph/0610322}}].

\bibitem{Bell:2007tv}
G.~Bell, {\it {NNLO vertex corrections in charmless hadronic B decays: Imaginary part}},  {\it Nucl. Phys.} {\bf B795} (2008) 1--26, [\href{http://arxiv.org/abs/0705.3127}{{\tt arXiv:0705.3127}}].

\bibitem{Jain:2007dy}
A.~Jain, I.~Z. Rothstein, and I.~W. Stewart, {\it {Penguin Loops for Nonleptonic B-Decays in the Standard Model: Is there a Penguin Puzzle?}},  \href{http://arxiv.org/abs/0706.3399}{{\tt arXiv:0706.3399}}.

\bibitem{Pilipp:2007mg}
V.~Pilipp, {\it {Hard spectator interactions in $B \to\pi \pi$ at order $\alpha_s^2$}},  {\it Nucl. Phys.} {\bf B794} (2008) 154--188, [\href{http://arxiv.org/abs/0709.3214}{{\tt arXiv:0709.3214}}].

\bibitem{Bell:2009nk}
G.~Bell, {\it {NNLO vertex corrections in charmless hadronic B decays: Real part}},  {\it Nucl. Phys.} {\bf B822} (2009) 172--200, [\href{http://arxiv.org/abs/0902.1915}{{\tt arXiv:0902.1915}}].

\bibitem{Beneke:2009ek}
M.~Beneke, T.~Huber, and X.-Q. Li, {\it {NNLO vertex corrections to non-leptonic B decays: Tree amplitudes}},  {\it Nucl. Phys. B} {\bf 832} (2010) 109--151, [\href{http://arxiv.org/abs/0911.3655}{{\tt arXiv:0911.3655}}].

\bibitem{Kim:2011jm}
C.~S. Kim and Y.~W. Yoon, {\it {Order $\alpha_s^2$ magnetic penguin correction for $B$ decay to light mesons}},  {\it JHEP} {\bf 11} (2011) 003, [\href{http://arxiv.org/abs/1107.1601}{{\tt arXiv:1107.1601}}].

\bibitem{Bell:2014zya}
G.~Bell and T.~Huber, {\it {Master integrals for the two-loop penguin contribution in non-leptonic B-decays}},  {\it JHEP} {\bf 12} (2014) 129, [\href{http://arxiv.org/abs/1410.2804}{{\tt arXiv:1410.2804}}].

\bibitem{Bell:2015koa}
G.~Bell, M.~Beneke, T.~Huber, and X.-Q. Li, {\it {Two-loop current-current operator contribution to the non-leptonic QCD penguin amplitude}},  {\it Phys. Lett.} {\bf B750} (2015) 348--355, [\href{http://arxiv.org/abs/1507.03700}{{\tt arXiv:1507.03700}}].

\bibitem{Huber:2015bva}
T.~Huber and S.~Kr{\"a}nkl, {\it {Two-loop master integrals for non-leptonic heavy-to-heavy decays}},  {\it JHEP} {\bf 04} (2015) 140, [\href{http://arxiv.org/abs/1503.00735}{{\tt arXiv:1503.00735}}].

\bibitem{Huber:2016xod}
T.~Huber, S.~Kr{\"a}nkl, and X.-Q. Li, {\it {Two-body non-leptonic heavy-to-heavy decays at NNLO in QCD factorization}},  {\it JHEP} {\bf 09} (2016) 112, [\href{http://arxiv.org/abs/1606.02888}{{\tt arXiv:1606.02888}}].

\bibitem{Bell:2020qus}
G.~Bell, M.~Beneke, T.~Huber, and X.-Q. Li, {\it {Two-loop non-leptonic penguin amplitude in QCD factorization}},  {\it JHEP} {\bf 04} (2020) 055, [\href{http://arxiv.org/abs/2002.03262}{{\tt arXiv:2002.03262}}].

\bibitem{Beneke:2020vnb}
M.~Beneke, P.~B\"oer, J.-N. Toelstede, and K.~K. Vos, {\it {QED factorization of non-leptonic $B$ decays}},  {\it JHEP} {\bf 11} (2020) 081, [\href{http://arxiv.org/abs/2008.10615}{{\tt arXiv:2008.10615}}].

\bibitem{Beneke:2021jhp}
M.~Beneke, P.~B\"oer, G.~Finauri, and K.~K. Vos, {\it {QED factorization of two-body non-leptonic and semi-leptonic B to charm decays}},  {\it JHEP} {\bf 10} (2021) 223, [\href{http://arxiv.org/abs/2107.03819}{{\tt arXiv:2107.03819}}].

\bibitem{Bordone:2020gao}
M.~Bordone, N.~Gubernari, T.~Huber, M.~Jung, and D.~van Dyk, {\it {A puzzle in $\bar{B}_{(s)}^0 \to D_{(s)}^{(*)+} \lbrace \pi^-, K^-\rbrace$ decays and extraction of the $f_s/f_d$ fragmentation fraction}},  {\it Eur. Phys. J. C} {\bf 80} (2020), no.~10 951, [\href{http://arxiv.org/abs/2007.10338}{{\tt arXiv:2007.10338}}].

\bibitem{Cai:2021mlt}
F.-M. Cai, W.-J. Deng, X.-Q. Li, and Y.-D. Yang, {\it {Probing new physics in class-I $B$-meson decays into heavy-light final states}},  {\it JHEP} {\bf 10} (2021) 235, [\href{http://arxiv.org/abs/2103.04138}{{\tt arXiv:2103.04138}}].

\bibitem{Meiser:2024zea}
S.~Meiser, D.~van Dyk, and J.~Virto, {\it {Towards a global analysis of the $b \to c\overline{u} q$ puzzle}},  {\it JHEP} {\bf 06} (2025) 019, [\href{http://arxiv.org/abs/2411.09458}{{\tt arXiv:2411.09458}}].

\bibitem{Fleischer:2010ca}
R.~Fleischer, N.~Serra, and N.~Tuning, {\it {Tests of Factorization and SU(3) Relations in B Decays into Heavy-Light Final States}},  {\it Phys. Rev. D} {\bf 83} (2011) 014017, [\href{http://arxiv.org/abs/1012.2784}{{\tt arXiv:1012.2784}}].

\bibitem{Endo:2021ifc}
M.~Endo, S.~Iguro, and S.~Mishima, {\it {Revisiting rescattering contributions to $\bar{B}_{(s)} \to D_{(s)}^{(\ast)}M$ decays}},  {\it JHEP} {\bf 01} (2022) 147, [\href{http://arxiv.org/abs/2109.10811}{{\tt arXiv:2109.10811}}].

\bibitem{Piscopo:2023opf}
M.~L. Piscopo and A.~V. Rusov, {\it {Non-factorisable effects in the decays $ {\bar{B}}_s^0\to {D}_s^{+}{\pi}^{-} $ and $ {\bar{B}}^0\to {D}^{+}{K}^{-} $ from LCSR}},  {\it JHEP} {\bf 10} (2023) 180, [\href{http://arxiv.org/abs/2307.07594}{{\tt arXiv:2307.07594}}].

\bibitem{Davies:2024vmv}
J.~Davies, S.~Schacht, N.~Skidmore, and A.~Soni, {\it {Reappraisal of $SU(3)$-flavor breaking in $B \to DP$}},  {\it Phys. Rev. D} {\bf 109} (2024), no.~11 113006, [\href{http://arxiv.org/abs/2403.04878}{{\tt arXiv:2403.04878}}].

\bibitem{Yaouanc:2024gjl}
A.~L. Yaouanc, J.~P. Leroy, and P.~Roudeau, {\it {The $a_1$ factorisation coefficient in ${\bar{B}^0} \to D^{(*)+}M^{-}$ and ${\bar{B}^0} \to D^{(*)+}D_s^{(*)-}$ decays: measurements versus theory}},  \href{http://arxiv.org/abs/2405.10111}{{\tt arXiv:2405.10111}}.

\bibitem{Bobeth:2014rda}
C.~Bobeth, U.~Haisch, A.~Lenz, B.~Pecjak, and G.~Tetlalmatzi-Xolocotzi, {\it {On new physics in $\Delta\Gamma_{d}$}},  {\it JHEP} {\bf 06} (2014) 040, [\href{http://arxiv.org/abs/1404.2531}{{\tt arXiv:1404.2531}}].

\bibitem{Brod:2014bfa}
J.~Brod, A.~Lenz, G.~Tetlalmatzi-Xolocotzi, and M.~Wiebusch, {\it {New physics effects in tree-level decays and the precision in the determination of the quark mixing angle $\gamma$}},  {\it Phys. Rev.} {\bf D92} (2015), no.~3 033002, [\href{http://arxiv.org/abs/1412.1446}{{\tt arXiv:1412.1446}}].

\bibitem{Lenz:2019lvd}
A.~Lenz and G.~Tetlalmatzi-Xolocotzi, {\it {Model-independent bounds on new physics effects in non-leptonic tree-level decays of B-mesons}},  {\it JHEP} {\bf 07} (2020) 177, [\href{http://arxiv.org/abs/1912.07621}{{\tt arXiv:1912.07621}}].

\bibitem{Iguro:2020ndk}
S.~Iguro and T.~Kitahara, {\it {Implications for new physics from a novel puzzle in $\bar{B}_{(s)}^0 \to D^{(\ast)+}_{(s)} \lbrace \pi^-, K^- \rbrace$ decays}},  {\it Phys. Rev. D} {\bf 102} (2020), no.~7 071701, [\href{http://arxiv.org/abs/2008.01086}{{\tt arXiv:2008.01086}}].

\bibitem{Bordone:2021cca}
M.~Bordone, A.~Greljo, and D.~Marzocca, {\it {Exploiting dijet resonance searches for flavor physics}},  {\it JHEP} {\bf 08} (2021) 036, [\href{http://arxiv.org/abs/2103.10332}{{\tt arXiv:2103.10332}}].

\bibitem{Fleischer:2021cct}
R.~Fleischer and E.~Malami, {\it {Using $B^0_s\to D_s^\mp K^\pm$ Decays as a Portal to New Physics}},  {\it Phys. Rev. D} {\bf 106} (2022), no.~5 056004, [\href{http://arxiv.org/abs/2109.04950}{{\tt arXiv:2109.04950}}].

\bibitem{Fleischer:2021cwb}
R.~Fleischer and E.~Malami, {\it {Revealing new physics in $B^{0}_{s}\rightarrow D_s^{\mp } K^{\pm }$ decays}},  {\it Eur. Phys. J. C} {\bf 83} (2023), no.~5 420, [\href{http://arxiv.org/abs/2110.04240}{{\tt arXiv:2110.04240}}].

\bibitem{Gershon:2021pnc}
T.~Gershon, A.~Lenz, A.~V. Rusov, and N.~Skidmore, {\it {Testing the Standard Model with CP asymmetries in flavor-specific nonleptonic decays}},  {\it Phys. Rev. D} {\bf 105} (2022), no.~11 115023, [\href{http://arxiv.org/abs/2111.04478}{{\tt arXiv:2111.04478}}].

\bibitem{Lenz:2022pgw}
A.~Lenz, J.~M\"uller, M.~L. Piscopo, and A.~V. Rusov, {\it {Taming new physics in $b \to c\bar{u}d(s)$ with $\tau(B^{+})/\tau(B_{d})$ and $a_{sl}^d$}},  {\it JHEP} {\bf 09} (2023) 028, [\href{http://arxiv.org/abs/2211.02724}{{\tt arXiv:2211.02724}}].

\bibitem{Panuluh:2024cxc}
A.~H. Panuluh, S.~Tanaka, and H.~Umeeda, {\it {$B_{(s)} \to D_{(s)}^{(*)}M$ decays in the presence of final-state interactions}},  {\it Phys. Rev. D} {\bf 111} (2025), no.~9 095020, [\href{http://arxiv.org/abs/2408.15466}{{\tt arXiv:2408.15466}}].

\bibitem{Atkinson:2024hqp}
O.~Atkinson, C.~Englert, M.~Kirk, and G.~Tetlalmatzi-Xolocotzi, {\it {Collider-flavour complementarity from the bottom to the top}},  {\it Eur. Phys. J. C} {\bf 85} (2025), no.~3 258, [\href{http://arxiv.org/abs/2411.00940}{{\tt arXiv:2411.00940}}].

\bibitem{Deng:2025znr}
W.-J. Deng, F.-M. Cai, X.-Q. Li, Y.~Shi, and Y.-D. Yang, {\it {Revisiting $B_c^- \to J/\psi(\eta_c)L^-$ decays within the SM and beyond in QCD factorization}},  {\it Phys. Rev. D} {\bf 112} (2025), no.~3 035013, [\href{http://arxiv.org/abs/2503.11279}{{\tt arXiv:2503.11279}}].

\bibitem{Araz:2026zlu}
J.~Y. Araz, C.~Englert, M.~Kirk, and G.~Tetlalmatzi-Xolocotzi, {\it {New insights into the $b\rightarrow c \bar{u}q$ puzzle through Top-Bottom synergies}},  \href{http://arxiv.org/abs/2604.25998}{{\tt arXiv:2604.25998}}.

\bibitem{Guo:2026oed}
X.~Guo, A.~H. Panuluh, H.~Umeeda, and J.~Zhu, {\it {$b\to c \bar u q$ decay and CP violating observables in the presence of new physics contributions}},  \href{http://arxiv.org/abs/2605.22088}{{\tt arXiv:2605.22088}}.

\bibitem{Ma:1998pi}
E.~Ma, M.~Raidal, and U.~Sarkar, {\it {Probing the exotic particle content beyond the standard model}},  {\it Eur. Phys. J. C} {\bf 8} (1999) 301--309, [\href{http://arxiv.org/abs/hep-ph/9808484}{{\tt hep-ph/9808484}}].

\bibitem{Manohar:2006ga}
A.~V. Manohar and M.~B. Wise, {\it {Flavor changing neutral currents, an extended scalar sector, and the Higgs production rate at the CERN LHC}},  {\it Phys. Rev. D} {\bf 74} (2006) 035009, [\href{http://arxiv.org/abs/hep-ph/0606172}{{\tt hep-ph/0606172}}].

\bibitem{Arnold:2009ay}
J.~M. Arnold, M.~Pospelov, M.~Trott, and M.~B. Wise, {\it {Scalar Representations and Minimal Flavor Violation}},  {\it JHEP} {\bf 01} (2010) 073, [\href{http://arxiv.org/abs/0911.2225}{{\tt arXiv:0911.2225}}].

\bibitem{Chen:2009xjb}
C.-H. Chen, {\it {Impact of colored scalars on $D^0 - \bar{D}^0$ mixing in diquark models}},  {\it Phys. Lett. B} {\bf 680} (2009) 133--136, [\href{http://arxiv.org/abs/0902.2620}{{\tt arXiv:0902.2620}}].

\bibitem{Han:2010rf}
T.~Han, I.~Lewis, and Z.~Liu, {\it {Colored Resonant Signals at the LHC: Largest Rate and Simplest Topology}},  {\it JHEP} {\bf 12} (2010) 085, [\href{http://arxiv.org/abs/1010.4309}{{\tt arXiv:1010.4309}}].

\bibitem{Grinstein:2011dz}
B.~Grinstein, A.~L. Kagan, J.~Zupan, and M.~Trott, {\it {Flavor Symmetric Sectors and Collider Physics}},  {\it JHEP} {\bf 10} (2011) 072, [\href{http://arxiv.org/abs/1108.4027}{{\tt arXiv:1108.4027}}].

\bibitem{Giudice:2011ak}
G.~F. Giudice, B.~Gripaios, and R.~Sundrum, {\it {Flavourful Production at Hadron Colliders}},  {\it JHEP} {\bf 08} (2011) 055, [\href{http://arxiv.org/abs/1105.3161}{{\tt arXiv:1105.3161}}].

\bibitem{Li:2013vlx}
X.-Q. Li, Y.-D. Yang, and X.-B. Yuan, {\it {Exclusive radiative B-meson decays within minimal flavor-violating two-Higgs-doublet models}},  {\it Phys. Rev. D} {\bf 89} (2014), no.~5 054024, [\href{http://arxiv.org/abs/1311.2786}{{\tt arXiv:1311.2786}}].

\bibitem{Cheng:2015lsa}
X.-D. Cheng, X.-Q. Li, Y.-D. Yang, and X.~Zhang, {\it {${B}_{s,d}\;-\;{\bar{B}}_{s,d}$ mixings and ${B}_{s,d}\;\to \;{{\ell }}^{+}{{\ell }}^{-}$ decays within the Manohar-Wise model}},  {\it J. Phys. G} {\bf 42} (2015), no.~12 125005, [\href{http://arxiv.org/abs/1504.00839}{{\tt arXiv:1504.00839}}].

\bibitem{Chang:2015rva}
Q.~Chang, P.-F. Li, and X.-Q. Li, {\it {${B_s^0}$ {\textendash} ${\bar{B}}_s^0$ mixing within minimal flavor-violating two-Higgs-doublet models}},  {\it Eur. Phys. J. C} {\bf 75} (2015), no.~12 594, [\href{http://arxiv.org/abs/1505.03650}{{\tt arXiv:1505.03650}}].

\bibitem{Cho:2017jym}
N.~Cho, X.-q. Li, F.~Su, and X.~Zhang, {\it {$K^0-\bar{K}^0$ mixing in the minimal flavor-violating two-Higgs-doublet models}},  {\it Adv. High Energy Phys.} {\bf 2017} (2017) 2863647, [\href{http://arxiv.org/abs/1705.07638}{{\tt arXiv:1705.07638}}].

\bibitem{Crivellin:2023saq}
A.~Crivellin and M.~Kirk, {\it {Diquark explanation of $b\to s\ensuremath{\ell}^+\ensuremath{\ell}^-$}},  {\it Phys. Rev. D} {\bf 108} (2023), no.~11 L111701, [\href{http://arxiv.org/abs/2309.07205}{{\tt arXiv:2309.07205}}].

\bibitem{Buchmuller:1986zs}
W.~Buchmuller, R.~Ruckl, and D.~Wyler, {\it {Leptoquarks in Lepton - Quark Collisions}},  {\it Phys. Lett. B} {\bf 191} (1987) 442--448. [Erratum: Phys.Lett.B 448, 320--320 (1999)].

\bibitem{Dorsner:2016wpm}
I.~Dor\v{s}ner, S.~Fajfer, A.~Greljo, J.~F. Kamenik, and N.~Ko\v{s}nik, {\it {Physics of leptoquarks in precision experiments and at particle colliders}},  {\it Phys. Rept.} {\bf 641} (2016) 1--68, [\href{http://arxiv.org/abs/1603.04993}{{\tt arXiv:1603.04993}}].

\bibitem{Han:2009ya}
T.~Han, I.~Lewis, and T.~McElmurry, {\it {QCD Corrections to Scalar Diquark Production at Hadron Colliders}},  {\it JHEP} {\bf 01} (2010) 123, [\href{http://arxiv.org/abs/0909.2666}{{\tt arXiv:0909.2666}}].

\bibitem{Pascual-Dias:2020hxo}
B.~Pascual-Dias, P.~Saha, and D.~London, {\it {LHC Constraints on Scalar Diquarks}},  {\it JHEP} {\bf 07} (2020) 144, [\href{http://arxiv.org/abs/2006.13385}{{\tt arXiv:2006.13385}}].

\bibitem{LHCb:2014rck}
{\bf LHCb} Collaboration, R.~Aaij et~al., {\it {Measurement of the ratio of $B_c^+$ branching fractions to $J/\psi\pi^+$ and $J/\psi\mu^+\nu_\mu$}},  {\it Phys. Rev. D} {\bf 90} (2014), no.~3 032009, [\href{http://arxiv.org/abs/1407.2126}{{\tt arXiv:1407.2126}}].

\bibitem{Denner:1992vza}
A.~Denner, H.~Eck, O.~Hahn, and J.~Kublbeck, {\it {Feynman rules for fermion number violating interactions}},  {\it Nucl. Phys. B} {\bf 387} (1992) 467--481.

\bibitem{Buras:2000if}
A.~J. Buras, M.~Misiak, and J.~Urban, {\it {Two loop QCD anomalous dimensions of flavor changing four quark operators within and beyond the standard model}},  {\it Nucl. Phys. B} {\bf 586} (2000) 397--426, [\href{http://arxiv.org/abs/hep-ph/0005183}{{\tt hep-ph/0005183}}].

\bibitem{Buras:2012gm}
A.~J. Buras and J.~Girrbach, {\it {Completing NLO QCD Corrections for Tree Level Non-Leptonic $\Delta F = 1$ Decays Beyond the Standard Model}},  {\it JHEP} {\bf 02} (2012) 143, [\href{http://arxiv.org/abs/1201.2563}{{\tt arXiv:1201.2563}}].

\bibitem{Fierz:1939zz}
M.~Fierz, {\it {Force-free particles with any spin}},  {\it Helv. Phys. Acta} {\bf 12} (1939) 3--37.

\bibitem{Buras:2020xsm}
A.~Buras, {\it {Gauge Theory of Weak Decays}}.
\newblock Cambridge University Press, 6, 2020.

\bibitem{Christensen:2008py}
N.~D. Christensen and C.~Duhr, {\it {FeynRules - Feynman rules made easy}},  {\it Comput. Phys. Commun.} {\bf 180} (2009) 1614--1641, [\href{http://arxiv.org/abs/0806.4194}{{\tt arXiv:0806.4194}}].

\bibitem{Alloul:2013bka}
A.~Alloul, N.~D. Christensen, C.~Degrande, C.~Duhr, and B.~Fuks, {\it {FeynRules 2.0 - A complete toolbox for tree-level phenomenology}},  {\it Comput. Phys. Commun.} {\bf 185} (2014) 2250--2300, [\href{http://arxiv.org/abs/1310.1921}{{\tt arXiv:1310.1921}}].

\bibitem{Kublbeck:1990xc}
J.~Kublbeck, M.~Bohm, and A.~Denner, {\it {Feyn Arts: Computer Algebraic Generation of Feynman Graphs and Amplitudes}},  {\it Comput. Phys. Commun.} {\bf 60} (1990) 165--180.

\bibitem{Hahn:2000kx}
T.~Hahn, {\it {Generating Feynman diagrams and amplitudes with FeynArts 3}},  {\it Comput. Phys. Commun.} {\bf 140} (2001) 418--431, [\href{http://arxiv.org/abs/hep-ph/0012260}{{\tt hep-ph/0012260}}].

\bibitem{Mertig:1990an}
R.~Mertig, M.~Bohm, and A.~Denner, {\it {FEYN CALC: Computer algebraic calculation of Feynman amplitudes}},  {\it Comput. Phys. Commun.} {\bf 64} (1991) 345--359.

\bibitem{Shtabovenko:2020gxv}
V.~Shtabovenko, R.~Mertig, and F.~Orellana, {\it {FeynCalc 9.3: New features and improvements}},  {\it Comput. Phys. Commun.} {\bf 256} (2020) 107478, [\href{http://arxiv.org/abs/2001.04407}{{\tt arXiv:2001.04407}}].

\bibitem{Shtabovenko:2023idz}
V.~Shtabovenko, R.~Mertig, and F.~Orellana, {\it {FeynCalc 10: Do multiloop integrals dream of computer codes?}},  {\it Comput. Phys. Commun.} {\bf 306} (2025) 109357, [\href{http://arxiv.org/abs/2312.14089}{{\tt arXiv:2312.14089}}].

\bibitem{Shtabovenko:2025lxq}
V.~Shtabovenko, {\it {FeynCalc 10.2 and FeynHelpers 2: Multiloop calculations streamlined}},  \href{http://arxiv.org/abs/2512.19858}{{\tt arXiv:2512.19858}}.

\bibitem{Patel:2015tea}
H.~H. Patel, {\it {Package-X: A Mathematica package for the analytic calculation of one-loop integrals}},  {\it Comput. Phys. Commun.} {\bf 197} (2015) 276--290, [\href{http://arxiv.org/abs/1503.01469}{{\tt arXiv:1503.01469}}].

\bibitem{Patel:2016fam}
H.~H. Patel, {\it {Package-X 2.0: A Mathematica package for the analytic calculation of one-loop integrals}},  {\it Comput. Phys. Commun.} {\bf 218} (2017) 66--70, [\href{http://arxiv.org/abs/1612.00009}{{\tt arXiv:1612.00009}}].

\bibitem{Tracas:1982gp}
N.~Tracas and N.~Vlachos, {\it {Two Loop Calculations in {QCD} and the $\Delta I = 1/2$ Rule in Nonleptonic Weak Decays}},  {\it Phys. Lett. B} {\bf 115} (1982) 419.

\bibitem{Aebischer:2018acj}
J.~Aebischer, A.~Crivellin, and C.~Greub, {\it {QCD improved matching for semileptonic B decays with leptoquarks}},  {\it Phys. Rev. D} {\bf 99} (2019), no.~5 055002, [\href{http://arxiv.org/abs/1811.08907}{{\tt arXiv:1811.08907}}].

\bibitem{Aebischer:2022aze}
J.~Aebischer and M.~Pesut, {\it {One-loop Fierz transformations}},  {\it JHEP} {\bf 10} (2022) 090, [\href{http://arxiv.org/abs/2208.10513}{{\tt arXiv:2208.10513}}].

\bibitem{ATLAS:2014wva}
{\bf ATLAS, CDF, CMS, D0} Collaboration, {\it {First combination of Tevatron and LHC measurements of the top-quark mass}},  \href{http://arxiv.org/abs/1403.4427}{{\tt arXiv:1403.4427}}.

\bibitem{Charles:2004jd}
{\bf CKMfitter Group} Collaboration, J.~Charles, A.~Hocker, H.~Lacker, S.~Laplace, F.~R. Le~Diberder, J.~Malcles, J.~Ocariz, M.~Pivk, and L.~Roos, {\it {CP violation and the CKM matrix: Assessing the impact of the asymmetric $B$ factories}},  {\it Eur. Phys. J. C} {\bf 41} (2005), no.~1 1--131, [\href{http://arxiv.org/abs/hep-ph/0406184}{{\tt hep-ph/0406184}}]. Updated results and plots available at: http://ckmfitter.in2p3.fr.

\bibitem{ckm2025}
{\bf CKMfitter Group} Collaboration {\it \href{http://ckmfitter.in2p3.fr/www/results/plots_summer25/ckm_res_summer25.html}{CKMfitter Summer 2025 results}} (2025).

\bibitem{Rosner:2015wva}
J.~L. Rosner, S.~Stone, and R.~S. Van~de Water, {\it {Leptonic Decays of Charged Pseudoscalar Mesons - 2015}},  \href{http://arxiv.org/abs/1509.02220}{{\tt arXiv:1509.02220}}.

\bibitem{Bharucha:2015bzk}
A.~Bharucha, D.~M. Straub, and R.~Zwicky, {\it {$B\to V\ell^+\ell^-$ in the Standard Model from light-cone sum rules}},  {\it JHEP} {\bf 08} (2016) 098, [\href{http://arxiv.org/abs/1503.05534}{{\tt arXiv:1503.05534}}].

\bibitem{Dimou:2012un}
M.~Dimou, J.~Lyon, and R.~Zwicky, {\it {Exclusive Chromomagnetism in heavy-to-light FCNCs}},  {\it Phys. Rev. D} {\bf 87} (2013), no.~7 074008, [\href{http://arxiv.org/abs/1212.2242}{{\tt arXiv:1212.2242}}].

\bibitem{Arthur:2010xf}
R.~Arthur, P.~A. Boyle, D.~Brommel, M.~A. Donnellan, J.~M. Flynn, A.~Juttner, T.~D. Rae, and C.~T.~C. Sachrajda, {\it {Lattice Results for Low Moments of Light Meson Distribution Amplitudes}},  {\it Phys. Rev. D} {\bf 83} (2011) 074505, [\href{http://arxiv.org/abs/1011.5906}{{\tt arXiv:1011.5906}}].

\bibitem{RQCD:2019osh}
{\bf RQCD} Collaboration, G.~S. Bali, V.~M. Braun, S.~B\"urger, M.~G\"ockeler, M.~Gruber, F.~Hutzler, P.~Korcyl, A.~Sch\"afer, A.~Sternbeck, and P.~Wein, {\it {Light-cone distribution amplitudes of pseudoscalar mesons from lattice QCD}},  {\it JHEP} {\bf 08} (2019) 065, [\href{http://arxiv.org/abs/1903.08038}{{\tt arXiv:1903.08038}}]. [Addendum: JHEP 11, 037 (2020)].

\bibitem{Chetyrkin:2000yt}
K.~G. Chetyrkin, J.~H. Kuhn, and M.~Steinhauser, {\it {RunDec: A Mathematica package for running and decoupling of the strong coupling and quark masses}},  {\it Comput. Phys. Commun.} {\bf 133} (2000) 43--65, [\href{http://arxiv.org/abs/hep-ph/0004189}{{\tt hep-ph/0004189}}].

\bibitem{Czakon:2004bu}
M.~Czakon, {\it {The Four-loop QCD beta-function and anomalous dimensions}},  {\it Nucl. Phys. B} {\bf 710} (2005) 485--498, [\href{http://arxiv.org/abs/hep-ph/0411261}{{\tt hep-ph/0411261}}].

\bibitem{Boyd:1994tt}
C.~G. Boyd, B.~Grinstein, and R.~F. Lebed, {\it {Constraints on form-factors for exclusive semileptonic heavy to light meson decays}},  {\it Phys. Rev. Lett.} {\bf 74} (1995) 4603--4606, [\href{http://arxiv.org/abs/hep-ph/9412324}{{\tt hep-ph/9412324}}].

\bibitem{Boyd:1995sq}
C.~G. Boyd, B.~Grinstein, and R.~F. Lebed, {\it {Model independent determinations of $ \bar{B}\to D (lepton), D^* (lepton)$ anti-neutrino form-factors}},  {\it Nucl. Phys. B} {\bf 461} (1996) 493--511, [\href{http://arxiv.org/abs/hep-ph/9508211}{{\tt hep-ph/9508211}}].

\bibitem{Boyd:1997kz}
C.~G. Boyd, B.~Grinstein, and R.~F. Lebed, {\it {Precision corrections to dispersive bounds on form-factors}},  {\it Phys. Rev. D} {\bf 56} (1997) 6895--6911, [\href{http://arxiv.org/abs/hep-ph/9705252}{{\tt hep-ph/9705252}}].

\bibitem{Cui:2023jiw}
B.-Y. Cui, Y.-K. Huang, Y.-M. Wang, and X.-C. Zhao, {\it {Shedding new light on $R(D_{(s)}^{(*)})$ and $|V_{cb}| $ from semileptonic $B_{(s)}\to D_{(s)}^{(*)}\ell\bar{\nu}_{\ell}$ decays}},  {\it Phys. Rev. D} {\bf 108} (2023), no.~7 L071504, [\href{http://arxiv.org/abs/2301.12391}{{\tt arXiv:2301.12391}}].

\bibitem{Bernlochner:2017jka}
F.~U. Bernlochner, Z.~Ligeti, M.~Papucci, and D.~J. Robinson, {\it {Combined analysis of semileptonic $B$ decays to $D$ and $D^*$: $R(D^{(*)})$, $|V_{cb}|$, and new physics}},  {\it Phys. Rev. D} {\bf 95} (2017), no.~11 115008, [\href{http://arxiv.org/abs/1703.05330}{{\tt arXiv:1703.05330}}]. [Erratum: Phys.Rev.D 97, 059902 (2018)].

\bibitem{Bernlochner:2022ywh}
F.~U. Bernlochner, Z.~Ligeti, M.~Papucci, M.~T. Prim, D.~J. Robinson, and C.~Xiong, {\it {Constrained second-order power corrections in HQET: $R(D^{(*)})$, $|V_{cb}|$, and new physics}},  {\it Phys. Rev. D} {\bf 106} (2022), no.~9 096015, [\href{http://arxiv.org/abs/2206.11281}{{\tt arXiv:2206.11281}}].

\bibitem{Harrison:2025kxm}
J.~Harrison, {\it {Update of HPQCD $B_c\to J/\psi$ Form Factors}},  {\it PoS} {\bf LATTICE2024} (2025) 257, [\href{http://arxiv.org/abs/2501.15180}{{\tt arXiv:2501.15180}}].

\bibitem{Bjorken:1988kk}
J.~D. Bjorken, {\it {Topics in $B$ Physics}},  {\it Nucl. Phys. B Proc. Suppl.} {\bf 11} (1989) 325--341.

\bibitem{Neubert:1997uc}
M.~Neubert and B.~Stech, {\it {Nonleptonic weak decays of $B$ mesons}},  {\it Adv. Ser. Direct. High Energy Phys.} {\bf 15} (1998) 294--344, [\href{http://arxiv.org/abs/hep-ph/9705292}{{\tt hep-ph/9705292}}].

\end{thebibliography}\endgroup

\end{document}